\documentclass[pre,11pt,nofootinbib]{revtex4}
\usepackage{amsmath,graphicx}
\usepackage{hyperref}
\hypersetup{colorlinks=true,linkcolor=blue,urlcolor=blue,citecolor=red}

\newcommand{\NDOF}{{N_{_{\rm DOF}}}}
\newcommand{\nb}{{N_{\rm boot}}}

\newcommand{\mc}{_{_{\rm MC}}}
\newcommand{\xb}{x^B}
\newcommand{\yb}{y^B}
\newcommand{\zb}{z^B}
\newcommand{\fb}{f^B}

\newcommand{\ie}{{\em i.e.}}

\begin{document}
\title{{\Large Everything you wanted to know about Data Analysis and Fitting 
but were afraid to ask} \\ 

}
\author{Peter Young}
\date{\today}

\begin{abstract}
These notes discuss, in a style intended for physicists,
how to average data and fit it to some functional form.
I try to make clear what is being calculated, what assumptions are being
made, and to give a derivation of results rather than just quote them. 
The aim is put a lot useful pedagogical material together in a
convenient place. 
\end{abstract}

\maketitle
\tableofcontents

\section{Introduction}
These notes describe how to average and fit numerical data that you have
obtained, presumably by some simulation.

%
%
%

Typically you will generate a set of values $x_i,\, y_i, \cdots,\, i = 1,
\cdots N$, where $N$ is the number of measurements. The first thing you will
want to do is to estimate various average values, and determine \textit{error
bars} on those estimates. As we shall see,
this is straightforward if one wants to compute a
single average, e.g. $\langle x \rangle$, but not quite so easy for more
complicated averages such as fluctuations in a quantity, $\langle x^2 \rangle
- \langle x \rangle^2$, or combinations of measured values such as $\langle y
\rangle / \langle x \rangle^2$.  Averaging of data will be discussed in
Sec.~\ref{sec:averages}.

Having obtained several good data points with error bars,
you might want to fit this data to some model.
Techniques for fitting data will be
described in the second part of these notes in Sec.~\ref{sec:fit}

I find that the books on these topics usually fall into one of two camps.
At one extreme,
the books for physicists don't discuss all that is needed and rarely
\textit{prove} the results that they quote. At the other extreme, 
the books for mathematicians
presumably prove everything but are written in a style of lemmas, proofs,
$\epsilon$'s and $\delta$'s, and unfamiliar notation, which is intimidating to
physicists. One exception, which finds a good middle ground,
is Numerical Recipes~\cite{press:92} and the
discussion of fitting given here is certainly influenced by Chap.\ 15 of that
book. In these notes I aim to be fairly complete and also to derive the
results I use, while the style is that of a physicist writing for physicists.
I also include scripts in python, perl, and gnuplot to perform certain
tasks in data analysis and fitting. 
For these reasons, these notes are perhaps rather lengthy.  Nonetheless, I
hope, that they will provide a useful reference.

\section{Averages and error bars}
\label{sec:averages}

\subsection{Basic Analysis}
\label{sec:basic}

A reference for the material in this subsection is the book by
Taylor~\cite{taylor:97}.

Suppose we have a set of 
data from a simulation, $x_i, \, (i = 1, \cdots, N)$, which
we shall refer to as a \textit{sample} of data.
This data will have some random noise so the $x_i$ are not all equal.
Rather they are governed by a distribution
$P(x)$, \textit{which we don't know}.

The distribution is normalized,
\begin{equation}
\int_{-\infty}^\infty P(x) \, d x = 1,
\end{equation}
and is usefully characterized by its moments, where
the $n$-th moment is defined by
\begin{equation}
\langle x^n \rangle = \int_{-\infty}^\infty x^n\, P(x) \, d x\, .
\end{equation}
We will denote the average 
\textit{over the exact distribution} by angular brackets.
Of particular interest are the first and second moments from which one forms
the mean $ \mu$ and variance $\sigma^2$, by
\begin{subequations}
\begin{align}
\mu &\equiv \langle x \rangle \label{xavexact} \\
\sigma^2 &\equiv \langle \, \left(x - \langle x\rangle\right)^2 \,
\rangle = \langle x^2 \rangle - \langle x \rangle^2
\, .
\label{sigma}
\end{align}
\end{subequations}
The term ``standard deviation''
is used for $\sigma$, the square root of the variance.

In this section we will estimate the mean
$\langle x \rangle$, and the uncertainty in our estimate,
from the $N$ data points $x_i$. The determination of more complicated
averages
and resulting error bars will be discussed in Sec.~\ref{sec:advanced}

In order to obtain error bars we need to assume that the data are uncorrelated with each
other. This is a crucial assumption, without which it is very difficult
to proceed. However, it is not always clear if the data points are truly
independent of each other; some correlations may be present but not
immediately obvious. Here, we take the usual approach of assuming that
even if there are some correlations, they are sufficiently weak
so as not to 
significantly perturb the results of the analysis. In Monte Carlo simulations,
measurements which differ by a sufficiently large number of Monte Carlo sweeps
will be uncorrelated. More
precisely the difference in sweep numbers should be greater 
than a ``relaxation time''.
This is exploited in the ``binning'' method in which the
data used in the analysis is not the individual measurements, but rather an
average over measurements during a range of
Monte Carlo sweeps, called a ``bin''.
If the bin size is greater than the relaxation time, results from adjacent
bins will be (almost) uncorrelated. A pedagogical treatment of binning has
been given by Ambegaokar and Troyer~\cite{ambegaokar:09}. Alternatively, one
can do independent Monte Carlo runs, requilibrating each time, and use, as 
individual data in the analysis, the average from each run.

The information \textit{from the data}
is usefully encoded in two parameters, the sample
mean $\overline{x}$ and the sample standard
deviation $s$ which are defined by\footnote{The
factor of $N$ is often replaced by $N-1$ in the expression
for the sample variance in Eq.~(\ref{sigmafromdata}).
We note, though, that
the final answer for the error bar on the mean,
Eq.~\eqref{finalans2} below,
will be independent of how the intermediate quantity $s^2$
is defined. The rationale for $N-1$ is that
the $N$ terms in Eq.~(\ref{sigmafromdata}) are not all
independent since $\overline{x}$, which depends on all the $x_i$, is
subtracted. Rather, as will be discussed more in the section on fitting,
Sec.~\ref{sec:fit}, there are really only $N-1$ independent variables (called
the ``number of degrees of freedom'' in the fitting context) and so dividing
by $N-1$ rather than $N$ also has a rational basis. Here we prefer to use
$N$.
}
\begin{subequations}
\begin{align}
\overline{x} & = {1 \over N} \sum_{i=1}^N x_i \, ,
\label{meanfromdata}
\\
s^2 & =  {1 \over N} \sum_{i=1}^N \left( x_i -
\overline{x}\right)^2 \, .
\label{sigmafromdata}
\end{align}
\end{subequations}

In statistics, notation is often confusing but crucial to
understand. Here, an average indicated by an over-bar,
$\overline{\cdots}$, is an average over the \textit{sample of $N$ data
points}. This is to be distinguished from an exact average over the
distribution $\langle \cdots \rangle$, as in Eqs.~(\ref{xavexact}) and
(\ref{sigma}).
The latter is, however, just a
theoretical construct since we \textit{don't know}
the distribution $P(x)$, only
the set of $N$ data points $x_i$ which have been sampled
from it.

Next we derive two simple results which will be useful later:
\begin{enumerate}
\item
The mean of the sum of $N$ independent variables \textit{with the
same distribution} is
$N$ times the mean of a single variable, and
\item
The variance of the sum of $N$ independent variables \textit{with
the same distribution} is $N$ times the variance of a single variable.
\end{enumerate}
The result for the mean is obvious since, defining $X = \sum_{i=1}^N x_i$,
\begin{equation}
\mu_X \equiv 
\langle X \rangle = \sum_{i=1}^N \langle x_i \rangle = N \langle x_i \rangle
\ \boxed{ = N \mu\, .}
\label{X}
\end{equation}
The result for the standard deviation needs a little more work:
\begin{subequations}
\begin{align}
\sigma_X^2 & \equiv \langle X^2 \rangle - \langle X \rangle^2  \\
&= \sum_{i,j=1}^N \left( \langle x_i x_j\rangle - \langle
x_i \rangle \langle x_j \rangle \right) \label{1} \\
& = \sum_{i=1}^N \left( \langle x_i^2 \rangle - \langle x_i
\rangle^2 \right) \label{2} \\
& = N \left(\langle x^2 \rangle - \langle x \rangle^2 \right)
\\
& \boxed{ = N \sigma^2  \, .}
\label{dXsq}
\end{align}
\end{subequations}
To get from Eq.~(\ref{1}) to Eq.~(\ref{2}) we note that, for $i \ne j$,
$\langle x_i x_j\rangle = \langle x_i \rangle \langle
x_j\rangle$ since $x_i$ and $x_j$ are assumed to be statistically
independent. (This is where the statistical independence of the data is
needed.) 

If the means and standard deviations are not all the same, then the above
results generalize to
\begin{subequations}
\begin{align}
\mu_X &= \sum_{i=1}^N \mu_i \, , \\
\sigma_X^2 &= \sum_{i=1}^N \sigma_i^2 \, .
\end{align}
\end{subequations}

\begin{center}
\begin{figure}
\includegraphics[width=9.5cm]{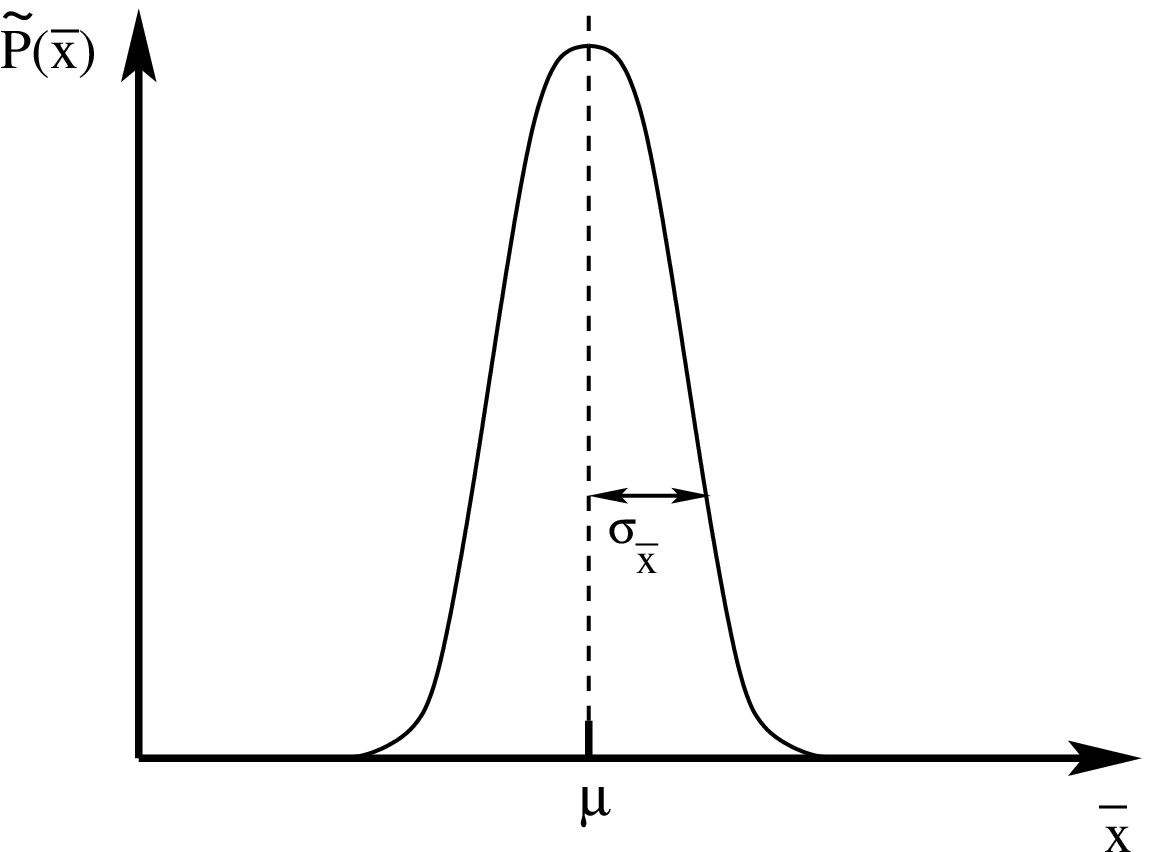}
\caption{
The distribution of results for the sample mean $\overline{x}$ obtained by
repeating the measurements of the $N$ data points $x_i$ many times. The
average of this distribution is $\mu$, the exact average value of $x$. The
mean, $\overline{x}$, obtained from one sample of data typically differs from 
$\mu$ by an amount of order $\sigma_{\overline{x}}$, the standard
deviation of the distribution $\widetilde{P}(\overline{x})$.
}
\label{Fig:distofmean}
\end{figure}
\end{center}

Now we describe an important thought experiment.
Let's \textit{suppose} that
we could repeat the set of $N$ measurements \textit{very many}
many times, each
time obtaining a value of the sample average $\overline{x}$. From these
results we could
construct a distribution, $\widetilde{P}(\overline{x})$,
for the sample average as shown in
Fig.~\ref{Fig:distofmean}.

If we do enough repetitions
we are effectively
averaging over the exact distribution.
Hence the average of the sample mean, $\overline{x}$, over very many
repetitions of the data, is given by
\begin{equation}
\langle \overline{x} \rangle = {1 \over N} \sum_{i=1}^N \langle x_i \rangle =
\langle x \rangle  \equiv \mu \, ,
\label{xav}
\end{equation}
i.e.~it is the exact average over the distribution of $x$,
as one would intuitively expect, see Fig.~\ref{Fig:distofmean}.
Eq.~\eqref{xav} also follows from Eq.~\eqref{X} by noting that $\overline{x} =
X/N$.

In fact, though, we have only the \textit{one} set of
data, so we can not determine $\mu $ exactly. However, 
Eq.~(\ref{xav}) shows that
\begin{equation}
\boxed{ \mbox{the best estimate of\ } \mu \mbox{ is }
\overline{x},}
\label{xbarest}
\end{equation}
i.e.~the sample mean,
since averaging the sample mean over many repetitions of the $N$ data
points gives the true mean
of the distribution, $\mu$.
An estimate like this, which gives the exact result if averaged over
many repetitions of the experiment, is said to be
\fbox{unbiased.}

We would also like an estimate of the uncertainty, or ``error bar'', in our
estimate of $\overline{x}$ for the exact average $\mu$.
We take $ \sigma_{\overline{x}}$, the standard deviation in $\overline{x}$
(obtained if one did many repetitions of the $N$ measurements),
to be the uncertainty, or error bar, in $\overline{x}$.
The reason is that $ \sigma_{\overline{x}}$ is
the width of the distribution
$\widetilde{P}(\overline{x})$, shown in Fig.~\ref{Fig:distofmean}, so a
\textit{single} estimate $\overline{x}$ typically differs from
the exact result $\mu$ by an amount of this order.
The variance $\sigma_{\overline{x}}^2$ is given by
\begin{equation}
\sigma_{\overline{x}}^2 \equiv \langle \overline{x}^2 \rangle
- \langle \overline{x} \rangle^2 = {\sigma^2 \over N}\, ,
\label{dxsq}
\end{equation}
which follows from Eq.~\eqref{dXsq} with $\overline{x} =X / N$.

The problem with Eq.~(\ref{dxsq}) is that
\textbf{we don't know $\sigma^2$} since it
is a function of the exact distribution $P(x)$.
We do, however, know the
\textit{sample} variance $s^2$, see
Eq.~(\ref{sigmafromdata}), and the average of this
over many repetitions of the $N$ data points,
is equal to $\sigma^2$ since
\begin{subequations}
\begin{align}
\langle s^2 \rangle & = 
{1 \over N} \sum_{i=1}^N \langle x_i^2 \rangle  -
{1 \over N^2} \sum_{i=1}^N \sum_{j=1}^N \langle x_i x_j \rangle
\label{3}  \\
& = \langle x^2 \rangle - {1 \over N^2} \left[ N(N-1) \langle x \rangle^2 +
N \langle x^2\rangle \right] \label{4}\\
& = {N-1 \over N}\,\left[\langle x^2 \rangle - \langle x \rangle^2 \right] \\
& = {N-1 \over N}\, \sigma^2 \, . 
\label{5}
\end{align}
\end{subequations}
To get from Eq.~(\ref{3}) to Eq.~(\ref{4}), we have separated the terms
with $i=j$ in the last term of Eq.~(\ref{3}) from those with $i \ne j$,
and used the fact that each of the $x_i$ is chosen from the same
distribution and is statistically independent of the others.
It follows from Eq.~(\ref{5}) that
\begin{equation}
\boxed{
\mbox{the best estimate of\ } \sigma^2 \mbox{ is }
{N \over N - 1} \,s^2 \, ,}
\label{sigmasamp}
\end{equation}
since averaging $s^2$ over many repetitions of
$N$ data points gives $\sigma^2$.
The estimate for $\sigma^2$ in
Eq.~(\ref{sigmasamp}) is
therefore unbiased. Note that the expression for $s^2$ in
Eq,~\eqref{meanfromdata} is a
sum of positive terms, so it is ``self-averaging'', which means that the
deviation of the result for one sample of $N$ data points from the average
over many data sets ($\sigma^2$ in this case) tends to zero for $N
\to \infty$.

Combining Eqs.~(\ref{dxsq}) and (\ref{sigmasamp}) gives
\begin{equation}
\boxed{
\mbox{the best estimate of\ } \sigma_{\overline{x}}^2 \mbox{ is }
{s^2  \over N-1}\; \, ,}
\label{errorbar}
\end{equation}
\fbox{since this estimate is also unbiased.}
We have now obtained, using only information from the data,
that the mean is given by
\begin{equation}
\boxed{
\mu = \overline{x}\; \pm \; \sigma_{\overline{x}} \, ,}
\end{equation}
where
\begin{equation}
\boxed{
\sigma_{\overline{x}} = {s \over \sqrt{N-1}}\, , }
\label{finalans}
\end{equation}
which we can write explicitly in terms of the data points as
\begin{equation}
\boxed{
\sigma_{\overline{x}} = \left[
{1 \over N(N-1)} \, \sum_{i=1}^N (x_i -  \overline{x})^2 \right]^{1/2} \, .}
\label{finalans2}
\end{equation}
Remember that $\overline{x}$ and $s$ are the mean and standard deviation
of the
(one set) of data that is available to us, see Eqs.~(\ref{meanfromdata})
and (\ref{sigmafromdata}).

As an example, suppose $N=5$ and the data points are 
\begin{equation}
x_i = 10, 11, 12, 13, 14,
\end{equation}
(not very random looking data it must be admitted!). Then, from
Eq.~(\ref{meanfromdata})
we have $\overline{x} = 12$, and from Eq.~(\ref{sigmafromdata})
\begin{equation}
s^2 = {1 \over 5} \, \left[(-2)^2 + (-1)^2 + 0^2 + 1^2 + 2^2\right] =
2 . 
\end{equation}
Hence, from Eq.~(\ref{finalans}),
\begin{equation}
\sigma_{\overline{x}} = {1 \over \sqrt{4}}\, \sqrt{2} = {1\over
\sqrt{2}}.
\end{equation}
so
\begin{equation}
\mu = \overline{x} \pm \sigma_{\overline{x}} =  12 \pm {1\over \sqrt{2}}.
\end{equation}

How does the error bar decrease with the number of statistically independent
data points $N$?
Equation (\ref{5})
shows that $s^2$ does not vary systematically with $N$, at large $N$ (where we
neglect the factor of $-1$ compared with $N$)
and so
from Eq.~(\ref{finalans}) we see that
\begin{quotation}
\fbox{
the error bar in the mean goes down like $1/\sqrt{N}$ for large $N$.}
\end{quotation}
Hence,
to reduce the error bar by a factor of 10 one needs
100 times as much data.
This is discouraging, but is a fact of life when
dealing with random noise.

For Eq.~(\ref{finalans}) to be really
useful we need to know the probability that the true answer
$\mu$
lies more than $\sigma_{\overline{x}}$ away from our estimate $\overline{x}$.
Fortunately, for large $N$, the central limit theorem, derived in Appendix
\ref{sec:clt}, tells us (for
distributions where the first two moments are finite) that
the distribution of $\overline{x}$ is a Gaussian. For this distribution we
know that the probability of finding a result more than one standard deviation
away from the mean is 32\%, more than two standard deviations is $4.5\%$ and
more than three standard deviations is $0.3\%$. Hence we expect that most of
the time $\overline{x}$ will be within $\sigma_{\overline{x}}$ of the correct
result $\mu$, and only occasionally
will be more than two times
$\sigma_{\overline{x}}$ from it. Even if $N$ is not very
large, so there are some deviations from the
Gaussian form, the above numbers are often a
reasonable guide.

However, as emphasized in appendix~\ref{sec:clt},
distributions which occur in nature typically have much more weight in the tails
than a Gaussian. As a result,
the weight in the tails of
the distribution \textit{of the sum} can also
be much larger than for a Gaussian even
for quite large values of $N$, see Fig.~\ref{Fig:converge_to_clt}. It follows that
the probability of an ``outlier'' can be much higher than that predicted for a
Gaussian distribution, as anyone who has invested in the stock market knows
well! 

We conclude this subsection by discussing the situation when there are several
random variables, $x, y, z, \cdots$, for which we generate a sample of data: $(x_i,
y_i, z_i, \cdots)$ with $i = 1, 2, \cdots, N$.  We indicate the means and standard
deviations of the different variables by suffices, i.e.
\begin{subequations}
\begin{align}
\mu_x &\equiv \langle x \rangle \, \\
\sigma^2_{x} &\equiv \langle x^2 \rangle - \langle x \rangle^2 \, ,
\label{sig_xx}
\end{align}
\end{subequations}
for averages over the exact distribution, and
\begin{align}
s_{x}^2 & \equiv  {1 \over N } \sum_{i=1}^N \left( x_i -
\overline{x}\right)^2 \, .
\label{s_xx}
\end{align}
for the sample variance. The main new feature is the appearance of
cross-correlations between different variables. One defines the ``covariance''
of $x$ and $y$ by
\begin{equation}
{\rm Cov}(x, y) \equiv \langle x y \rangle - \langle x \rangle \langle y \rangle
= \langle \left(\, x - \langle x \rangle \right)\,\left(y - \langle y \rangle\right)\, \rangle \, .
\end{equation}
It is convenient to have a more compact notation for the covariance, analogous
to that in Eq.~\eqref{sig_xx} for the variance. I use the
notation $\sigma^2_{xy}$ for the covariance of $x$ and $y$, i.e.
\begin{equation}
\sigma_{xy}^2 \equiv 
\langle \left(\, x - \langle x \rangle \right)\,\left(y - \langle y \rangle\right)\, \rangle,\, .
\label{sig_xy}
\end{equation}
This notation is not ideal since there is no guarantee that the
covariance $\sigma^2_{xy}$ is
positive.\footnote{One should therefore think of $\sigma_{xy}^2$ as a single quantity, rather
than the square of something, just as $\chi^2$, discussed extensively in the
section on fitting below, is never regarded as the square of an object called
$\chi$. Admittedly, though, $\chi^2$ can not be negative.}
The standard notation is to write
the covariance
of $x$ and $y$ as $\sigma_{xy}$ (no
square), but I find this even more confusing. 

By analogy to
Eq.~\eqref{sig_xy} I write the sample
covariance of $x$ and $y$ as
\begin{equation}
s_{xy}^2 \equiv  {1 \over N } \sum_{i=1}^N
\left( x_i - \overline{x}\right) \, \left(y_i - \overline{y}\right)\, .
\label{s_xy}
\end{equation}



\subsection{Advanced Analysis}
\label{sec:advanced}

In Sec.~\ref{sec:basic} we learned how to estimate a simple average, such as
$\mu_x \equiv \langle x \rangle$,
plus the error bar in that quantity, from a set of data
$x_i$. Trivially this method also applies to a \textit{linear} combination of
different averages, $\mu_x, \mu_y, \cdots$ etc.
However, we often need more complicated, \textit{non-linear} functions of
averages. One example is
the fluctuations in a quantity, i.e.\ $\langle x^2 \rangle -
\langle x \rangle^2$. Another example is a dimensionless combination of
moments, which gives information about the \textit{shape} of a
distribution independent of its overall scale. Such quantities are very
popular in finite-size scaling (FSS) analyses since the FSS form is simpler
than for quantities with dimension. An popular example, first proposed by
Binder, is $\langle x^4 \rangle / \langle x^2 \rangle^2$, which is known as the
``kurtosis'' (frequently a factor of 3 is subtracted to make it zero for a
Gaussian).

Hence, in this section we consider how to determine 
\textit{non-linear functions} of averages of
one or more variables,
$f(\mu_y, \mu_z, \cdots)$,
where
\begin{equation}
\mu_y \equiv \langle y \rangle \, ,
\end{equation}
etc.  For
example, the two quantities mentioned in the previous paragraph correspond to
\begin{equation}
f(\mu_y, \mu_z) = \mu_y - \mu_z^2 \, ,
\end{equation}
with $y=x^2$ and $z = x$ and
\begin{equation}
f(\mu_y, \mu_z) = {\mu_y \over \mu_z^2} \, ,
\end{equation}
with $y = x^4$ and $z = x^2$.

The natural estimate of $f(\mu_y, \mu_z)$ from the
sample data is clearly $f(\overline{y}, \overline{z} )$. However, it will take
some more thought to estimate the error bar in this quantity. The traditional
way of doing this is called ``error propagation'', described in
Sec.~\ref{sec:traditional} below and Ch.~3 of Ref.~\cite{taylor:97}.
However, it is now more
common to use either ``jackknife'' or ``bootstrap'' procedures, described in
Secs.~\ref{sec:jack} and \ref{sec:boot}.
At the price of some additional computation, which is no
difficulty when done on a modern computer (though it would have been tedious in the
old days when statistics calculations were done by hand), these methods
automate the calculation of the error bar.

In addition,
the estimate of $f(\mu_y, \mu_z)$
turns out to have some \textit{bias} if $f$ is a non-linear function.
Usually this is small effect because it is
order $1/N$, see for example Eq.~\eqref{bias} below,
whereas the statistical error is of order $1/\sqrt{N}$. Since 
$N$ is usually large, the bias is generally much less than the statistical
error and so can generally be neglected. In any case, the jackknife and
bootstrap methods also enable one to eliminate the leading ($\sim 1/N$)
contribution to the bias in a automatic fashion.

\subsubsection{Traditional method}
\label{sec:traditional}
First we will discuss the traditional method, known as error
propagation~\cite{taylor:97},
to compute the error bar and
bias. We expand $f(\overline{y}, \overline{z})$ about $f(\mu_y, \mu_z)$ up to
second order in the deviations:
\begin{equation}
f(\overline{y}, \overline{z}) = f(\mu_y, \mu_z) 
+ (\partial_{\mu_y}f)\, \delta_{\overline{y}}
+ (\partial_{\mu_z}f)\, \delta_{\overline{z}}
+ {1\over 2}\, (\partial^2_{\mu_y\mu_y}f)\, \delta_{\overline{y}}^2
+ (\partial^2_{\mu_y\mu_z}f)\, \delta_{\overline{y}} \delta_{\overline{z}}
+ {1\over 2}\, (\partial^2_{\mu_z\mu_z}f)\, \delta_{\overline{z}}^2
+ \cdots \, ,
\label{expand}
\end{equation}
where
\begin{equation}
\delta_{\overline{y}} = \overline{y} - \mu_y ,
\end{equation}
etc.

The terms of first order in the $\delta's$ in Eq.~\eqref{expand} give the
leading contribution to the error, but would average to zero if the procedure were
to be repeated many times. However, the terms of second order do not average
to zero and so give the leading contribution to the bias. We now estimate that
bias.

Averaging
Eq.~\eqref{expand} over many repetitions, and noting that
\begin{equation}
\langle \delta_{\overline{y}}^2 \rangle = \langle \overline{y}^2 \rangle - \langle
\overline{y} \rangle^2 \equiv \sigma_{\overline{y}}^2 , \quad
\langle \delta_{\overline{z}}^2 \rangle = \langle \overline{z}^2 \rangle - \langle
\overline{z} \rangle^2 \equiv \sigma_{\overline{z}}^2 , \quad
\langle \delta_{\overline{y}} \delta_{\overline{z}} \rangle =
\langle \overline{y}\, \overline{z} \rangle - \langle \overline{y} \rangle
\langle \overline{z} \rangle \equiv \sigma_{\overline{y}\,\overline{z}}^2 ,
\end{equation}
we get
\begin{equation}
\langle f(\overline{y}, \overline{z})\rangle - f(\mu_y, \mu_z) =
  {1\over 2}\, (\partial^2_{\mu_y\mu_y}f)\, \sigma_{\overline{y}}^2
+ (\partial^2_{\mu_y\mu_z}f)\, \sigma_{\overline{y}\,\overline{z}}^2
+ {1\over 2}\, (\partial^2_{\mu_z\mu_z}f)\, \sigma_{\overline{z}}^2  + \cdots\, .
\label{df}
\end{equation}
As shown in Eq.~\eqref{errorbar}
$\sigma_{\overline{y}}^2$
is $(N-1)^{-1}$ times
the average sample variance 
$\langle s_{y}^2 \rangle$.
Furthermore, as noted below Eq.~\eqref{sigmasamp}, $s_y^2$ is self
averaging, which means that the difference between the value of $s_y^2$ from one
data set and the average over all data sets, $\sigma_y^2$, tends to zero for
$N \to \infty$. Hence we can replace $\sigma_{\overline{y}}^2$ by $(N-1)^{-1}
s_y^2$, and similarly replace $\sigma_{\overline{z}}^2$ by $(N-1)^{-1} s_z^2$. 
In the same way, we can replace
$\sigma_{\overline{y}\,\overline{z}}^2$ by $(N-1)^{-1}$ times
$s_{y z}^2$, the sample \textit{covariance} of $y$ and $z$, defined in
Eq.~\eqref{s_xy}.
Hence, from Eq.~\eqref{df}, we have
\begin{equation}
f(\mu_y, \mu_z) 
=
\langle f(\overline{y}, \overline{z})\rangle 
- {1\over (N-1)}\,
\left[{1\over 2}\, (\partial^2_{\mu_y\mu_y}f)\, s_{y}^2
+ (\partial^2_{\mu_y\mu_z}f)\, s_{y z}^2
+ {1\over 2}\, (\partial^2_{\mu_z\mu_z}f)\, s_{z}^2 \right] + \cdots\, .
\label{bias2}
\end{equation}
The leading contribution to the bias is the $1/(N-1)$ term.
It follows from Eq.~\eqref{bias2}
that if one wants to eliminate the leading contribution to the bias
one should
\begin{equation}
\boxed{
\mbox{estimate } f(\mu_y,\mu_z)\ \mbox{ from }
f(\overline{y}, \overline{z}) -  {1\over (N-1)}\,
\left[{1\over 2}\, (\partial^2_{\mu_y\mu_y}f)\, s_{y}^2
+ (\partial^2_{\mu_y\mu_z}f)\, s_{y z}^2
+ {1\over 2}\, (\partial^2_{\mu_z\mu_z}f)\, s_{z}^2 \right].}
\label{bias}
\end{equation}
As claimed earlier, the bias correction is of order $1/N$. Note that it
vanishes if $f$
is a linear function, as shown in Sec.~\ref{sec:basic}.
The generalization to functions of more than two
averages, $f(\mu_y, \mu_z, \mu_w, \cdots)$, is obvious. 

Next we discuss the leading \textit{error} in using
$f(\overline{y}, \overline{z})$ as an estimate for
$f(\mu_y, \mu_z)$. This comes from the terms linear in the $\delta$'s in
Eq.~\eqref{expand}.
Just including these terms we have
\begin{subequations}
\begin{align}
\langle f(\overline{y}, \overline{z}) \rangle &=
f(\mu_y, \mu_z) \, , \\
\langle\, f^2(\overline{y}, \overline{z})\, \rangle &=
f^2(\mu_y, \mu_z) + 
(\partial_{\mu_y}f)^2 \, \langle \delta_{\overline{y}}^2 \rangle
+ 2(\partial_{\mu_y}f)\, (\partial_{\mu_z}f) \,
\langle \delta_{\overline{y}} \delta_{\overline{z}} \rangle
+ (\partial_{\mu_z}f)^2 \, \langle \delta_{\overline{z}}^2 \rangle  \, .
\end{align}
\end{subequations}
Hence
\begin{align}
\sigma_f^2 
&\equiv 
\langle\, f^2(\overline{y}, \overline{z})\, \rangle -
\langle f(\overline{y}, \overline{z}) \rangle^2
\nonumber \\
&= 
(\partial_{\mu_y}f)^2 \, \langle \delta_{\overline{y}}^2 \rangle
+ 2(\partial_{\mu_y}f)\, (\partial_{\mu_z}f) \,
\langle \delta_{\overline{y}} \delta_{\overline{z}} \rangle
+ (\partial_{\mu_z}f)^2 \, \langle \delta_{\overline{z}}^2 \rangle  \, .
\end{align}
As above, we use $s_{y y}^2 / (N-1)$ as an estimate of $\langle
\delta_{\overline{y}}^2 \rangle$ and similarly for the other terms. Hence 
\begin{equation}
\boxed{
\mbox{the best estimate of }
\sigma_f^2
\mbox{ is } 
  {1 \over (N-1)}\, 
  (\partial_{\mu_y}f)^2 \, s_{y}^2 
+ 2(\partial_{\mu_y}f)\, (\partial_{\mu_z}f) \,
s_{y z}^2 
+ (\partial_{\mu_z}f)^2 \, s_{z}^2 \, .}
\label{sigma_f}
\end{equation}
This estimate is unbiased to leading order in $N$. Note that we need to keep
track not only of fluctuations in $y$ and $z$, characterized by their
variances $s_{y}^2$ and
$s_{z}^2$, but also cross correlations
between $y$ and $z$, characterized by their covariance $s_{y z}^2$.

Hence, still to leading order in $N$, we get
\begin{equation}
\boxed{f(\mu_y, \mu_z) = f(\overline{y}, \overline{z}) \pm \sigma_f\, ,}
\end{equation}
where we estimate the error bar $\sigma_f$ from Eq.~\eqref{sigma_f} which shows
that it is of order $1/\sqrt{N}$.
The generalization to functions of more than two
averages is obvious.

Note that in the simple case studied in Sec.~\ref{sec:basic} where
$f(\mu_x)$ is a linear function, $f =\mu_x$, Eq.~\eqref{bias2} tells us
that there is no bias, which is correct,
and Eq.~\eqref{sigma_f} gives an expression for the error bar
which agrees with Eq.~\eqref{finalans}.

In Eqs.~\eqref{bias} and \eqref{sigma_f} we need to keep track how errors
in the individual quantities like $\overline{y}$ propagate to the estimate of
the function $f$. This requires inputting by hand the various partial
derivatives into the analysis program, and keeping track of all the variances
and covariances. In the next two sections we see how
\textit{resampling} the data automatically takes account of error propagation
without needing to input the partial derivatives and keep
track of variances and covariances. There are two resampling approaches,
called
jackknife and bootstrap, and each provide a \textit{fully automatic} method of
determining error bars and bias.

\subsubsection{Jackknife}
\label{sec:jack}
We define the $i$-th jackknife estimate, $y^J_i\, (i = 1,2, \cdots, N)$
to be the average over all data in the sample
\textit{except the point} $i$, i.e.
\begin{equation}
y^J_i \equiv {1 \over N-1}\, \sum_{j \ne i} y_j\, 
\ \left( = \overline{y} + {1 \over N-1}\, (\overline{y} - y_i) \right)
\, .
\label{yJ_def}
\end{equation}
We also define corresponding jackknife estimates of the function $f$ (again for
concreteness we will assume that $f$ is a function of just 2 averages but the
generalization will be obvious):
\begin{equation}
f^J_i \equiv f(y^J_i, z^J_i) \, .
\label{fJi}
\end{equation}
In other words, we use the jackknife values, $y^J_i, z^J_i$, rather than the
sample means, $\overline{y}, \overline{z}$, as the arguments of $f$. For
example a jackknife estimate of
the Binder ratio $\langle x^4 \rangle / \langle x^2 \rangle^2$ is
\begin{equation}
f^J_i = {(N-1)^{-1} \sum_{j, (j \ne i)} x_j^4 \over
\left[(N-1)^{-1} \sum_{j,(j \ne i)}
x_j^2\right]^2 }
\end{equation}
The overall
jackknife estimate of $f(\mu_ y, \mu_z)$ is then the average over the $N$
jackknife estimates
$f_i^J$:
\begin{equation}
\boxed{
\overline{f^J} \equiv {1 \over N} \sum_{i=1}^N f_i^J \, .}
\label{fJ}
\end{equation}
It is straightforward to show that if $f$ is a linear function of $\mu_y$ and
$\mu_z$ then $\overline{f^J} = f(\overline{y},\overline{z})$, i.e. the
jackknife and standard averages are identical, see e.g.~Eq.~\eqref{yJ_def}.
However, when $f$ is not a
linear function, so there is bias, there \textit{is}
a difference, and we will now show the resampling carried out in the jackknife
method can be used to determine bias and error bars in an
automated way.

We proceed as for
the derivation of Eq.~\eqref{bias2}, which we now write as
\begin{equation}
f(\mu_y, \mu_z) = \langle f(\overline{y},\overline{z}) \rangle
- {A \over N} - {B\over N^2} + \cdots, 
\end{equation}
where $A$ is the term in rectangular brackets in Eq.~\eqref{bias2}, and we
have added the next order correction. The jackknife data sets have $N-1$
points with the same distribution as the $N$ points in the actual
distribution, and so the bias in the
jackknife average will be of the same form, with the same values
of $A$ and $B$, but with $N$ replaced by $N-1$, i.e.
\begin{equation}
f(\mu_y, \mu_z) = \langle \overline{f^J} \rangle -
{A \over N-1} - {B \over (N-1)^2} \cdots \, .
\end{equation}
We can therefore eliminate the leading contribution to the bias by forming an
appropriate linear combination of $f(\overline{y},\overline{z})$ and
$\overline{f^J}$, namely
\begin{equation}
f(\mu_y, \mu_z) = N \langle f(\overline{y},\overline{z}) \rangle
- (N-1) \langle \overline{f^J} \rangle + O\left({1\over N^2}\right) \, .
\end{equation}
It follows that, to eliminate the leading bias without computing partial
derivatives, one should
\begin{equation}
\boxed{
\mbox{estimate }
f(\mu_y, \mu_z) \mbox{ from }
N f(\overline{y},\overline{z}) 
- (N-1) \overline{f^J} \, .
}
\label{bias_elim}
\end{equation}
The bias is then of order $1/N^2$.  However, as mentioned earlier, bias is
usually not a big problem because, even without eliminating the leading 
contribution, the bias is of order $1/N$ whereas the statistical
error is of order $1/\sqrt{N}$ which is much bigger if $N$ is large. In most
cases, therefore,
$N$ is sufficiently large that one can use \textit{either} the usual
average $f(\overline{y}, \overline{z})$, or the jackknife average
$\overline{f^J}$ in Eq.~\eqref{fJ}, to estimate $f(\mu_y, \mu_z)$, since the
difference between them will be much smaller than the statistical error. In
other words, elimination of the leading bias using Eq.~\eqref{bias_elim} is
usually not necessary.


Next we show that the jackknife method gives error bars, which agree with
Eq.~\eqref{sigma_f} but without the need to explicitly keep track of the partial
derivatives and the variances and covariances. 

We define the variance of the jackknife averages by
\begin{equation}
s^2_{f^J} \equiv \overline{\left(f^J\right)^2} - \left(
\overline{f^J} \right)^2 \, ,
\label{sigmafJ}
\end{equation}
where
\begin{equation}
\overline{\left(f^J\right)^2} = {1 \over N} \sum_{i=1}^N \left(f_i^J\right)^2
\, .
\end{equation}
Using Eqs.~\eqref{fJi} and \eqref{fJ}, we
expand $\overline{f^J}$ away from the exact result $f(\mu_y, \mu_z)$. Just
including the leading contribution gives
\begin{align}
\overline{f^J} - f(\mu_y, \mu_z) &= {1 \over N} \sum_{i=1}^N \left[
(\partial_{\mu_y} f)\, (y_i^J - \mu_y) +
(\partial_{\mu_z} f)\, (z_i^J - \mu_z) \right] \nonumber \\
&= {1 \over N(N-1)} \sum_{i=1}^N \left[
(\partial_{\mu_y} f)\, \left\{N(\overline{y} - \mu_y) - (y_i - \mu_y) \right\}
+
(\partial_{\mu_z} f)\, \left\{N(\overline{z} - \mu_z) - (z_i - \mu_z) \right\}
\right] \nonumber \\
&= (\partial_{\mu_y} f)\, (\overline{y} - \mu_y) +
(\partial_{\mu_z} f)\, (\overline{z} - \mu_z) \, ,
\label{fJ-f}
\end{align}
where we used Eq.~\eqref{yJ_def}.
Similarly we find
\begin{align}
\overline{\left(f^J\right)^2 } &= {1 \over N} \sum_{i=1}^N \left[
f(\mu_y, \mu_z) + (\partial_{\mu_y} f)\, (y_i^J - \mu_y) +
(\partial_{\mu_z} f)\, (z_i^J - \mu_z) \right]^2 \nonumber \\
&= f^2(\mu_y, \mu_z) + 2 f(\mu_y, \mu_z) \, \left[ (\partial_{\mu_y} f)\,
(\overline{y} - \mu_y) + (\partial_{\mu_z} f)\, (\overline{z} - \mu_z) \right] \nonumber \\
&\quad + 
(\partial_{\mu_y} f)^2\, \left[(\overline{y} - \mu_y)^2 + {s_{y}^2 \over (N-1)^2}\right] +
(\partial_{\mu_z} f)^2\, \left[(\overline{z} - \mu_z)^2 + {s_{z}^2 \over (N-1)^2}\right] 
\nonumber \\
&\qquad + 2(\partial_{\mu_y} f)(\partial_{\mu_z} f)\,\left[(\overline{y} - \mu_y) (\overline{z} - \mu_z) +
{s_{yz}^2 \over (N-1)^2}\right]
\, .
\end{align}
Hence, from Eqs.~\eqref{sigmafJ} and \eqref{fJ-f},
the variance in the jackknife estimates is
given by
\begin{equation}
s^2_{f^J} = {1 \over (N-1)^2} \, \left[
(\partial_{\mu_y} f)^2\, s_{y}^2 +
(\partial_{\mu_z} f)^2\, s_{z}^2 
+ 2(\partial_{\mu_y} f)(\partial_{\mu_z} f) s_{yz}^2\right] \, ,
\end{equation}
which is just $1/(N-1)$ times $\sigma_f^2$, the estimate of the square of the error bar in
$f(\overline{y}, \overline{z})$ given in Eq.~\eqref{sigma_f}.
Hence 
\begin{equation}
\boxed{
\mbox{the jackknife estimate for } \sigma_f \mbox{ is } \sqrt{N-1} \,
s_{f^J}\, .}
\label{error_jack}
\end{equation}
Note that this is directly obtained from the jackknife estimates without
having to put in the partial derivatives by hand. Note too that the
$\sqrt{N-1}$ factor is in the \textit{numerator} whereas the factor of
$\sqrt{N-1}$ in Eq.~\eqref{finalans} is in the \textit{denominator}. Intuitively the
reason for this difference 
is that the jackknife estimates are very close, much closer than the error 
in the means, since they would all
be equal except that each one omits just one data point. 

If $N$ is very large, roundoff errors could become significant from having to
subtract large, almost equal, numbers to get the error bar
from the jackknife method. It is then advisable to group the $N$ data points into
$N_\text{group}$ groups (or ``bins'') of data and take, as individual
data points in the
jackknife analysis, the average of the data in each group. The above results
clearly go through with $N$ replaced by $N_\text{group}$.

To summarize this subsection, to estimate $f(\mu_y, \mu_z)$ one can use either
$f(\overline{y}, \overline{z})$ or the jackknife average $\overline{f^J}$ in
Eq.~\eqref{fJ}. The error bar in this estimate, $\sigma_f$, is related to the
standard deviation in the jackknife estimates $s_{f^J}$ by
Eq.~\eqref{error_jack}.

\subsubsection{Bootstrap}
\label{sec:boot}

The bootstrap, like the jackknife, is a resampling of the $N$ data points. A
brief discussion, in the context of data analysis is given in
Ref.~\cite{newman:99}. 
Whereas jackknife considers $N$ new data sets, each of containing all
the original
data points minus one, bootstrap uses $\nb$ data sets each containing $N$ points
obtained by random (Monte Carlo) sampling of the original set of $N$ points.
During the Monte Carlo sampling, the probability that a data point is picked is
$1/N$ irrespective of whether it has been picked before. (In the statistics
literature this is called picking from a set ``with replacement''.)
Hence a given data
point $x_i$ will, \textit{on average}, appear once in each Monte Carlo-generated
data set, but
may appear not at all, or twice, and so on.
The probability that $x_i$ appears $n_i$ times is close to a Poisson
distribution with mean unity. However, it is not exactly Poissonian
because of the constraint in Eq.~(\ref{constraint}) below. It turns out
that we shall need
to include the deviation from the Poisson distribution even for large $N$.
We shall use the term ``bootstrap'' to
denote the Monte Carlo-generated
data sets.

More precisely, let us suppose that the number of times $x_i$ appears in a
bootstrap data set is $n_i$. Since each bootstrap dataset
contains exactly $N$ data points, we have the constraint
\begin{equation}
\sum_{i=1}^N n_i = N \, .
\label{constraint}
\end{equation}
Consider one of the $N$ variables $x_i$. Each time we generate an element in a
bootstrap dataset the probability that it is $x_i$ is $1/N$, which we will
denote by $p$.  From
standard probability theory, the probability
that $x_i$ occurs $n_i$ times is given by a binomial
distribution
\begin{equation}
P(n_i) = {N! \over n_i! \, (N - n_i)!} \, p^{n_i} (1-p)^{N -n_i} \, . 
\end{equation}
The mean and standard deviation of a binomial distribution are given by
\begin{align}
[ n_i ]\mc & = N p = 1 \, ,
\label{nimc} \\
{[ n_i^2 ]\mc} - [n_i]\mc^2 & = N p (1 - p)  = 1 - {1 \over N} \, ,
\label{epsi_epsi}
\end{align}
where $[ \dots ]\mc$ denotes an exact average over bootstrap samples
(for a fixed original data set $x_i$). For $N \to\infty$, the binomial
distribution goes over to a Poisson distribution, for which
the factor of $1/N$ in Eq.~(\ref{epsi_epsi}) does not appear.
We assume that $\nb$ is sufficiently large that the bootstrap average we
carry out
reproduces this result with sufficient accuracy.
Later, we will discuss what values for $\nb$ are sufficient in practice.
Because of the constraint in Eq.~(\ref{constraint}), $n_i $ and $n_j$
(with $i \ne j$) are not independent and we find, by squaring
Eq.~\eqref{constraint} and using Eqs.~\eqref{nimc} and \eqref{epsi_epsi},
that
\begin{equation}
[ n_i n_j ]\mc - [ n_i ]\mc [ n_j ]\mc = - {1 \over N} \qquad (i \ne j)\, .
\label{epsi_epsj}
\end{equation}

First of all we just consider the simple average $\mu_x \equiv \langle x
\rangle$, for which, of course, the standard methods in Sec.~\ref{sec:basic}
suffice, so bootstrap is not necessary.
However, this will show how to get averages
and error bars in a simple case, which we will then generalize to
non-linear functions of averages.

We denote the average of $x$ for a given bootstrap data set by 
$\xb_\alpha$, where $\alpha$ runs from 1 to $\nb$, \ie
\begin{equation}
\xb_\alpha = {1 \over N}  \sum_{i=1}^N n_{i,\alpha} x_i  \, .
\end{equation}
We  then
compute the bootstrap average of the mean of $x$ and the
bootstrap variance in the mean, by averaging over all the bootstrap data sets.
We assume that $\nb$ is large enough for
the bootstrap average to be exact, so we can use Eqs.~(\ref{epsi_epsi}) and
(\ref{epsi_epsj}). The result is
\begin{eqnarray}
\label{xb}
\overline{\xb} \equiv
{1 \over \nb} \sum_{\alpha=1}^\nb \xb_\alpha 
& = & {1\over N} \sum_{i=1}^N [n_i]\mc x_i =  {1\over N}
\sum_{i=1}^N x_i  = \overline{x} \\
s^2_{\xb}
\equiv 
\overline{\left(\xb\right)^2} - \left(\overline{\xb}\right)^2
& = & {1\over N^2}
\left(1 - {1\over N}\right) \sum_i
x_i^2 - {1 \over N^3} \sum_{i \ne j} x_i x_j
= {1\over N} \left(\overline{x^2} - {\overline{x}}^2\right) = {s^2 \over N}\, ,
\label{sigmab}
\end{eqnarray}
where
\begin{equation}
\overline{\left(\xb\right)^2} \equiv {1 \over \nb} \sum_{\alpha=1}^\nb
\left[ \left(\xb_\alpha\right)^2\right]\mc \, .
\end{equation}

We now average Eqs.~(\ref{xb}) and (\ref{sigmab}) over many repetitions of the
original data set $x_i$. Averaging Eq.~(\ref{xb}) gives
\begin{equation}
\langle \overline{\xb} \rangle = \langle \overline{x} \rangle = \langle x
\rangle \equiv \mu_x \, .
\end{equation}
This shows that the bootstrap average $\,\overline{\xb}\, $
is an unbiased estimate of the
exact average $\mu_x$. Averaging Eq.~(\ref{sigmab}) gives
\begin{equation}
\left\langle s^2_{\xb} \right\rangle
= {1 \over N}\, \left\langle s^2 \right\rangle
= {N-1 \over N^2} \sigma^2 =
{N-1 \over N} \sigma^2_{\overline{x}} \, ,
\end{equation}
where we used Eqs.~(\ref{5}) and (\ref{dxsq}). Since
$\sigma_{\overline{x}}$ is the uncertainty in the sample mean, we see that
\begin{equation}
\boxed{\mbox{the bootstrap estimate of }\sigma_{\overline{x}} \mbox{ is }
\sqrt{N \over N-1}\, s_{\xb} \, .}
\label{sigmaxb}
\end{equation}
Remember that $\sigma_{\xb}$ is
the standard deviation of the bootstrap data sets.
Usually $N$ is sufficiently large
that the square root in Eq.~(\ref{sigmaxb}) can be replaced by unity.

As for the jackknife, these results can be generalized to finding the error
bar in some possibly non-linear function,
$f(\mu_y, \mu_z)$, rather than for $\mu_x$.
One computes the bootstrap estimates for $f(\mu_y, \mu_z)$, which are
\begin{equation}
\fb_\alpha = f(\yb_\alpha, \zb_\alpha) \, .
\end{equation}
In other words, we use the bootstrap values, $\yb_\alpha, \zb_\alpha$,
rather than the
sample means, $\overline{y}, \overline{z}$, as the arguments of $f$.
The final bootstrap estimate for $f(\mu_y, \mu_z)$ is the average of these, \ie
\begin{equation}
\boxed{ \overline{\fb} = {1 \over \nb} \sum_{\alpha=1}^\nb \fb_\alpha \, .}
\label{fb}
\end{equation}
Following the same methods in the jackknife section, one obtains the error bar,
$\sigma_f$, in $f(\mu_y, \mu_z)$. The result is
\begin{equation}
\boxed{\mbox{the bootstrap estimate for } \sigma_f \mbox{ is }
\sqrt{N \over N-1} \,\, s_{\fb},} 
\label{sigmafb}
\end{equation}
where
\begin{equation}
\boxed{
s^2_{\fb} = 
\overline{\left(\fb\right)^2} - \left(\overline{\fb}\right)^2 \, ,}
\end{equation}
is the variance of the bootstrap estimates. Here
\begin{equation}
\overline{\left(\fb\right)^2} \equiv {1 \over \nb} \sum_{\alpha=1}^\nb
\left(\fb_\alpha\right)^2 \, .
\end{equation}
Usually $N$ is large enough that the factor of $\sqrt{N/(N-1)}$ is
Eq.~\eqref{sigmafb} can be replaced by unity.
Equation (\ref{sigmafb}) corresponds to the result Eq.~(\ref{sigmaxb})
which we derived for the special case of $f = \mu_x$.

Again,
following the same path as in the jackknife section, it is straightforward to
show that the bias of the estimates in Eqs.~(\ref{fb}) and (\ref{sigmafb}) is
of order $1/N$ and so vanishes
for $N\to\infty$. However, if $N$ is not too large it may be
useful to eliminate the leading contribution to the bias in the mean, as
we did for jackknife in Eq.~(\ref{bias_elim}). The result is that one should
\begin{equation}
\boxed{\mbox{estimate } f(\mu_y, \mu_z) \mbox{ from }
2 f(\overline{y}, \overline{z}) - \overline{\fb} \, .}
\label{improved_boot}
\end{equation}
The bias in Eq.~(\ref{improved_boot})
is of order $1/N^2$, whereas $f(\overline{y}, \overline{z})$
and $\overline{\fb}$ each
have a bias of order $1/N$. However, it is not normally necessary to eliminate
the bias since, if $N$ is large, the bias is much smaller than the statistical
error.

I have not systematically studied the values of $\nb$ that are needed in
practice to get accurate estimates for the error. It seems that
$\nb$ in the range 100 to 500 is typically chosen, and this seems to be adequate
irrespective of how large $N$ is.

It is sometimes stated, e.g.~\cite{newman:99}, that the bootstrap method can
give error bars correctly even when there are correlations in the data. This
is not so. If one applies bootstrap to the direct average of a set of data, it
simply reproduces the results of the standard analysis. Bootstrap is useful
both to get error bars when one is looking at combination of averages of the
data, and to get confidence limits when the noise on the data is not Gaussian,
see Sec.~\ref{sec:resample}.  Unfortunately, bootstrap does not work miracles
and cannot give correct error bars for correlated data. 

To summarize this subsection, to estimate $f(\mu_y, \mu_z)$ one can either use
$f(\overline{y}, \overline{z})$, or the bootstrap average in Eq.~\eqref{fb}, and
the error bar in this estimate, $\sigma_f$, is related to the standard
deviation in the bootstrap estimates by Eq.~\eqref{sigmafb}.

\subsubsection{Jackknife or Bootstrap?}
\label{sec:jorb}
The jackknife approach involves less calculation than bootstrap, and is fine
for estimating combinations of moments of the measured quantities.
Furthermore, identical results are obtained each time jackknife is run on the
same set of data, which is not the case for bootstrap.  However, the range of
the jackknife estimates is  very much smaller, by a factor of $\sqrt{N}$
for large $N$, than
the scatter of averages which would be obtained from individual data sets, see
Eq.~\eqref{error_jack}.  By contrast, for bootstrap, $\sigma_{\fb}$, which
measures the deviation of the bootstrap estimates $\fb_\alpha$ from the result
for the single actual data set $f(\overline{y}, \overline{z})$, \textit{is
equal to} $\sigma_f$, the deviation of the average of a single data set from
the exact result $f(\mu_y,\mu_z)$ (if we replace the factor of $N/(N-1)$ by
unity, see Eq.~\eqref{sigmafb}).  This is the main strength of the bootstrap
approach; it samples the full range of the distribution of the sample
distribution.
Hence, if you want to generate data which covers the full range
then should use bootstrap. This is useful in fitting, see for example,
Sec.~\ref{sec:resample}. However, if you just want to generate error bars on
combinations of moments quickly and easily, then use jackknife.

\section{Fitting data to a model}
\label{sec:fit}

A good reference for the material in this section is Chapter 15 of Numerical
Recipes~\cite{press:92}.

Frequently we are given a set of data points $(x_i, y_i), i = 1, 2,
\cdots, N$, with corresponding error bars, $\sigma_i$,
through which we would like to fit to a smooth function $f(x)$. The
function could be straight line (the simplest case), a higher order
polynomial, or a more complicated function. The fitting function will depend
on $M$ ``fitting parameters'', $a_\alpha$ and we would like the ``best'' fit obtained by
adjusting these parameters. We emphasize that a fitting procedure
should not only
\begin{enumerate}
\item
\label{give_params}
give the values of the fit parameters, but also
\item
\label{give_errors}
provide error
estimates on those parameters, and 
\item
\label{gof}
provide a measure of how good the fit is. 
\end{enumerate}
If the result of part~\ref{gof} is that the fit is very poor, the results of
parts~\ref{give_params} and \ref{give_errors} are probably meaningless.

The definition of ``best'' is not unique. However, the most useful
choice, and the one nearly always taken, is ``least squares''. For this case,
one minimizes the weighted sum of the squares of the difference between the
observed $y$-value, $y_i$,
and the fitting function evaluated at $x_i$.
The weight of each point depends on its error bar, since
the fit should be more tightly bound to points with smaller error bars than to
those with large error bars.
The quantity to be minimized, called  $\chi^2$
(``chi-squared''),\footnote{$\chi^2$ should be thought
of as a single variable rather than the square of something called
$\chi$. This notation is standard.} 
is defined by
\begin{equation}
\boxed{
\chi^2 = \sum_{i=1}^N \left( \, {y_i - f(x_i) \over \sigma_i } \, \right)^2. }
\label{chisq}
\end{equation}
A big advantage of least squares over other definitions of ``best'' fit is
that for a linear model (see below) the equations which determine the fit
parameters are themselves \textit{linear}. We should mention that 
Eq.~\eqref{chisq} implicitly assumes that the data points are uncorrelated. A
generalization of the least squares method which is
applicable, in principle, to correlated data
is given later in Eqs.~\eqref{min_gen}--\eqref{Uv_corr}.

Often we assume that the distribution of the errors is Gaussian, since,
according to the central limit theorem discussed in Appendix \ref{sec:clt}, the
sum of $N$ independent random variables has a Gaussian distribution (under
fairly general conditions) if $N$ is large. However, distributions which occur
in nature usually have more weight in the ``tails'' than a
Gaussian, and as a result, even for moderately large values of $N$, the probability of an
``outlier'' might be much bigger than expected from a Gaussian, see
Fig.~\ref{Fig:converge_to_clt}. 

If the errors \textit{are} distributed with a Gaussian distribution, 
and if $f(x)$ has the \textit{exact} values of the fit parameters, then
$\chi^2$ in
Eq.~\eqref{chisq} is a sum of squares of $N$ random
variables with a Gaussian distribution with mean zero and
standard deviation unity. However, when we have minimized the value of
$\chi^2$ with respect to the $M$ fitting parameters $a_\alpha$ the terms are
not all independent. It turns out, see Appendix \ref{sec:NDF}, that, at least
for a linear model (which we define below), the
distribution of $\chi^2$ at the minimum is that of the sum of the squares of
$N-M$ (not $N$) Gaussian random variables with zero mean and standard deviation
unity\footnote{Although this result is only valid if the fitting model
is linear in the
parameters, it is usually taken to be a reasonable approximation for
non-linear models as well.}.  We call $N-M$ the ``number of
degrees of freedom'' ($\NDOF$). The $\chi^2$ distribution is 
discussed in
Appendix \ref{sec:Q}. The formula for it is Eq.~\eqref{chisq-dist}.

The simplest problems are where
the fitting function is a \textit{linear function of the parameters}. We
shall call this a \textit{linear model}.
Examples are a straight line ($M=2$),
\begin{equation}
y = a_0 + a_1 x \, ,
\label{sl}
\end{equation}
and an ($M$$-1$)-th order polynomial,
\begin{equation}
y = a_0 + a_1 x + a_2 x^2 + \cdots + a_{M-1} x^{M-1} 
= \sum_{\alpha=0}^{M-1} a_\alpha x^\alpha  \, ,
\label{poly}
\end{equation}
where the parameters to be adjusted are the $a_\alpha$.
(Note that we are \textit{not}
stating here that $y$ has to be a linear function of $x$, only of the fit
parameters $a_\alpha$.)

An example where the fitting function depends \textit{non}-linearly on the
parameters is
\begin{equation}
y = a_0 x^{a_1} + a_2 \, .
\end{equation}

Linear models are fairly simply because, as we shall see below, the parameters
are determined by \textit{linear} equations, which, in general, have a unique
solution that can be found by straightforward methods. However, for
fitting functions which are non-linear functions of the parameters, the
resulting equations are \textit{non-linear} which may have many solutions or
none at all, and so are much less straightforward to solve.
We shall discuss fitting to both linear and non-linear models in these notes.

Sometimes a non-linear model can be transformed into a linear model by a
change of variables. For example, if we want to fit to
\begin{equation}
y = a_0 x^{a_1} \, ,
\end{equation}
which has a non-linear dependence on $a_1$,
taking logs gives
\begin{equation}
\ln y = \ln a_0 + a_1 \ln x \, ,
\end{equation}
which is a \textit{linear} function of the parameters $a'_0 = \ln a_0$
and $a_1$. Fitting a
straight line to a log-log plot is a very common procedure in science and
engineering. However, it should be noted that transforming the data does not
exactly take Gaussian errors into Gaussian errors, though the difference will
be small if the errors are ``sufficiently small''. For the above log transformation
this means $\sigma_i / y_i \ll 1$, i.e.\ the \textit{relative} error is much less
than unity. 

\subsection{Fitting to a straight line}
To see how least squares fitting
works, consider the simplest case of a straight line
fit, Eq.~(\ref{sl}), for which we have to minimize
\begin{equation}
\chi^2(a_0, a_1) = \sum_{i=1}^N \left({\, y_i - a_0 - a_1 x_i\, 
\over \sigma_i} \right)^2 \, ,
\label{chisq_sline}
\end{equation}
with respect to $a_0$ and $a_1$. Differentiating $\chi^2$ with respect to
these parameters and setting the results to zero gives
\begin{subequations}
\label{sline}
\begin{align}
a_0\, \sum_{i=1}^N {1 \over \sigma_i^2} + a_1\, \sum_{i=1}^N
{x_i\over\sigma_i^2} &= \sum_{i=1}^N
{y_i\over \sigma_i^2} ,
\label{da0}\\
a_0\, \sum_{i=1}^N {x_i\over \sigma_i^2} +a_1\, \sum_{i=1}^N
{x_i^2\over \sigma_i^2} &=
\sum_{i=1}^N {x_i y_i \over \sigma_i^2}  .
\label{da1}
\end{align}
\end{subequations}
We write this as
\begin{subequations}
\begin{align}
U_{00} \, a_0 + U_{01} \, a_1 &= v_0 , \\
U_{10} \, a_0 + U_{11} \, a_1 &= v_1 ,
\end{align}
\label{lssl}
\end{subequations}
where
\begin{align}
&\boxed{U_{\alpha\beta} = \sum_{i=1}^N {x_i^{\alpha + \beta}\over
\sigma_i^2}, } \quad \mbox{and}
\label{Uab}  \\
&\boxed{v_\alpha = \sum_{i=1}^N{ y_i\,  x_i^\alpha \over
\sigma_i^2 \, }. } \label{v}
\end{align}
The matrix notation, while an overkill here,
will be convenient later when we do a general polynomial
fit. Note that $U_{10} = U_{01}$. (More generally, later on, $U$ will be a
symmetric matrix).
Equations (\ref{lssl}) are two linear equations in two unknowns.
These can be solved by
eliminating one variable, which immediately gives an equation for the
second one. The solution can also be determined from
\begin{equation}
\boxed{
a_\alpha = \sum_{\beta=0}^{M-1} \left(U^{-1}\right)_{\alpha\beta} \, v_\beta , }
\label{soln}
\end{equation}
(where we have temporarily generalized to a polynomial of order $M-1$).
For the straight-line fit, the inverse of the $2\times 2$ matrix
$U$ is given, according to standard rules,
by
\begin{equation}
U^{-1} = {1 \over \Delta} \, 
\begin{pmatrix}
U_{11} &  -U_{01} \\
-U_{01} &  U_{00} 
\end{pmatrix}
\label{Uinv}
\end{equation}
where
\begin{equation}
\boxed{
\Delta =  U_{00} U_{11} - U_{01}^2 ,}
\label{Delta}
\end{equation}
and we have noted that $U$ is symmetric so
$U_{01} = U_{10}$. The solution for $a_0$ and
$a_1$ is therefore given by
\begin{subequations}
\begin{align}
&\boxed{a_0 = {U_{11}\, v_0 - U_{01}\, v_1 \over \Delta}, } \\
&\boxed{a_1 = {-U_{01}\, v_0 + U_{00}\, v_1 \over \Delta}. } 
\end{align}
\label{soln_sl}
\end{subequations}
We see that it is straightforward to determine
the slope, $a_1$, and the intercept, $a_0$, of the fit from
Eqs.~(\ref{Uab}), (\ref{v}), (\ref{Delta}) and (\ref{soln_sl})
using the $N$ data points $(x_i,y_i)$, and their error bars $\sigma_i$.

\subsection{Fitting to a polynomial}

Frequently we need to fit to a higher order polynomial than a
straight line, in which case we minimize
\begin{equation}
\chi^2(a_0,a_1,\cdots,a_{M-1}) =
\sum_{i=1}^N \left({y_i - \sum_{\alpha=0}^{M-1} a_\alpha x_i^\alpha \over
\sigma_i} \right)^2 
\label{chisq_poly}
\end{equation}
with respect to the $M$ parameters
$a_\alpha$. Setting to zero the derivatives of $\chi^2$ with
respect to the $a_\alpha$ gives
\begin{equation}
\boxed{
\sum_{\beta=0}^{M-1} U_{\alpha\beta}\, a _\beta = v_\alpha ,}
\label{lspoly}
\end{equation}
where $U_{\alpha\beta}$ and $v_\alpha$ have been defined in Eqs.~(\ref{Uab}) and
(\ref{v}).
Eq.~(\ref{lspoly}) represents
$M$ \textit{linear} equations, one for each value of $\alpha$. Their
solution is again given by Eq.~(\ref{soln}), i.e.\ it
is expressed in terms of
the inverse matrix $U^{-1}$, which is called the \textit{covariance matrix}.

\subsection{Error Bars}
\label{sec:error_bars}

In addition to the best fit values of the parameters we also need to determine
the error bars in those values.
Interestingly, this information is \textit{also} contained in the
covariance matrix $U^{-1}$.

\begin{center}
\begin{figure}
\includegraphics[width=7.5cm]{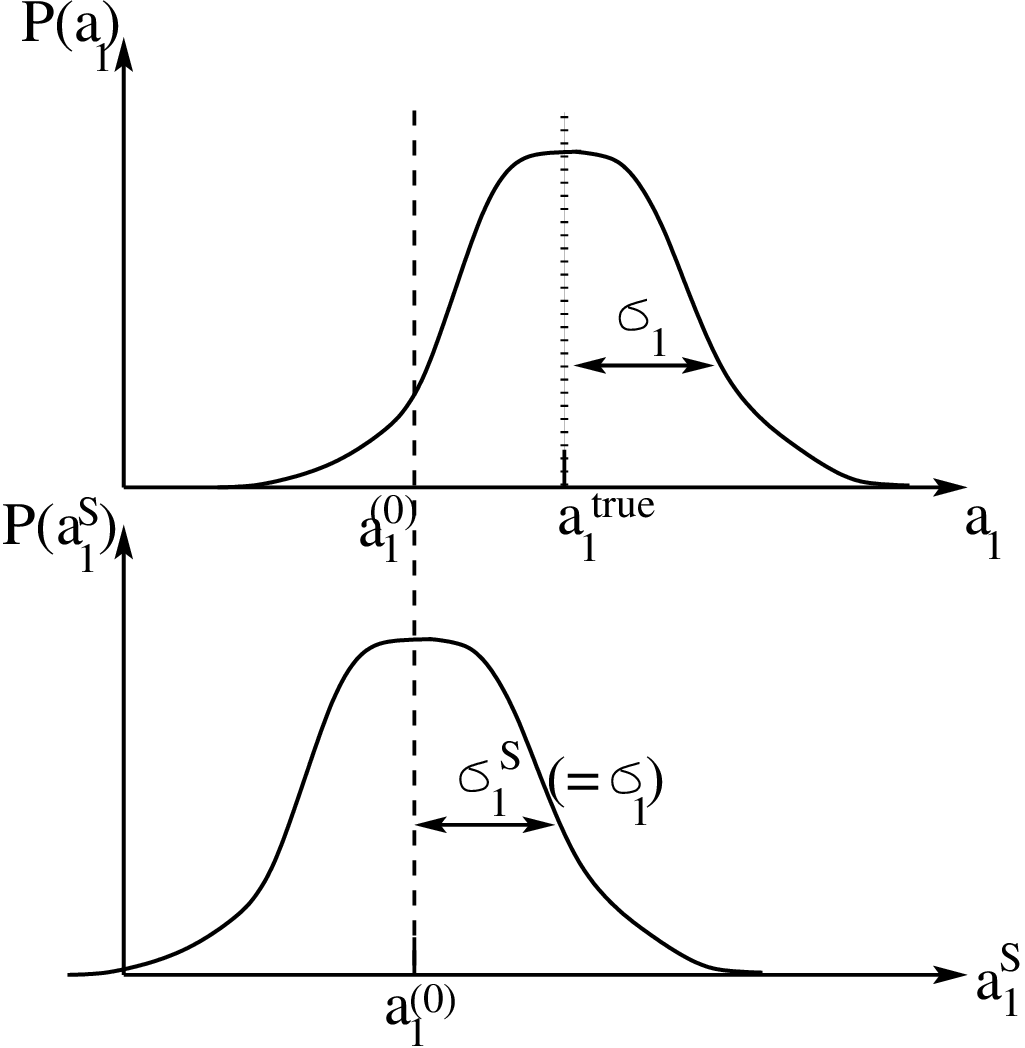}
\caption{
The top figure shows the distribution of one of the fit parameters $a_1$ if
one could obtain many real data sets. The distribution has standard deviation
$\sigma_1$ about the true value $a_1^\text{true}$ and is Gaussian if the noise
on the data is Gaussian. In fact, however, we have only one
actual data set which has fit
parameter $a_1^{(0)}$, and this typically lies within about $\sigma_1$ of
$a_1^\text{true}$. Hence we \fbox{define the error bar on the estimate of
$a_1^\text{true}$ to be 
$\sigma_1$.} However, we \fbox{cannot calculate $\sigma_1$ directly
from the distribution of $a_1$}
because we have only one the one value, $a_1^{(0)}$. However, we can generate
many \textit{simulated} data sets from the one actual set and hence \fbox{we can estimate
the standard deviation, $\sigma_1^S$,} of
the distribution of the resulting fit parameter $a_1^S$, which is shown in
the lower figure.
This distribution is centered about the value from the actual
data, $a_1^{(0)}$, and has
standard deviation, $\sigma_1^S$. The important point is that
if one assumes a linear model then one can show that
$\boxed{\sigma_1^S = \sigma_1 ,}$ see text.
Even if the model is non linear, 
one usually assumes that the difference in the standard deviations
is sufficiently small that one can still equate the true error bar with
the standard deviation from the simulated data sets. We
emphasize that \fbox{the error bar
quoted by fitting programs is actually $\sigma_1^S$,} and this is assumed to
equal $\sigma_1$.
Furthermore, as
shown in Appendices
\ref{sec:proof} and \ref{sec:proof2}, if the noise on the data is Gaussian (and
the model is linear) both the distributions in this figure are also
Gaussian.
}
\label{Fig:distofa1}
\end{figure}
\end{center}

First of all, we explain the significance of error bars in fit parameters.
We assume that the data is described by a model with a particular set of
parameters $\vec{a}^\text{true}$ which, unfortunately, we don't know. If we
were, somehow, to have many real data sets each one would give a different
set of fit parameters $\vec{a}^{(i)}, i = 0, 1, 2, \cdots$,
because of noise in the data,
\textit{clustered about the
true set} $\vec{a}^\text{true}$. Projecting on
to a single fit parameter, $a_1$ say, there will be a distribution of values
$P(a_1)$ centered on $a_1^\text{true}$ with standard deviation $\sigma_1$, see
the top part of Fig.~\ref{Fig:distofa1}.  Typically
the value of $a_1$ obtained from our \textit{one actual data set}, $a_1^{(0)}$,
will lie within about $\sigma_1$ of $a_1$.
Hence we define
the error bar to be $\sigma_1$.

Unfortunately, we can't determine the error bar this way because we have only
one actual
data set, which we denote here by $y_i^{(0)}$ to distinguish it from other
data sets that we will introduce. Our actual data set gives 
one set of fit parameters, which we call $\vec{a}^{(0)}$. Suppose,
however, we were
to generate many \textit{simulated} data sets from of the one which is
available to us, by
generating random values (possibly with a Gaussian distribution though this
won't be necessary yet) centered at the $y_i$ with standard deviation
$\sigma_i$.
Fitting each simulated dataset would give 
different values for $\vec{a}$, \textit{clustered now about} $\vec{a}^{(0)}$, 
see the bottom part of Fig.~\eqref{Fig:distofa1}.
We now come to
an important, but rarely discussed, point:

\begin{quotation}
\noindent For a linear model 
the standard deviation of the fit parameters of these simulated
data sets about $\vec{a}^{(0)}$,
is equal to the standard deviation of the
fit parameters of real data sets $\vec{a}$ about
$\vec{a}^\text{true}$. The latter is what we \textit{really} want to know (since
it is our estimate of the error bar on $\vec{a}^\text{true}$) but can't
determine
directly. See Fig.~\ref{Fig:distofa1} for an illustration.
This result is also applicable to a non-linear model if
it can be 
represented by an effective linear model for the needed range of parameters
about the minimum of $\chi^2$.
Furthermore, we show in
Appendices \ref{sec:proof} and \ref{sec:proof2} that if
the noise on the data
is Gaussian (and the model is linear),
the two distributions in Fig.~\eqref{Fig:distofa1} are also both
Gaussian.
\end{quotation}

We shall now prove this result.
As stated above, to derive the error bars in the fit parameters we
take simulated values 
of the data points, $y_i^S$, which vary by some amount $\delta
y_i^S$ about $y_i^{(0)}$, i.e.\ $\delta y_i^S = y_i^S - y_i^{(0)}$,
with a standard deviation given by the error bar
$\sigma_i$.
%
The fit
parameters of this simulated data set,
$\vec{a}^S$, then deviate from $\vec{a}^{(0)}$ by an
amount $\delta \vec{a}^S$ where
\begin{equation}
\delta a_\alpha^S = \sum_{i=1}^N {\partial a_\alpha \over \partial y_i}\, \delta
y_i^S\, .
\label{deltaaS}
\end{equation}
Averaging over fluctuations in the $y_i^S$ we get the variance of
$a_\alpha^S$ to be
\begin{equation}
\left(\sigma_\alpha^S\right)^2
\equiv \langle \left(\delta a_\alpha^S\right)^2 \rangle = \sum_{i=1}^N
\sigma_i^2 \, \left( {\partial a_\alpha \over \partial y_i} \right)^2  \, ,
\label{sigma_alpha}
\end{equation}
since $\langle \left(\delta y_i^S\right)^2 \rangle = \sigma_i^2$, and
the data points $y_i$ are statistically independent.
Writing Eq.~\eqref{soln} explicitly in terms of the data values,
\begin{equation}
a_\alpha = \sum_\beta \left(U^{-1}\right)_{\alpha\beta} \sum_{i=1}^N { y_i\,
x_i^\beta \over \sigma_i^2 \, } \, ,
\end{equation}
and noting that $U$ is independent of the $y_i$, we get
\begin{equation}
{\partial a_\alpha \over \partial y_i} = \sum_\beta
\left(U^{-1}\right)_{\alpha\beta} {x_i^\beta \over \sigma_i^2} \, .
\end{equation}
Substituting into Eq.~\eqref{sigma_alpha} gives
\begin{equation}
\left(\sigma_\alpha^S \right)^2 = \sum_{\beta, \gamma} \left(U^{-1}\right)_{\alpha\beta}
\left(U^{-1}\right)_{\alpha\gamma} \left[
\sum_{i=1}^N {x_i^{\beta + \gamma} \over \sigma_i^2}
\right]
\, .
\end{equation}
The term in rectangular brackets is just $U_{\beta\gamma}$, and since
$U$ is given by Eq.~\eqref{Uab} and is symmetric, the last
equation reduces to
\begin{equation}
\left(\sigma_\alpha^S \right)^2 = 
\left(U^{-1}\right)_{\alpha\alpha} \, .
\label{error_params_s}
\end{equation}
Recall that $\sigma_\alpha^S$ is
the standard deviation of the fitted parameter values about the $\vec{a}^{(0)}$
when constructing simulated data sets from the one set of data that is
available to us.

However, the error bar is defined to be the
standard deviation the fitted parameter values would have relative to
$a_\alpha^\text{true}$ if we could average over many actual data sets. To determine
this quantity
we simply repeat the above calculation
with $\delta y_i = y_i - y_i^\text{true}$ in which $y_i$ is the value of the
$i$-th data point in one of the actual data sets. Since $U$ is a constant (for
a linear model) equations \eqref{deltaaS} to \eqref{error_params_s} go through unchanged
simply omitting the superscript $S$'s. The (unknown) values of
$y_i^\text{true}$ never enter.
In other words
\begin{equation}
\boxed{ \sigma_\alpha^2 = 
\left(U^{-1}\right)_{\alpha\alpha} \, ,}
\label{error_params}
\end{equation}
which shows that $\sigma^s_\alpha = \sigma_s$ for a linear model. However,
this equality does not hold precisely for fitting to a non-linear model.
We have therefore showed that the diagonal elements of the covariance matrix
$U^{-1}$ give the square of the errors bar in the fit parameters.

In addition to error bars, we
also need a parameter to describe the quality of the fit. A useful quantity
is the probability that, given the fit, the data could have occurred with a
$\chi^2$ greater than or equal to the value found. This is generally denoted by $Q$
and, as shown in Appendix \ref{sec:Q}, is given by
\begin{equation}
Q =  {1 \over \Gamma(N_\text{DOF}/2)} \, \int_{\chi^2/2}^\infty\,
y^{(N_\text{DOF}/2)-1}\,  e^{-y} \, dy\, , 
\label{Q_expression2}
\end{equation}
assuming the data have Gaussian noise. Here $N_\text{DOF} \equiv N -M$ is the
number of degrees of freedom.
Note that 
the effects of \textit{non-Gaussian} statistics is to increase the
probability of outliers, so fits with a fairly small value of $Q$, say around
$0.01$, may be considered acceptable. However, fits with a \textit{very} small
value of $Q$ should not be trusted and the values of the fit parameters are
probably meaningless in these cases. 

\begin{center}
\begin{figure}
\includegraphics[width=10cm]{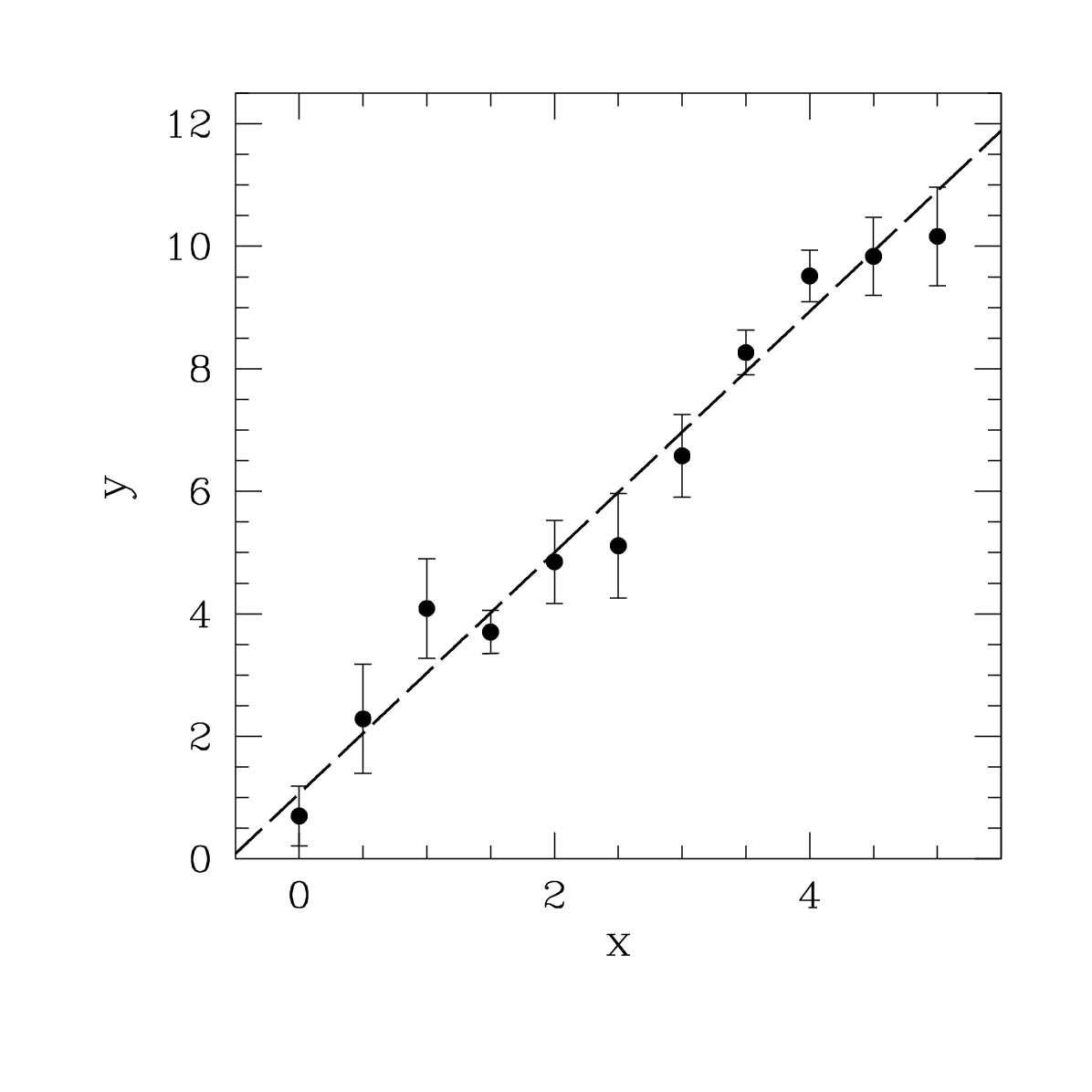}
\caption{An example of a straight-line fit to a set of data with error bars.}
\label{fig:slinefit}
\end{figure}
\end{center}

For the case of a straight line fit, the
inverse of $U$ is given explicitly in Eq.~(\ref{Uinv}).
Using this information, and the values of
$(x_i, y_i, \sigma_i)$ for the data in Fig.~\ref{fig:slinefit},
the fit parameters (assuming a straight line fit) are
\begin{align}
a_0 &= 0.84 \pm 0.32 , \\
a_1 &= 2.05 \pm 0.11 ,
\end{align}
in which the error bars on the fit parameters on $a_0$ and $a_1$, which
are denoted by $\sigma_0$ and
$\sigma_1$, are determined from Eq.~(\ref{error_params}).
The data was generated by starting with $y = 1 + 2x$ and then
adding some noise with zero mean. Hence the fit should be consistent
with $y = 1 +2x$ within the error bars, and it is. The value of $\chi^2$
is 7.44 so $\chi^2/\NDOF = 7.44 / 9 = 0.866$ and the quality of fit
parameter, given by Eq.~\eqref{Q_expression2}, is $Q = 0.592$ which is good.

The off-diagonal elements of the covariance matrix $U^{-1}$ are also useful
since they contain information about
correlations between the fitted parameters. More precisely, one can show,
following the
lines of the above derivation of $\sigma_\alpha^2$, that the correlation
of fit parameters $\alpha$ and $\beta$, known mathematically as their
``covariance'', is given by the appropriate off-diagonal element of 
the covariance matrix,
\begin{equation}
\text{Cov}(\alpha, \beta)  \equiv \langle \delta a_\alpha \, \delta a_\beta
\rangle = \left(U^{-1}\right)_{\alpha\beta} \, .
\label{Covab}
\end{equation}
The correlation coefficient, $r_{\alpha\beta}$, which is a dimensionless 
measure of the correlation between $\delta
a_\alpha$ and $\delta a_\beta$
lying between $-1$ and 1, 
is given by
\begin{equation}
r_{\alpha\beta} = {\text{Cov}(\alpha, \beta) \over \sigma_\alpha \sigma_\beta}
\, .
\label{rab}
\end{equation}
A good fitting program should output the correlation coefficients as well as the
fit parameters, their error bars, the value of $\chi^2/\NDOF$, and the
goodness of fit parameter $Q$.

So far we have considered a polynomial fit, which is a particular case of a
linear model.
If we fit to a \textit{general} linear model, writing
\begin{equation}
f(x) = \sum_{\alpha=1}^M a_\alpha \, X_\alpha(x) ,
\label{general_lin}
\end{equation}
where $X_1(x), X_2(x), \cdots, X_M(x)$ are fixed functions of $x$ called basis
functions, $\chi^2$ is given by
\begin{equation}
\chi^2 = \sum_{i=1}^N \left({y_i - \sum_{\alpha=1}^M
a_\alpha X_\alpha(x_i) \over \sigma_i} \right)^2 \, ,
\label{chisq_gen}
\end{equation}
and the matrix $U$ is given by
\begin{equation}
\boxed{
U_{\alpha\beta} = \sum_{i=1}^N {X_\alpha(x_i)\, X_\beta(x_i) \over \sigma_i^2}
\, .}
\label{Uab_general}
\end{equation}
Similarly, the quantities $v_\alpha$ in Eq.~\eqref{v} become
\begin{equation}
\boxed{
v_\alpha = \sum_{i=1}^N {y_i\, X_\alpha(x_i) \over \sigma_i^2}
\, ,}
\label{v_general}
\end{equation}
and, as before, the best fit parameters are given by the
solution of the $M$ linear equations
\begin{equation}
\boxed{ \sum_{\beta=1}^M U_{\alpha\beta}\, a_\beta = v_\alpha \, , }
\label{lin_eq}
\end{equation}
for $\alpha= 1, 2, \cdots, M$, namely by Eq.~\eqref{soln}.

For a linear model, $\chi^2$ is a quadratic function of the fit parameters and
so the elements of the ``\textit{curvature matrix}''\footnote{It is
conventional to include the factor of $1/2$.}, $(1/2)\, \partial^2 \chi^2 /
\partial {a_\alpha}\partial {a_\beta}$ are constants, independent of the values
of the fit parameters. In fact, we see from
Eqs.~\eqref{chisq}, \eqref{general_lin}
and \eqref{Uab_general} that
\begin{equation}
{1\over 2}\, { \partial^2 \chi^2 \over
\partial {a_\alpha} \partial {a_\beta}} = U_{\alpha \beta} \, ,
\label{curv}
\end{equation}
so \textit{the curvature matrix is equal to $U$}, given by 
Eq.~\eqref{Uab_general} (Eq.~\eqref{Uab} for a
polynomial fit).
Note that for a linear
model the curvature matrix $U$ is a constant, independent of the fit
parameters and data values. However, $U$ is not constant for a non-linear model.

\subsection{Interpolating}
\label{sec:interpol}

In a physics context it usually the fitting parameters \textit{per se} which
are of interest. However, in other contexts where fitting is performed, for
example machine learning~\cite{bishop:06}, one is less interested in
the fitting model and more interested in predicting the
value of the function for a new value of $x$. We give a brief discussion of
this here.

We assume a general linear model as in Eq.~\eqref{general_lin}.
Clearly the best estimate
for $y$ at some value $x$ is the fitting function with the optimized fitting
parameters evaluated at $x$, i.e.
\begin{equation}
y(x) = \sum_{\alpha=1}^M a_\alpha \, X_\alpha(x) ,
\label{yx}
\end{equation}
and the error bar (squared) is given by
\begin{align}
\sigma_y^2 &= \sum_{\alpha, \beta}\, \langle \delta a_\alpha \delta a_\beta\rangle
\, X_\alpha(x) X_\beta(x) \nonumber \\
&= 
\sum_{\alpha, \beta}\, \left(U^{-1}\right)_{\alpha\beta}
\, X_\alpha(x) X_\beta(x)
\, ,
\end{align}
where we used Eqs.~\eqref{error_params} and \eqref{Covab}.

It is instructive to substitute for the optimized fitting parameters into
Eq.~\eqref{yx}, i.e.
\begin{align}
y(x) & = \sum_{\alpha=1}^M  X_\alpha(x) \sum_{\beta = 1}^M
\left(U^{-1}\right)_{\alpha\beta} 
\sum_{i=1}^N {y_i X_\beta(x_i)  \over \sigma_i^2}  \nonumber \\
&= \sum_{i=1}^N k(x, x_i)\, y_i \, ,
\label{interp}
\end{align}
where the \textit{kernel} $k(x, x_i)$ is given by
\begin{equation}
k(x, x_i) = \sum_{\alpha, \beta}{X_\alpha(x) \left(U^{-1}\right)_{\alpha\beta}
X_\beta(x_i) \over \sigma_i^2} \, .
\end{equation}
The kernel is independent of the $y_i$ and
has the simple, intuitive property that
\begin{equation}
\sum_{i=1}^N k(x, x_i) = 1.
\label{kernel_sum}
\end{equation}
Clearly Eq.~\eqref{kernel_sum} is correct if the $y_i$ in Eq.~\eqref{interp}
are constant. In fact, Eq.~\eqref{kernel_sum} is correct quite generally 
as long as one of the basis functions is a constant (the usual
situation). In this case
$X_1(x)$, say, is equal to 1 so $\sum_iX_\beta(x_i)/\sigma_i^2 = 
U_{0 \beta}$ according to Eq.~\eqref{Uab_general}. Hence
\begin{equation}
\sum_{i=1}^N k(x, x_i) = \sum_{\alpha, \beta}
X_\alpha(x) \left(U^{-1}\right)_{\alpha\beta} U_{\beta 1}
= \sum_\alpha X_\alpha(x)\, \delta_{\alpha 1} = 1 \, ,
\end{equation}
which is Eq.~\eqref{kernel_sum}.

Presumably the data close to $x$ have the
larger weight in Eq.~\eqref{kernel_sum}.  It would be interesting to examine
this. 
I expect that Eq.~\eqref{kernel_sum} is
approximately true even if one of the basis functions is not a constant.

\subsection{Fitting to a non-linear model}
\label{sec:nlmodel}

As for linear models, one minimizes $\chi^2$ in Eq.~\eqref{chisq}. The difference is that the
resulting equations are non-linear so there might be many solutions or non at
all. Techniques for solving the coupled non-linear equations invariably require specifying
an initial
value for the variables $a_\alpha$. The most common method for
fitting to non-linear models is the 
Levenberg-Marquardt (LM) method, see e.g.\ Numerical Recipes~\cite{press:92}.
Implementing the Numerical Recipes code for LM
is a little complicated because it requires the user to provide a routine for
the derivatives of $\chi^2$ with respect to the fit parameters
as well as for $\chi^2$ itself, and to check for convergence. Alternatively,
one can use the fitting routines in the \texttt{scipy} package of
\texttt{python} or use \texttt{gnuplot}. But see the comments in
Appendix \ref{sec:ase} about getting the error bars in the parameters correct,
which apply when fitting to linear as well as non-linear models. Gnuplot and
scipy scripts for fitting to a non-linear model are given in Appendix
\ref{sec:scripts}.

One difference from fitting to a linear model is that the curvature matrix,
defined by the LHS of Eq.~\eqref{curv}, is not constant but is a function of
the fit parameters. Hence it is no longer true that the standard deviations of
the two distributions in Fig.~\ref{Fig:distofa1} are equal. However, it still
generally assumed that the difference is small enough to be unimportant and
hence that
the covariance matrix, which is now defined to be
the inverse of the curvature matrix \textit{at the minimum of
$\chi^2$}, still gives information about error bars on
the fit parameters. This is discussed more in the next two
subsections, in which we
point out, however, that a more detailed analysis is needed if the model is
non-linear and the spread of fitted parameters is sufficiently large that it
\textit{cannot} be represented by an effective linear model, i.e.\ $\chi^2$ is not well
fitted by a parabola over the needed range of parameter values.

As a reminder:
\begin{itemize}
\item
The \textit{curvature matrix}
is defined in general by the LHS of
Eq.~\eqref{curv}, which,
for a linear model, is equivalent to
Eq.~\eqref{Uab_general} (Eq.~\eqref{Uab} for a polynomial fit.)
\item
The \textit{covariance matrix} is the inverse of the curvature matrix. For a
linear model this matrix is constant, independent of the fit parameters or data
values. However, for a non-linear model this is no longer true and we are then
interested in the covariance matrix at the minimum of $\chi^2$.
Its diagonal elements give error bars on the fit parameters according
to Eq.~\eqref{error_params} (but see the caveat in the previous paragraph for
non-linear models) and its off-diagonal elements give correlations between fit
parameters according to Eqs.~\eqref{Covab} and \eqref{rab}.
\end{itemize}

\subsection{Confidence limits}
\label{sec:conf_limits}

In the last two subsections we showed that the diagonal elements of the
covariance matrix give an error bar on the fit parameters.  In this section we
extend the notion of error bar to embrace the concept of a ``confidence limit''.

There is a theorem~\cite{press:92} which states that, for a linear model,
if we take simulated data sets assuming Gaussian noise in the data about
the actual data points, and
compute the fit parameters $\vec{a}^{S(i)}, i = 1, 2, \cdots$ for each data
set,
then the probability distribution
of the $\vec{a}^S$ is given by the
multi-variable Gaussian distribution
\begin{equation}
\boxed{
P(\vec{a}^S) \propto \exp\left(-{1 \over 2} \, \sum_{\alpha, \beta}
\delta a_\alpha^S\, U_{\alpha\beta}\, \delta a_\beta^S \right) \, ,}
\label{theorem}
\end{equation}
where $\delta \vec{a}^S \equiv \vec{a}^{S(i)} - \vec{a}^{(0)}$,
$U$, given by Eq.~\eqref{Uab_general},
is the curvature matrix which can also be
defined in terms of the second derivative of
$\chi^2$
according to Eq.~\eqref{curv}, and $\vec{a}^{(0)}$ is the fit to the
actual data set. A proof of this result
is given in Appendix \ref{sec:proof}. It applies for a linear model
with Gaussian noise, and also
for a non-linear model if the uncertainties in the parameters do not extend
outside a region where an effective linear model could be used.
(In the latter case
one still
needs a non-linear routine to \textit{find} the best parameters).
Note that for a
non-linear model, $U$ is not a constant and it
is the curvature \textit{at the minimum} of $\chi^2$ which has to be put into
Eq.~\eqref{theorem}.

From Eq.~\eqref{curv}
the change in $\chi^2$ as the parameters are varied away from the minimum is
given by
\begin{equation}
\Delta \chi^2 \equiv \chi^2(\vec{a}^{S}) -
\chi^2(\vec{a}^{(0)}) = \sum_{\alpha, \beta}
\delta a_\alpha^S\, U_{\alpha\beta}\, \delta a_\beta^S \, ,
\label{Dchisq}
\end{equation}
in which the $\chi^2$ are all evaluated from the single
(actual) data set $y_i^{(0)}$.
Equation \eqref{theorem} can therefore be written as
\begin{equation}
P(\vec{a}^S) \propto \exp\left(-{1 \over 2} \Delta \chi^2 \right) \, .
\label{P_dalpha}
\end{equation}
We remind the reader that in deriving Eq.~\eqref{P_dalpha}
we have assumed the noise in the data is Gaussian
and that either the model is linear or, if non-linear, 
the uncertainties in the parameters do not
extend outside a region where an effective linear model could be used.

Hence the probability of a particular
deviation,
$\delta \vec{a}^S$,
of the fit parameters in a simulated data set away from the
parameters in the \textit{actual} data set,
depends on how much this change increases $\chi^2$ (evaluated from
the actual data set) away from the minimum.
In general a ``confidence limit'' is the range of fit parameter values such
that $\Delta \chi^2$ is less than some specified value. The simplest case,
and the only one we discuss here, is the variation of \textit{one} variable
at a time, though multi-variate confidence limits can also be defined, see
Numerical Recipes~\cite{press:92}.

We therefore consider the change in $\chi^2$ when one variable, $a_1^S$ say, is
held at a specified value,
and all the others $(\beta = 2, 3,\cdots, M)$ are varied in order
to minimize $\chi^2$. Minimizing $\Delta \chi^2$ in Eq.~\eqref{Dchisq} with
respect to $a_\beta^S$ gives
\begin{equation}
\sum_{\gamma=1}^M U_{\beta\gamma}\, \delta a_\gamma^S = 0 , \qquad (\beta = 2, 3, \cdots,M) \, .
\end{equation}
The corresponding sum for $\beta = 1$, namely $\sum_{\gamma=1}^M U_{1\gamma}\,
\delta a_\gamma^S$, is not zero because $\delta a_1$ is fixed. It will be some
number, $c$ say. Hence we can write
\begin{equation}
\sum_{\gamma=1}^M U_{\alpha\gamma}\, \delta a_\gamma^S = c_\alpha, \qquad
(\alpha = 1, 2, \cdots,M) \, ,
\end{equation}
where $c_1 = c$ and $c_\beta = 0\, (\beta \ne 1)$. The solution is
\begin{equation}
\delta a_\alpha^S = \sum_{\beta=1}^M \left(U^{-1}\right)_{\alpha\beta} c_\beta
 = \left(U^{-1}\right)_{\alpha1} c \, .
\label{aalpha}
\end{equation}
For $\alpha = 1$ this gives
\begin{equation}
c = \delta a_1^S / \left(U^{-1}\right)_{11} \, .
\label{c}
\end{equation}
Substituting Eq.~\eqref{aalpha} into Eq.~\eqref{Dchisq}, and using
Eq.~\eqref{c} we find that $\Delta \chi^2 $ is related to $\left(\delta a_1^S\right)^2$
by
\begin{equation}
\Delta \chi^2 = {(\delta a_1^S)^2 \over \left(U^{-1}\right)_{11} } .
\label{Dchi2}
\end{equation}
(Curiously, the coefficient of $(\delta a_1)^2$  is one over the $11$
element of the inverse of $U$, rather than $U_{11}$ which is how it appears in
Eq.~\eqref{Dchisq} in which the $\beta \ne 1$ parameters are free rather than adjusted to
minimize $\chi^2$.)

From Eq.~\eqref{P_dalpha} we finally get
\begin{equation}
P(a_1^S) \propto \exp\left(-{1 \over 2} \,
{(\delta a_1^S)^2  \over\sigma_1^2}\right) \, , 
\label{Pa1S}
\end{equation}
where
\begin{equation}
\sigma_1^2 = \left(U^{-1}\right)_{11} \, .
\end{equation}

As shown in Appendices \ref{sec:proof} and \ref{sec:proof2},
Eqs.~\eqref{theorem}, ~\eqref{P_dalpha} and \eqref{Pa1S}
also apply, under the same conditions (linear model and Gaussian noise
on the data) to the probability for $\delta a_1 \equiv
a_1^\text{true} - a_1^{(0)} $, where we remind the reader that
$a_1^{(0)}$ is the fit parameter obtained from the actual data, and
$a_1^\text{true}$ is the exact value.
In other words the probability of the true value is given by
\begin{equation}
\boxed{
P(\vec{a}^\text{true}) \propto \exp\left(-{1 \over 2} \Delta \chi^2 \right) \, ,}
\label{P_dalphatrue}
\end{equation}
where
\begin{equation}
\Delta \chi^2 \equiv \chi^2(\vec{a}^\text{true}) -
\chi^2(\vec{a}^{(0)})  \, ,
\end{equation}
in which we remind the reader that both values of $\chi^2$ are evaluated from
the single set of data available to us, $y_i^{(0)}$.
Projecting onto a single parameter, as above, gives
\begin{equation}
\boxed{
P(a_1^\text{true}) \propto \exp\left(-{1\over 2}\,
{(\delta a_1)^2  \over \sigma_1^2}\right) \, , }
\label{Pa1}
\end{equation}
so $\langle \left(\delta a_1\right)^2 \rangle = \sigma_1^2 =
\left(U^{-1}\right)_{11}$,
in agreement with what we found earlier in Eq.~\eqref{error_params}.  We
emphasize that Eqs.~\eqref{P_dalphatrue} and~\eqref{Pa1} assume
Gaussian noise on the data points, and that
either the model is linear or, if non-linear, the range of uncertainty in
the parameters is small enough that a description in terms of an effective
linear model is satisfactory.

However we have done more than recover our earlier result,
Eq.~\eqref{error_params}, by more complicated means since we have gained
\textit{additional} information.
From the properties of a Gaussian distribution we now know that,
from Eq.~\eqref{Pa1}, the probability
that $a_\alpha$ lies within one standard deviation $\sigma_\alpha$ of the
value which minimizes $\chi^2$ is 68\%, the probability of its being within
two standard deviations is 95.5\%, and so on.
Furthermore, from Eq.~\eqref{P_dalphatrue}, we see that
\begin{quotation}
\noindent \textit{if a single fit parameter is one
standard deviation away from its value at the minimum of $\chi^2$ (the other
fit parameters being varied to minimize $\chi^2$), then
$\Delta \chi^2 = 1$.}
\end{quotation}

This last sentence, and the corresponding equations
Eqs.~\eqref{P_dalphatrue} and \eqref{Pa1},
are not valid for a non-linear
model if the uncertainties of the parameters extends outside the range where
an effective linear model can be used.
In this situation, to get confidence limits, one should do a bootstrap
resampling of the data, as discussed in the next subsection. Even for a linear
model one needs bootstrap resampling to get confidence limits if the noise on
the data is non-Gaussian.

\begin{center}
\begin{figure}
\includegraphics[width=8cm]{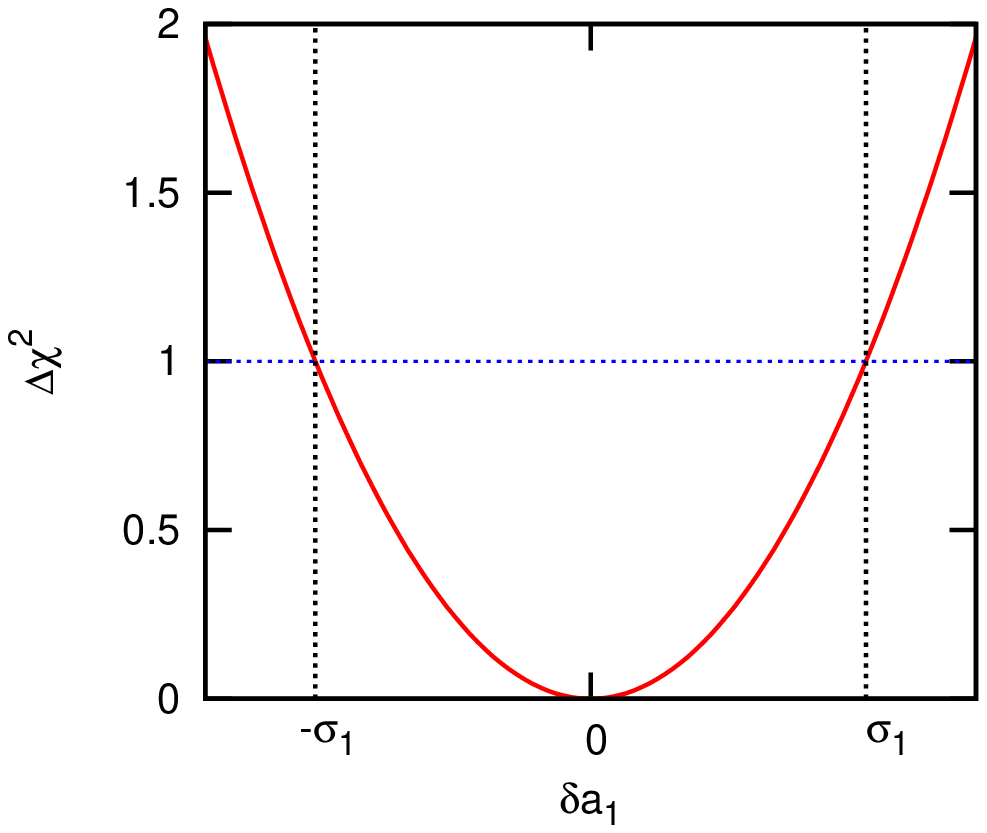}
\includegraphics[width=8cm]{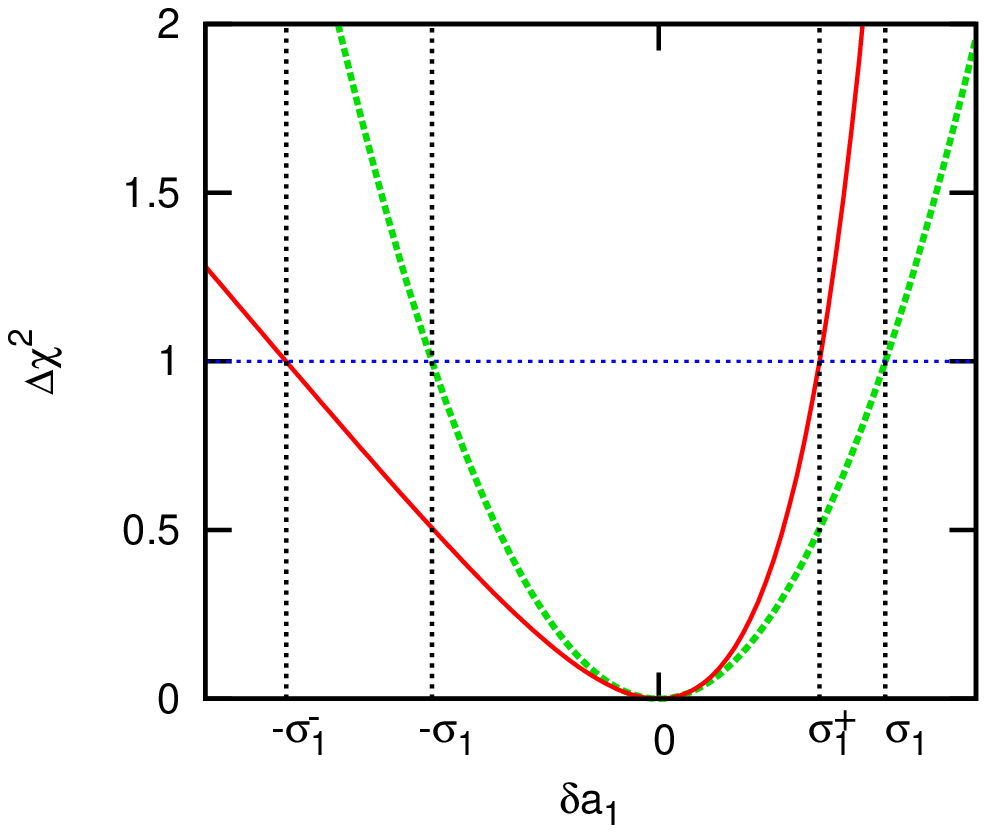}
\caption{{\bf Left:} The change in $\chi^2$ as a fit parameter $a_1$ is varied away from the
value that minimizes $\chi^2$ for a \textit{linear} model. The shape is a parabola for which $\Delta
\chi^2=1$ when $\delta a = \pm \sigma_1$, where $\sigma_1$ is the error
bar.\\
{\bf Right:} The solid curve is a sketch of the change in $\chi^2$ for a
\textit{non-linear} model. The curve is no longer a parabola and is 
not even symmetric. The dashed curve is a parabola which fits the solid curve at
the minimum. The fitting program only has information about the
\textit{local} behavior
at the minimum and so gives an error range $\pm \sigma_1$ at which the value
of the parabola is $1$. However, the parameter $a_1$ is clearly more tightly
constrained on the plus side than on the minus side, so a better way to
determine the error range is to look \textit{globally} and locate
the values of $\delta a_1$ where $\Delta \chi^2 = 1$.
This gives an error bar $\sigma_1^+$ on the plus side, and a different error
bar, $\sigma_1^-$, on the minus side, both of which are different from
$\sigma_1$.}
\label{fig:chi2}
\end{figure}
\end{center}

However, if one
is not able to resample the data we argue that it is better to take the range
where $\Delta \chi^2 \le 1$ as an error bar for each parameter rather
than the error bar
determined from the curvature of $\chi^2$ at the minimum, see
Fig.~\ref{fig:chi2}. The
left hand plot is for a linear model, for which the curve of $\Delta \chi^2$
against $\delta a_1$ is 
exactly a parabola, and the right hand plot is a sketch for a
non-linear model, for which it is not a parabola though it has a quadratic
variation about the minimum shown by the dashed curve.
For the linear case, the values of $\delta a_1$ where
$\Delta \chi^2 = 1$ are the \textit{same} as the values $\pm \sigma_1$, where
$\sigma_1$ is the standard error bar obtained from
the \textit{local} curvature in the vicinity of the minimum.
However, for
the non-linear case, the values of $\delta a_1$ where
$\Delta \chi^2 = 1$ are \textit{different} from $\pm \sigma_1$, and indeed the
values on the
positive and negative sides, $\sigma_1^+$ and $\sigma_1^-$, are not equal.
For the data Fig.~\ref{fig:chi2},
it is clear that the value of $a_1$ is more tightly
constrained on the positive side than the negative side, and so it is better
to give the error bars as $+\sigma_1^+$ and $-\sigma_1^-$, obtained
from the range where $\Delta \chi^2 \le 1$, rather the symmetric range $\pm
\sigma_1$. While it is plausible that the range where
$\Delta \chi^2 \le 1$ is a reasonable estimate of the uncertainty in the fit
parameter, one can not assign a precise confidence limit to it.
If possible, error bars and a
confidence limit should be obtained from an alternative approach, a
bootstrap resampling of
the data as discussed in the next section.

\subsection{Confidence limits by resampling the data}
\label{sec:resample}

Each data point $(x_i, y_i)$ with its error bar $\sigma_i$, comes
from averaging over $N_i$ values of ``raw data'' whose mean is $y_i$ and whose
standard deviation gives $\sigma_i$ according to Eq.~\eqref{finalans}.
If one has access to this raw data, one can do a bootstrap resampling of it
to obtain:
\begin{itemize}
\item
Confidence limits for a linear model if the noise is non-Gaussian.
\item
Both error bars and confidence limits for a non-linear model in which the
range of parameter uncertainty extends outside the region where an effective
linear model is applicable.
\end{itemize}

As discussed in Sec.~\ref{sec:boot},
one generates bootstrap data
sets in which the data points have values $y_{i,\alpha}^B$, where $\alpha$
runs from 1 to $N_\text{boot}$, the number of bootstrap data sets.
The distribution of the $y_{i,\alpha}^B$ for a given $i$
has a standard deviation equal
to the estimate of standard deviation on the mean of the actual data set,
i.e.~$\sigma_i$,
see
Eq.~\eqref{sigmafb} (replacing the factor of $\sqrt{N/(N-1)}$ by unity which 
is valid since
$N$ is large in practice). Hence, if we fit each of the
$N_\text{boot}$ bootstrap data
sets, the scatter of the fitted parameter values will be a
measure of the uncertainty in the values from the single \textit{actual} dataset.
Forming a histogram of the values of a single
parameter we can obtain a confidence interval within which 68\%, say, of the
bootstrap datasets lie (16\% missing on either side) and interpret this
range as a 68\% confidence limit for the actual parameter value. The
justification for this interpretation has been discussed in the statistics
literature, see e.g.\ the references in
Ref.\ \cite{press:92}, but I'm not able to go into the
details here. The method can be applied to both linear and non-linear models,
and does not assume Gaussian noise.

Unfortunately, use of the bootstrap procedure to get error bars in fits to
non-linear models does not yet seem to be a standard procedure in the
statistical physics community.

If one does not have access to the ``raw" data, but is confident that the
noise is close to Gaussian,
another possibility, which is useful for non-linear models, is
is to generate \textit{simulated} data sets, assuming
Gaussian noise on the
$y_i$ values with standard deviation given by the error bars
$\sigma_i$. Each simulated dataset is fitted and the distribution of fitted
parameters is determined. This corresponds to the analytical approach in 
Appendix \ref{sec:proof}
but without the assumption that the model can be represented
by an effective linear one over of the needed parameter range.

\subsection{A tale of two probabilities. When can one rule out a fit?}
\label{sec:lin_or_quad}

In this section we assume that the noise on the data is Gaussian.
We have, so far, considered two different probabilities.
Firstly, as discussed in Appendix \ref{sec:Q}, the value of $\chi^2$ is typically
in the range $\NDOF \pm \sqrt{2 \NDOF}$. 
The quality of fit parameter $Q$ is the probability that,
\textit{given the fit}, the data could have this value of $\chi^2$ or greater,
and is given mathematically by Eq.~\eqref{Q_expression2}. It varies from 
unity when 
$\chi^2 \ll \NDOF - \sqrt{2 \NDOF}$ to zero when $\chi^2 \gg
\NDOF + \sqrt{2 \NDOF}$.
In other words \fbox{$Q$ only changes substantially if $\chi^2$ changes
by an amount of order $\sqrt{\NDOF}$.}
We emphasize that 
$Q$ is the probability of the data given the fit.

Secondly, in the context of error bars and confidence limits, we have
discussed, in Eqs.~\eqref{P_dalphatrue} and
\eqref{Pa1}, the probability that a fit parameter, $a_1$ say,
takes a certain value relative to the optimal one. Equation
\eqref{P_dalphatrue} tells us that the \fbox{relative probability of two fits 
changes substantially when
$\chi^2$ varies by unity.} Note that
Eqs.~\eqref{P_dalphatrue} and
\eqref{Pa1} refer to the relative probabilities of two fits, given the
data.

\begin{center}
\begin{figure}
\includegraphics[width=8cm]{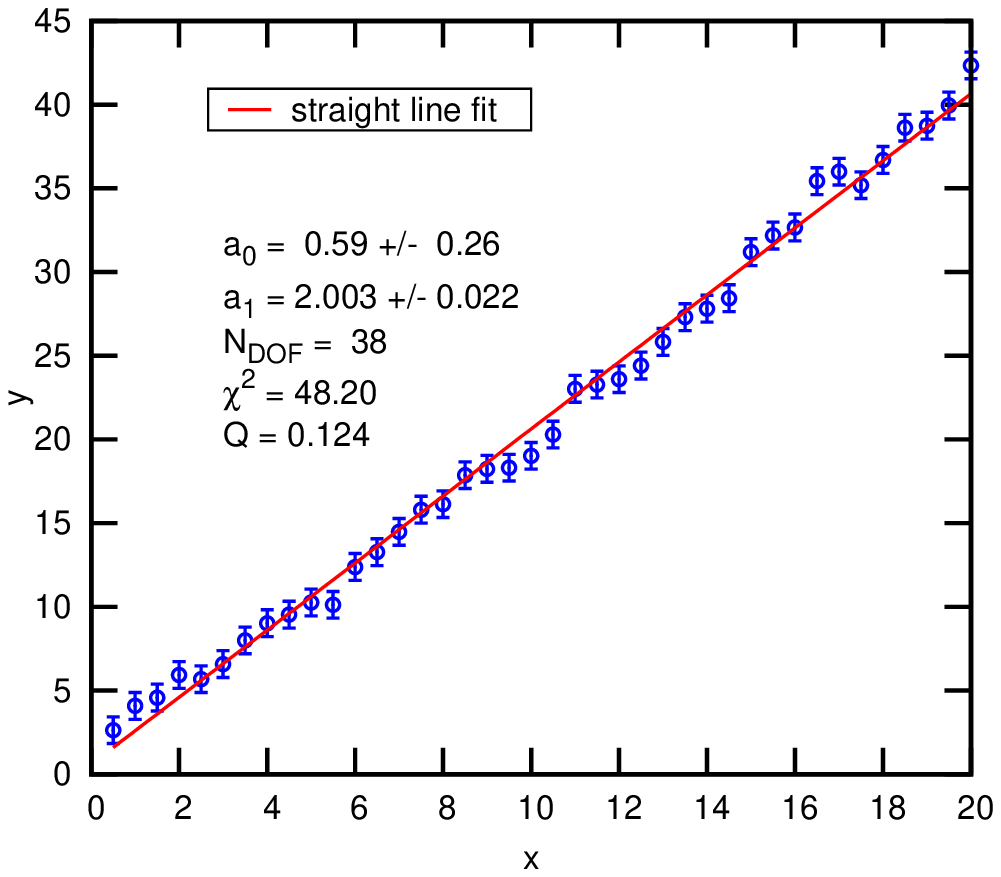}
\includegraphics[width=8cm]{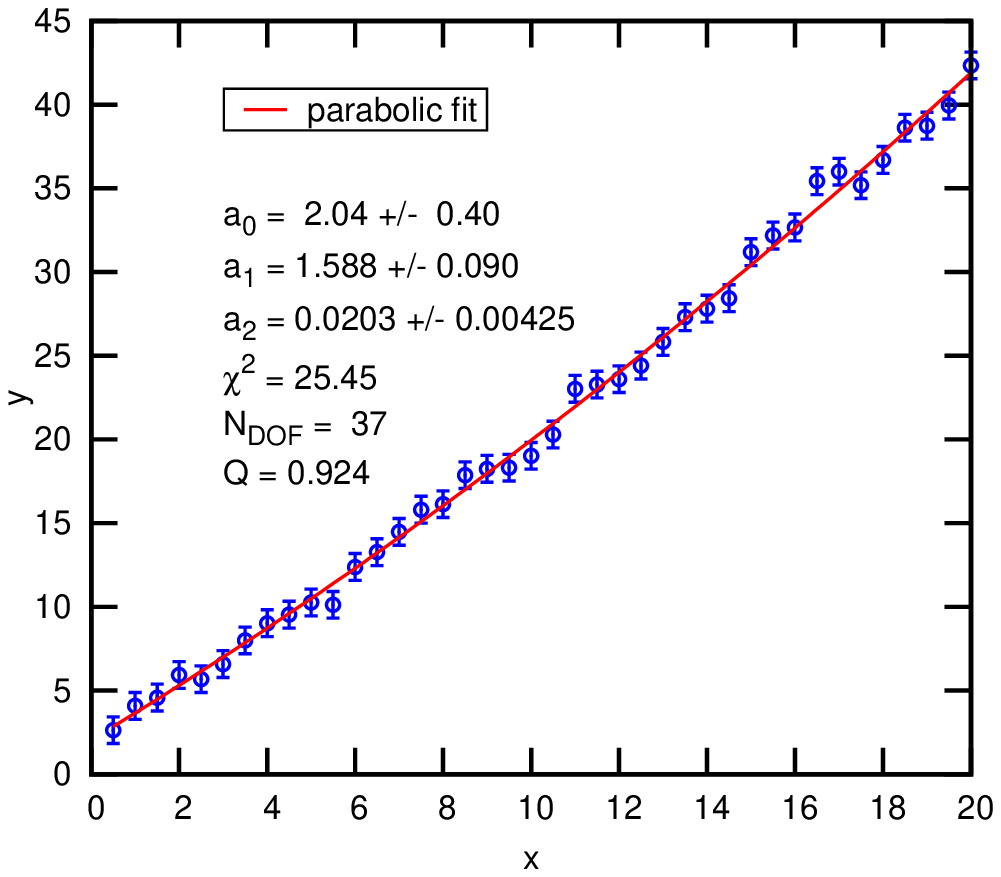}
\caption{{\bf Left:}
A straight-line fit to a data set. The value of $Q$ is reasonable. However,
one notices that the data is systematically above the fit for small $x$ and
for large $x$ while it is below the fit for intermediate $x$. This is unlikely
to happen by random chance. This remark is made more precise in the right
figure.
\\
{\bf Right:} 
A parabolic fit to the same data set. The value of $Q$ is larger than for the
straight-line fit, but the main result is that the coefficient of the
quadratic term is about 5 $\sigma$ away from zero, showing that the straight-line fit in the
left panel is much less likely than the parabolic fit.
}
\label{fig:lin_or_quad}
\end{figure}
\end{center}

At first it seems curious that $Q$ needs a much bigger change in $\chi^2$ to
change significantly than does the relative probability of two fits
($\sqrt{\NDOF}$ rather than 1).
While there is no mathematical
inconsistency, since the two probabilities refer to different situations (one
is the probability of the data given the fit and the other is the relative
probability of two fits given the data), it is useful to understand this
difference intuitively.

We take, as an example, a problem where we want to know whether the data
can be modeled by a straight line, or whether a quadratic term needs to be
included as well. A set of data is shown in Fig.~\ref{fig:lin_or_quad}.

Looking at the left panel one sees that the data more or less agrees with the
straight-line fit ($Q=0.124$). However, one also sees systematic trends: the data is
too high for
small $x$ and for high $x$, and too low for intermediate $x$.
\fbox{The probability that
this trend would occur by chance is very low.} Chi-squared just sums up the
contributions from each data point and is insensitive to any systematic trend
in the deviation of the data from the fit. Hence the value of $\chi^2$, in
itself, does not tell us that this data is unlikely to be represented by a
straight line. It is only when we add another parameter in the fit which
corresponds to those correlations, that we realize the straight-line model is
relatively very unlikely.
In this case, the extra parameter is the coefficient of $x^2$, and the
resulting parabolic
fit is shown in the right figure.

The qualitative comments in the last paragraph
are made more precise by the parameters of the fits.
The straight-line fit gives $a_0 = 0.59 \pm 0.26, a_1 = 2.003 \pm 0.022$
with $\chi^2 = 48.2, Q = 0.124$, whereas the parabolic fit gives
$a_0 = 2.04 \pm 0.40, a_1 =  1.588 \pm 0.090, a_2 = 0.0203 \pm 0.00425$
with $\chi^2=25.5, Q = 0.924$. The actual parameters 
which were used to generate the data are
$a_0 = 2, a_1 = 1.6, a_2 = 0.02$, and there is Gaussian noise with standard
deviation equal to $0.8$. Hence the parameters of the quadratic fit are
correct, but those of the linear fit are not. 
Although the quality of fit factor $Q$ for the straight-line
fit is reasonable, the quadratic fit strongly excludes having the 
fit parameter $a_2$ equal to zero,
since zero is $4.78$ standard deviations away
from the best value. As shown in Eq.~\eqref{erfc}, the probability of a
$4.78$-sigma deviation or greater for a Gaussian distribution
is $\text{erfc}(4.78/\sqrt{2}) \simeq 1.78\times
10^{-6}$, which is tiny.
Thus a careful analysis correctly concludes that the straight-line fit is unlikely
to be correct.

From the figures we see that
difference in $\chi^2$ between the quadratic fit and the straight-line
fit (in which we force $a_2 = 0$) is $22.8$, which should equal 
$(0.0203 / 0.00425)^2$
according to Eqs.~\eqref{Dchi2} and \eqref{error_params},
and indeed it does. This provides a useful check on the
parameters computed by the fit program (gnuplot in this case).

The moral of this tale is that a reasonable value of $Q$ does not, in itself,
ensure that you have the \textit{right} model. Another model might be 
more probable. To quote from Ch.~14 of Numerical
Recipes~\cite{press:92},\footnote{I find the use of the word ``null'' in the
quote to be confusing. It is, however, common usage in the statistics
literature. The brackets round it are mine.}
\begin{quotation}
\noindent ``If a statistic falls in a \textit{reasonable} part of the
distribution, you must not make the mistake of concluding that the (null)
hypothesis is ``verified'' or ``proved''.
That is the curse of statistics. It can never prove things, only disprove
them!''
\end{quotation}
We will discuss more the question of determining the right model (model
selection) in the next section. 

\subsection{Model selection (i.e.\ how to avoid over-fitting):
Maximum Likelihood versus Bayes}
\label{sec:model_selection}

Apart from the last subsection we have \textit{assumed} that the model is given, 
and our goal is to obtained the best fit parameters of that model. In the last
subsection we started a discussion of how to compare \textit{different}
models, which is necessary if the correct model is not known. 

\begin{center}
\begin{figure}[!tb]
\includegraphics[width=8cm]{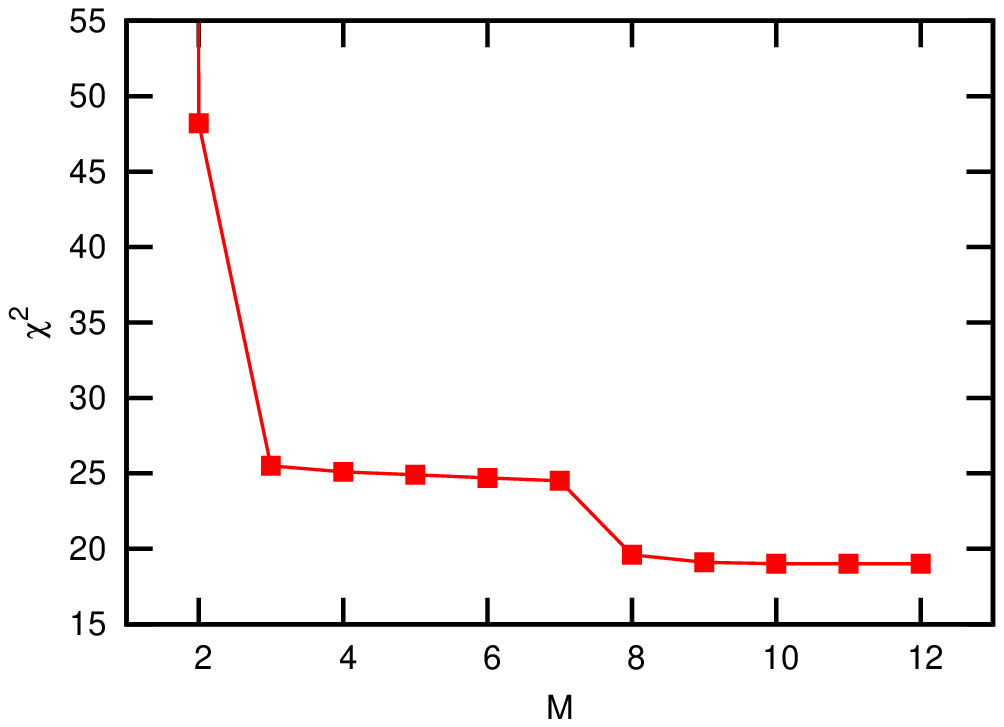}
\includegraphics[width=8cm]{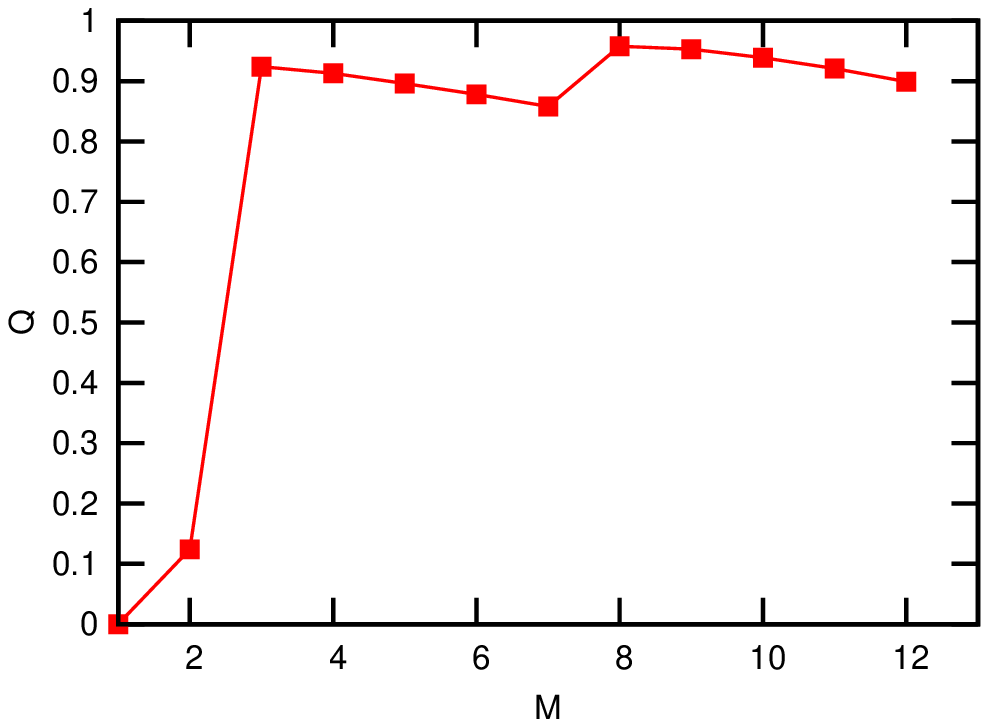}
\caption{
{\bf Left:}
A plot of $\chi^2$ against the number of fit parameters for
polynomial fits to the data in Fig.~\ref{fig:lin_or_quad}. (\textit{Note:} the
order of the polynomial is $M-1$.)  The values of $\chi^2$ decrease
monotonically with $M$, as expected.  There is a rapid drop in going from $M =
2$ to $M=3$, but then a much more gradual decrease. This is consistent with
the fact that the data was generated from a parabolic function plus Gaussian
noise.  The reason for the subsidiary drop in going from $M=7$ to 8 is
unclear. Perhaps, by coincidence, the noise in the data has generated a
noticeable $x^7$ component.
{\bf Right}
The goodness of fit parameter $Q$, see Ref.~\cite{press:92} and
Eq.~\eqref{Q_expression2},
for different values of $M$.
Although the value of $\chi^2$ monotonically decreases with increasing $M$ as
shown in the left panel, the goodness of fit parameter involves $\chi^2$
\textit{per degree of freedom}, where $N_\text{DOF} = N-M$,
and so there is a penalty be paid as the number of fit
parameters increases, since this decreases the number of degrees of freedom.
As a result, $Q$ has a peak at $M=3$, a parabola,
(which indeeds corresponds to the 
function which generated the data) and then (slowly)
decreases. Based on this data, one would come to the (correct) conclusion the
data should be fitted to a parabola. For $M > 3$, which is
the region of over-fitting, the variation of $Q$ with $M$
is non-monotonic, at least in this case.
}
\label{fig:chi2vM}
\end{figure}
\end{center}

For the data shown in Fig.~\ref{fig:lin_or_quad} we showed that the value of
$\chi^2$ for the parabolic fit ($M=3$) is much smaller than that of the
straight-line fit ($M=2$), and we argued that it is therefore preferred. This
is correct since the data was indeed obtained from a parabolic function (plus
noise).  However, suppose we consider higher order fits. Since the fit is
obtained by minimizing $\chi^2$, it is clear that $\chi^2$ can only decrease
with the number of fit parameters $M$. (Here we consider polynomial fits for
which
the order of the polynomial is $M-1$.)  This is illustrated by the left panel
line in Fig.~\ref{fig:chi2vM}, which shows $\chi^2$ against the number of fit
parameters $M$ for $2 \le M \le 12$ for the data in
Fig.~\ref{fig:lin_or_quad}.

\begin{center}
\begin{figure}[!tb]
\includegraphics[width=11cm]{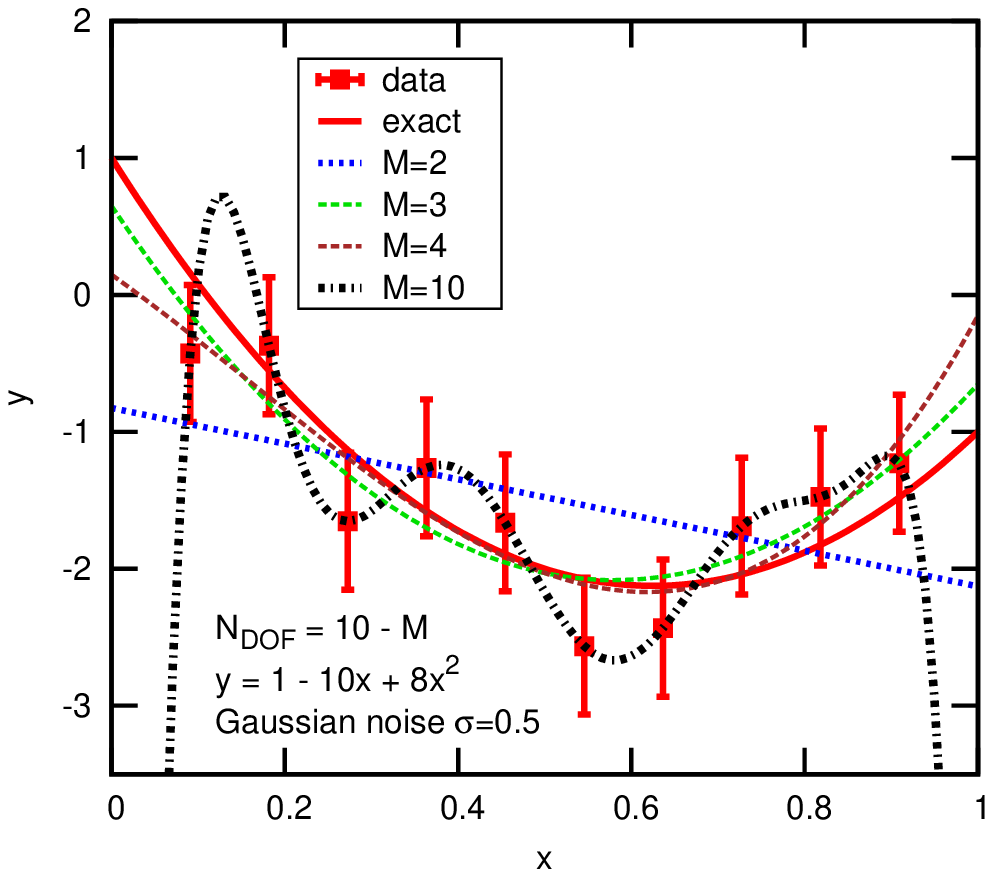}
\caption{
The data points are obtained by adding Gaussian noise with standard deviation
0.5 to the parabolic function $y = 1 - 10x + 8x^2$ (shown by the solid line).
Also shown are
polynomial fits with $M = 2, 3, 4$ and $10$ parameters. The
values of $\chi^2$ for all values of $M$ from 2 through 10 are shown in
the left panel of Fig.~\ref{fig:N10b}.
For $M = 10\ (=N)$ the fit goes perfectly through
the points. However, unphysical oscillations are seen in the fitting function
for $M=10$,
clear evidence of over fitting. Furthermore the fit parameters get very large
when $M$ increases, as shown in Table~\ref{tab:params1}.\\
}
\label{fig:N10}
\end{figure}
\end{center}

Based on the plot of $\chi^2$ in the left panel Fig.~\ref{fig:chi2vM}, 
should we say that a 9-th order polynomial ($M=10$), say,
is to be preferred to
a quadratic fit because it has a smaller $\chi^2$? Intuitively we would say
``no" because we feel that a fit with a smaller number of parameters is
more likely to be correct than a fit with
a larger number, if the quality of the fits is very similar.
Clearly if the number of fit parameters is equal to the number of data points,
40 here, then the fit will go perfectly through the points. However, in
practice this can lead to large oscillations between the points, in order to
force the curve to exactly fit the data. This unphysical result is called
``\textit{over-fitting}''.  In addition, the fit parameters become very large
when over-fitting.

The problem of
determining the correct model (the order of polynomial in the present example)
is called ``\textit{model selection}''.

\begin{table}[!tb]
\begin{tabular}{|c || c | c | c | c | c || c|}
\hline
           &$\quad M=2\quad$&$\quad M=3\quad$ &$\quad M=4\quad$&$\quad M=7\quad$ &\quad $M = 10\quad$ &$\quad$exact\\
\hline\hline
$a_0$ &  -0.826       & 0.650            & 0.148          &   -2.206       & -31.02   & 1\\
$a_1$ &  -1.303       & -9.42            & -4.52          &   44.14        & 740.3    & -10\\
$a_2$ &               & 8.12             & -3.57          &   -364.9       & -6329.6  & 8\\
$a_3$ &               &                  & 7.79           &   1304.1       &24157.5   & \\
$a_4$ &               &                  &                &  -2407.2       &-35884.6  & \\
$a_5$ &               &                  &                &  2210.4        &-29162.7  & \\
$a_6$ &               &                  &                &  -789.6        &185757    & \\
$a_7$ &               &                  &                &                &-275351   & \\
$a_8$ &               &                  &                &                & 183538   & \\
$a_9$ &               &                  &                &                & -47449.3 & \\
\hline
\end{tabular}
\caption{
Fit parameters $a_\alpha$ for
polynomial fits to the data shown in Fig.~\ref{fig:N10} for different numbers
of fit parameters $M$.  For the larger values of $M$ the fit parameters are
very large indicating severe over-fitting. (\textit{Note:} the order of the
polynomial is $M-1$.)
\label{tab:params1}
}
\end{table}

We shall apply different approaches to the model selection
problem to two sets of data:
that in Fig.~\ref{fig:lin_or_quad} and also the smaller data set with just ten
points shown in Fig.~\ref{fig:N10}.  Table~\ref{tab:params1} shows the
parameters for all possible polynomial fits to the data in Fig.~\ref{fig:N10}.

We will discuss two approaches to the problem of model selection and
over-fitting.  These correspond to the
the two basic approaches to
statistics, the one more familiar to most physicists is called
``frequentist'', and the other is called ``Bayesian''. 
The frequentist
approach is called ``maximum likelihood'' in the context of fitting, and is
just the least-square method described
in the earlier part of these notes. We explain in 
subsection~\ref{sec:max_like} below
why last-squares is
called maximum likelihood and how one can approach model selection
within this approach. However, a fully systematic, maximum likelihood
method for model selection does not
appear to have been developed.
The alternative Bayesian approach is described in
in Sec.~\ref{sec:Bayes}. It has been argued elsewhere, e.g.~\cite{bishop:06},
to provide provides a systematic,
robust method of model selection which avoids over-fitting. However, we shall
find some reservations about this approach when we apply it to actual data.


\subsubsection{Maximum likelihood}
\label{sec:max_like}

First we will show that the ``frequentist'' approach to statistics, called
maximum likelihood in the context of fitting, corresponds to the least-squares
approach discussed up to now in these notes.
In the frequentist approach to statistics, we determine the probability of
a particular event in a random process
by repeating the process many times and dividing the
number of times the specified event occurred by the total number of events. In
the limit when the number of events tends to infinity this ratio tends to the
actual probability. This is to be distinguished from the ``Bayesian''
approach, discussed in the next subsection,
in which, in addition to the data, we include 
\textit{extra} information in the form
of a ``prior'' distribution for some parameters.

In curve fitting, we only have one set of data, but, as we have
repeatedly emphasized in these notes, we obtain unbiased estimates of fit parameters
and their uncertainties by a thought experiment in which we consider the results
that \textit{would} be obtained if we could obtain many data sets. 
In particular, the error bars in the fit parameters come precisely from the scatter that would be
obtained by repeating the fit on many data sets, \textit{assuming that the fit
from the one set of data that we actually have is correct}.  If, as we shall
do in the rest of this section, we assume Gaussian noise on the data, the error bars 
determine the whole probability
distribution of the data.

\fbox{Thus the frequentist approach gives 
the probability of the data given the fit.} This seems a bit strange. We would
really like to know what is the probability that the set of fitted parameters
in correct. However, as stated by Numerical Recipes~\cite{press:92} (implicity assuming
the frequentist approach)
\begin{quotation}
\noindent ``$\ldots$ there is no statistical universe of models from which the
parameters are drawn. There is just one model, the correct one, and a statistical
universe of data sets that are drawn from it!"
\end{quotation}

Thus, as mentioned earlier in these notes, in the frequentist approach,
we take the probability of the data
given the fit, as a measure of whether the fit parameters are likely to be
right.
If we assume that the fit is correct, and the noise on the data is Gaussian
with variance $\sigma_i^2$, then this probability is
\begin{equation}
\boxed{
P(\{y\}) = 
{1 \over (2 \pi)^{N/2} \left(\prod_{i=1}^N\sigma_i\right)} \, \exp\left[-{1 \over 2}
\sum_{i=1}^N
\left({y_i - \sum_\alpha a_\alpha X_\alpha(x_i) \over \sigma_i} \right)^2
\right]  \, .}
\label{max_like}
\end{equation}
For ease of notation, we have written this for a linear model, but the
generalization to a non-linear model is obvious.

A sensible approach, then, is to maximize this probability, which is known as
``maximum likelihood'' method. It is equivalent to maximizing the expression
in the exponential in Eq.~\eqref{max_like},
which is just $(-1/2)$ times $\chi^2$, and hence to
minimizing $\chi^2$.\textit{Thus, in the context of fitting, maximum
likelihood is equivalent to the standard least-squares approach.}

We can use the result that least-squares is equivalent to maximum likelihood
to formulate the
least squares method for correlated data. This is described in
Appendix~\ref{corr_data}.

Now we now that least-squares corresponds to maximum likelihood
(frequentist) approach to statistics, we can use it to tackle the problem of
overfitting.

The intuition that the simplest model which fits the data is to be preferred
can be inferred from the results for $\chi^2$ in
in Fig.~\ref{fig:chi2vM}
since $\chi^2$
decreases \textit{considerably}
in going from $M=2$ (a straight line) to 3 (a parabola), but then decreases
only \textit{slightly}
for larger $M$. We might therefore conclude ``\textit{by eye}''
that underlying model is a parabola. But how can we select the right model in a
systematic manner?

Intuitively, we would like to add a penalty to
$\chi^2$ which increases with $M$ and then look at the minimum of the
resulting quantity as a function of $M$.  One way to do this is to consider
the quality of fit factor $Q$~\cite{press:92}, given in
Eq.~\eqref{Q_expression2} since this involves the value of $\chi^2$
\textit{per degree of freedom},
so if $\chi^2$ does not decrease by much on increasing $M$
by one, the value of $\chi^2$ per degree of freedom will increase, so
$Q$ can decrease. This is shown in the right part of
Fig.~\eqref{fig:chi2vM}
where a peak in $Q$ is seen at $M=3$ (the correct value). At larger
values of $M$, in the region of over-fitting, the variation of $Q$ with $M$ is
non-monotonic, at least in this case.



\begin{center}
\begin{figure}[!tb]
\includegraphics[width=8cm]{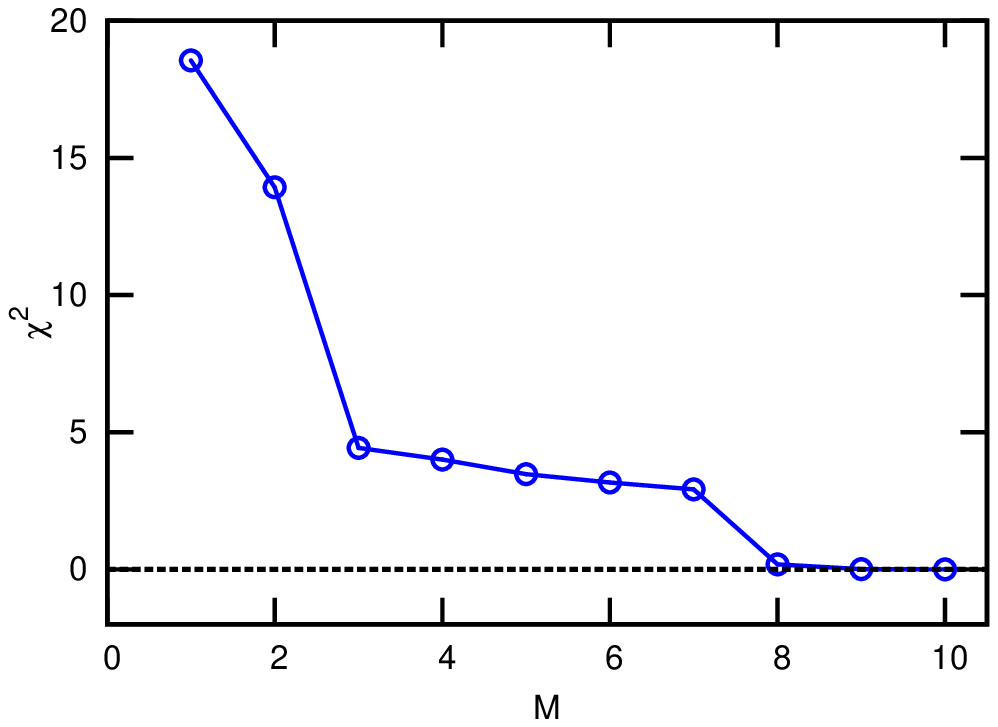}
\includegraphics[width=8cm]{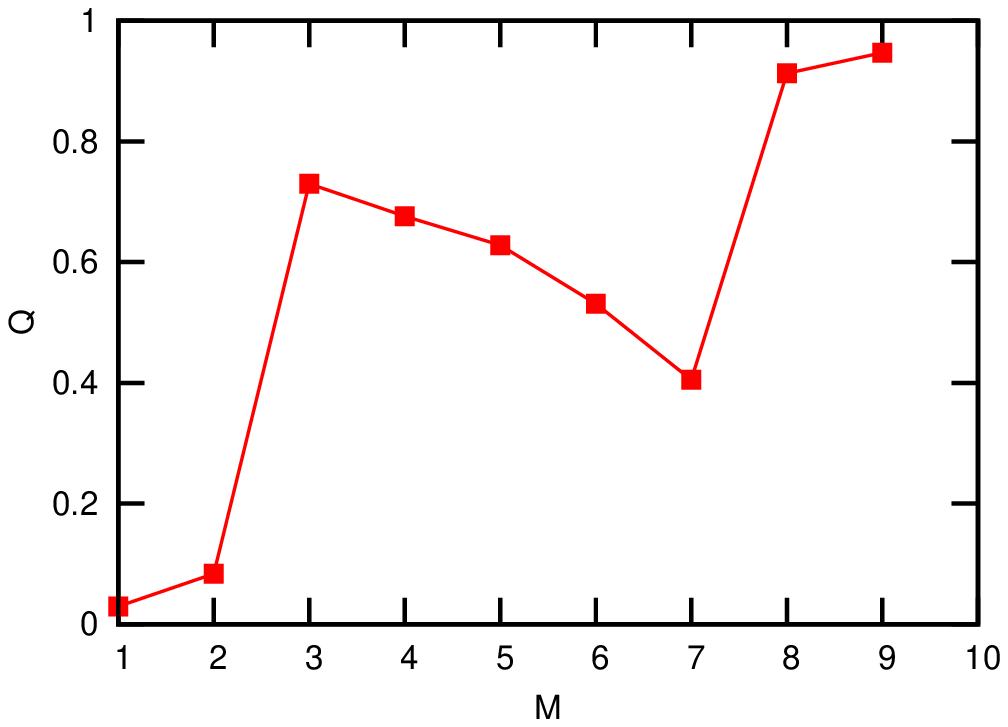}
\caption{
{\bf Left:}
A plot of $\chi^2$ versus $M$ for the data in Fig.~\ref{fig:N10}.
This shows a rapid drop in going
from $M=2$ to $3$ but then continues to decrease on further increasing $M$.
However, the data is being over-fitted in this region.
{\bf Right:}
A plot of the confidence of fit factor $Q$ for the data shown in
Fig.~\ref{fig:N10}.  This shows a peak at $M=3$, followed by a gradual
decrease.  (For still larger values of $M$, $Q$ increases again, presumably 
connected with
the fact that the fit is perfect for $M = N$.)
}
\label{fig:N10b}
\end{figure}
\end{center}

We have also analyzed the data in Fig.~\ref{fig:N10} which
has just 10 data points.  The data are from a quadratic function, 
$f(x) = 1 - 10 x + 8 x^2$, plus Gaussian noises with standard deviation 0.5.
The value of $\chi^2$, shown in the left part of
Fig.~\ref{fig:N10b}, decreases considerably when $M$ in increases
from 2 and 3, but decreases much more slowly for the next few values of $M$.
Of course, $\chi^2 = 0$ for $M = N\  (= 10)$. Fit parameters for different
values of $M$ are shown in Table~\ref{tab:params1} and the corresponding values
of $Q$ are shown in the right part of Fig.~\ref{fig:N10b}. A clear peak is seen
for $M=3$, the parabolic fit, indicating, again correctly, that a 
parabola is the
right function to model the data.

\subsubsection{A Bayesian approach}
\label{sec:Bayes}

The Bayesian approach heavily uses ``conditional'' probability distributions,
so we start by explaining these.
Consider two random variables $x$ and $y$.
We write $P(X, Y)$ as the probability that $x$
has the value $X$ and $y$ has the value $Y$, and
also write $P(X)$ as the probability
that $x$ has value $X$, irrespective of the value of $y$.
Clearly $P(X)$ can be obtained by
\textit{summing} $P(X, Y)$ over $Y$, i.e.
\begin{equation}
P(X) = \sum_Y P(X, Y) \, .
\label{PX}
\end{equation}
We also
need \textit{conditional} probabilities, such as $P(X|Y)$, which is the probability that $x$
has value $X$ \textit{given} that $y$ has value $Y$, and $P(Y|X)$ which 
is the probability that $y$ has value $Y$ given that $x$ has value $X$.  

We can relate $P(X, Y)$ to conditional probabilities
in two different ways. Firstly, we can determine the
probability that $x$ has value $X$, and 
multiply this by the conditional
probability of $Y$ given $X$. Alternatively, we can do the same thing with $X$ and $Y$
interchanged. Thus we have
\begin{equation}
P(X, Y) = P(Y|X)\, P(X)  =  P(X|Y) P(Y) \, .
\label{PXY}
\end{equation}
If we divide by $P(X)$ we get 
\begin{equation}
\boxed{P(Y|X) = {P(X|Y)\, P(Y) \over P(X)} \, , }
\label{bayes}
\end{equation}
which is known as \textit{Bayes' theorem} and plays a central role in
the Bayesian approach to statistics.
From Eqs.~\eqref{PX} and \eqref{PXY} we have
\begin{equation}
P(X) = \sum_Y P(X|Y)\, P(Y) ,
\end{equation}
so the denominator in
Eq.~\eqref{bayes} is the normalizing constant which ensures that the sum of
the conditional probabilities on the left hand side over all values of $Y$ is
equal to one.  We can therefore
also write Bayes' theorem in the form
\begin{equation}
P(Y|X) = {P(X|Y)\, P(Y) \over \sum_{Y'} P(X|Y')\, P(Y')} \, .
\end{equation}

In the Bayesian approach to statistics, Eq.~\eqref{bayes} is interpreted in
the following way. $P(Y)$ is taken to be some knowledge about $Y$ which has been acquired
before the statistical analysis is performed, and is called the ``\textit{prior}''. Data
is then used
to determine $P(X|Y)$, the probability of $X$ given this value of $Y$. Bayes'
theorem then gives $P(Y|X)$, the probability of $Y$ now that the
statistical information about
$X$ has been included, and is called the ``\textit{posterior}'' distribution.
If, later on, more information is available, say that a third variable $z$ has
value $Z$, then $P(Y|X)$ can be taken as the prior, and the posterior
distribution for $y$, including information about \textit{both} $x$ and $z$ is given
according to Bayes' theorem by
\begin{equation}
P(Y|X, Z) = {P(Z|Y,X)\, P(Y|X) \over P(Z|X)} \, , 
\label{PXYZ}
\end{equation}
where $P(Y|X, Z)$ is the conditional probability for $Y$ given both that $x=X$
and $z=Z$, and the denominator is equal to $\sum_Y P(Z|Y,X)\, P(Y|X)$. If
Eq.~\eqref{PXYZ} looks confusing, note that $X$ is fixed throughout and could
be omitted from the notation, in which case Eq.~\eqref{PXYZ} just Bayes' theorem,
Eq.~\eqref{bayes}, for $Y$ and $Z$.

Bayesian statistics is sometimes regarded with suspicion on the grounds that
the prior distributions used seem either to be subjective or to be determined by
mathematical convenience rather than physical intuition. Nonetheless, Bayesian
methods can be very useful in many situations,
and the problem of model selection in
fitting is claimed~\cite{bishop:06}
to be one of them. Recall from Fig.~\ref{fig:N10}
and Table \ref{tab:params1} that the problem with maximum likelihood methods
is that they prefer complicated models which over-fit the data, give
oscillatory behavior between the data points, and produce unphysically large
values for the fit parameters.

\begin{quotation}
\noindent Thus, we have the notion that
simpler models are better, and this is precisely the sort of additional
information that is included in Bayesian analysis in the form of a prior.
Hence a Bayesian analysis is natural for the model selection problem.
\end{quotation}

Our goal is to determine the best ``model'', ${\mathcal M}$, to fit the data.
Specifying a model requires specifying a functional form 
and the number of fit parameters $M$. Since we 
stick to polynomials here we can indicate a model simply by specifying
the number of parameters $M$. We want to compute the relative
probabilities of different values of $M$,
given the data, i.e.\ $P(M | D)$, where the symbol $D$ indicates our
set of data.

To do this we need 
some additional information to prevent over-fitting. We
therefore introduce a prior for the fit parameters, and a mathematically
convenient choice is a Gaussian,
\begin{equation}
P(F|{\gamma}) =
\left({\gamma \over  2 \pi}\right)^{M/2} \exp\left[-{\gamma\over 2}\, \sum_{\alpha=1}^M a_\alpha^2
 \right] \, ,
\label{PFGauss}
\end{equation}
in which we write $F$ to symbolically indicate the fitting parameters.  The
quantity $\gamma$ is called a ``\textit{hyperparameter''} because it controls
the parameters of the fit.  Eq.~\eqref{PFGauss} gives the probability
of the fit parameters given a particular value for the hyperparameter.
For simplicity we have taken the same value of
$\gamma$ for each of the fit parameters. Clearly Eq.~\eqref{PFGauss} can be
criticized for the reason mentioned above, namely that it is chosen for
mathematical convenience rather than any real prior information.  Nonetheless,
we shall see that it serves the purpose of penalizing over-fitting. 

The probability of $M$ given the data is obtained by summing over all
possible values of $\gamma$ so we have
\begin{equation}
P(M|D) = \sum_\gamma P(\gamma| D) 
 = \sum_\gamma {P(D | \gamma) P(\gamma) \over P(D)}\, ,
\label{PMD}
\end{equation}
where we used Bayes' equation to get the last equality.
The denominator is a constant independent of the model or fit parameters and
will be ignored from now on. What do we take for $P(\gamma)$? Since we have no
information on it, one might think of setting it to a constant. However, a
constant distribution between $0$ and $\infty$ is not normalizable, so we
have to put in bounds. This is tricky because we have no idea of
$\gamma$ is very large or very small or in between. One does better by
taking $P(\log \gamma)$ to be constant because, even though one still needs
bounds,
one can cover a huge range of magnitudes. Hence we take $P(\gamma) \propto
1/\gamma$.  We expect that $P(D | \gamma)$ will be fairly sharply peaked at
some value $\hat{\gamma}$ and that this value will dominate the sum in
Eq.~\eqref{PMD}. Hence we have
\begin{equation}
P(M|D) \propto {P(D | \hat{\gamma}) \over\hat{\gamma} } \, ,
\label{PMD'}
\end{equation}
and so the ratio of probabilities for two different values of $M$ is
\begin{equation}
{P(M_1|D) \over P(M_2|D)} =  {P_{M_1}(D | \hat{\gamma}_{M_1}) \over
P_{M_2}(D | \hat{\gamma}_{M_2})} \, {\hat{\gamma}_{M_2} \over
\hat{\gamma}_{M_1}} \, ,
\label{rat_P}
\end{equation}
where we temporarily indicate that the distribution $P(D|\gamma)$ depends on
the number of fit parameters $M$.
(The ratio of the $\gamma$ values in Eq.~\eqref{rat_P}
comes from our choice of a uniform
distribution for $P(\ln\gamma)$ and is often ignored.)

Hence our goal is to determine $P(D | \gamma)$ and maximum it
with respect to $\gamma$.  Now $P(D | \gamma)$ 
is obtained by \textit{summing} over all
possible values for the fit parameters, i.e.
\begin{equation}
P(D| \gamma) = \sum_{F} P(D| F) \, P(F | \gamma) \, .
\label{PDmu_a}
\end{equation}

The probability of the data given the fit, $P(D|F)$, is
given by the maximum likelihood result in
Eq.~\eqref{max_like}, namely 
\begin{equation}
P(D|F) = 
{1 \over (2 \pi)^{N/2} \left(\prod_{i=1}^N\sigma_i\right)} \, \exp\left[-{1 \over 2}
\sum_{i=1}^N
\left({y_i - \sum_\alpha a_\alpha X_\alpha(x_i) \over \sigma_i} \right)^2
\right]  \, .
\label{max_like2}
\end{equation}
Multiplying this by Eq.~\eqref{PFGauss} and substituting into
Eq.~\eqref{PDmu_a} we get
\begin{equation}
P(D|\gamma) = \left({\gamma \over  2 \pi}\right)^{M/2} \, 
{1 \over (2 \pi)^{N/2} \left(\prod_{i=1}^N\sigma_i\right)} \,
\prod_{\alpha=1}^M \left( \int_{-\infty}^\infty d a_\alpha\right)\, \exp\left[-{1 \over 2}
E_0(\{a_\alpha\}) \right] \, ,
\label{PDmu}
\end{equation}
where the cost function, $E_0(\{a_\alpha\})$, is given by
\begin{equation}
\boxed{
E_0(\{a_\alpha\}) =
\sum_{i=1}^N \left({y_i - \sum_{\alpha=1}^M a_\alpha X_\alpha(x_i) \over \sigma_i} \right)^2 +
\gamma\sum_\alpha a_\alpha^2 \, .}
\label{E_0}
\end{equation}
The first term in $E_0(\{a_\alpha\})$
is just $\chi^2$ and the second term has the effect of
a ``\textit{regularizer}'' which penalizes
fits with large parameter values. Over-fitting 
leads to very large parameter values, see Table \ref{tab:params1} for an example,
so the second term in Eq.~\eqref{E_0} acts as to suppress
over-fitting, as desired.


We need to find a minimum of Eq.~\eqref{E_0} with respect to the fit parameters
(equivalent to the
maximum of the exponential in Eq.~\eqref{PDmu}), which is
straightforward because it is quadratic function of
the parameters (a mathematical advantage of the Gaussian prior). The 
solution for the parameters is
still given by Eq.~\eqref{soln}, with the $v_\alpha$ still given by
Eq.~\eqref{v_general}, and the only change is that $U_{\alpha\beta}$ now has an extra
term involving $\gamma$, 
\begin{equation}
\boxed{
U_{\alpha\beta} = \sum_{i=1}^N {X_\alpha(x_i)\, X_\beta(x_i) \over \sigma_i^2}
+ \gamma \, \delta_{\alpha\beta}
\, .}
\label{Uab_reg}
\end{equation}

\noindent\fbox{We will write the values of
fit parameters at the minimum of $E_0$ as
$\hat{a}_\alpha$.} \\
\noindent From Eq.~\eqref{soln} these are given by
\begin{equation}
\boxed{
\hat{a}_\alpha = \sum_{\beta=1}^M \left(U^{-1}\right)_{\alpha\beta}\, v_\beta \, ,}
\label{ahat}
\end{equation}
where now $U$ is given by Eq.~\eqref{Uab_reg} and $v$ by
Eq.~\eqref{v_general}.
These maximize the exponential in Eq.~\eqref{PDmu} and are called
the \textit{maximum posterior} (or MAP) values of the fit parameters.

The expectation
value of $a_\alpha$ using the Gaussian weight in Eq.~\eqref{PDmu} is just the
optimal value  $\hat{a}_\alpha$.
To perform the integrals in Eq.~\eqref{PDmu} we expand the
$a_\alpha$ about $\hat{a}_\alpha$ and perform the resulting Gaussian integrals
by completing the square. The result is
\begin{equation}
P(D|\gamma) = {\gamma^{M/2} \over (\det U)^{1/2}} \, 
{1 \over (2 \pi)^{N/2} \left(\prod_{i=1}^N\sigma_i\right)} \,
\exp\left[-{1 \over 2} E_0(\{\hat{a}_\alpha\}) \right] \, ,
\label{PDmu2}
\end{equation}
see the discussion below Eq.~\eqref{eq1} for an explanation of where the
determinant comes from.
Because $\gamma$ in Eq.~\eqref{Uab_reg} only appears proportional to the identity
matrix, the eigenvectors of $U$ are independent of $\gamma$ and so the eigenvalues
can be written as
$\lambda_I = \lambda_I^{(0)} + \gamma$, where $\lambda_I^{(0)}$
is the eigenvalue in the unregularized case with $\gamma = 0$. We use an
uppercase Roman letter to label one of the $M$ eigenvalues.
Hence, from Eq.~\eqref{PDmu2}, 
the log of the probability of the data given the
hyperparameter $\gamma$ is given by
\begin{equation}
\boxed{
\ln P(D|\gamma) = -{1 \over 2}\, \overline{E}(\{\hat{a}_\alpha\}, \gamma) , }
\label{lnPDgamma}
\end{equation}
where the cost function, $\overline{E}(\{\hat{a}_\alpha\}, \gamma)$,
which we have to minimize with respect to
$\gamma$, is given by
\begin{align}
\overline{E}(\{\hat{a}_\alpha\}, \gamma)  & =
E_0(\{\hat{a}_\alpha\}) + 
\sum_{I=1}^M
\ln \left({\lambda_I^{(0)} + \gamma  \over \gamma}\right)
+ \sum_{i=1}^N \ln(2\pi\sigma_i^2) \label{E_E0}\\
& \boxed{
=
\sum_{i=1}^N \left({y_i - \sum_{\alpha=1}^M \hat{a}_\alpha X_\alpha(x_i)
\over \sigma_i} \right)^2 + \gamma \,\sum_{\alpha=1}^M
\hat{a}_\alpha^2 +
\sum_{I=1}^M
\ln \left({\lambda_I^{(0)} + \gamma  \over \gamma}\right)
+ \sum_{i=1}^N \ln(2\pi\sigma_i^2) \, , }
\label{Ebar}
\end{align}
where we have written $\det U = \prod_I (\lambda_I^{(0)} + \gamma)$.
We should mention that
the eigenvalues $\lambda_I^{(0)}$  are \textit{independent} of both the
parameters  $\hat{a}_\alpha$ and also $\gamma$ since they come from
diagonalizing the matrix $U$ in Eq.~\eqref{Uab_reg} without the $\gamma$ term.
In the machine learning literature
$P(D|\gamma) \equiv \exp(-\overline{E}(\{\hat{a}_\alpha\}, \gamma)/2)$ is
called the \textit{evidence function}~\cite{bishop:06}.

We now discuss each of the terms in the cost function $E$ in Eq.~\eqref{Ebar}.
\begin{enumerate}
\item
The first term is just $\chi^2$.
\item
Intuitively we want to add to $\chi$
a term which increases with
$M$ to penalize over-fitting.  This role is played by the the second and third
terms in Eq.~\eqref{Ebar}, which are each the sum of $M$ factors.
The second term is the regularizer which was discussed after
Eq.~\eqref{E_0}. 
\item
The third term is the most interesting. From Eq.~\eqref{PFGauss}
we see that
$1/\gamma$ is the variance of the prior distribution for each of the
$a_\alpha$. Further, $1/(\lambda_I^{(0)} + \gamma)$, being the $I$-th
eigenvalue of the
covariance matrix, is the variance of the posterior distribution of the
linear combination of fit parameters
corresponding to the $I$-th eigenvector.
We'll call this $\sigma_I^2$.
Since we chose the same value of
$\gamma$ for each of the fit parameters, the variance of the prior
distribution for eigenvector $I$
is also $1/\gamma$ for all $I$.
We'll denote this by $\left(\sigma_I^{(P)}\right)^2$.
We therefore use the following definitions,
\begin{equation}
\sigma_I^{(P)} = {1 \over \gamma^{1/2}}, \qquad 
\sigma_I = {1 \over \left(\gamma + \lambda_I^{(0)}\right)^{1/2}}, \qquad 
\sigma_I^{(0)} = {1 \over \left(\lambda_I^{(0)}\right)^{1/2}}, 
\quad (I= 1, 2, \cdots, M)\, ,
\end{equation}
in which we also define $\sigma_I^{(0)}$ to be the standard
deviation in the estimate of the $I$-th eigenvector for
$\gamma=0$, see Fig.~\ref{fig:post_prior} for an illustration.
Hence the third
term in Eq.~\eqref{Ebar} can be written as
\begin{equation}
2 \sum_{I=1}^M \ln \left({\sigma^{(P)}_I \over \sigma_I}\right) \, .
\label{sig_PrPo}
\end{equation}
Equivalently, from Eqs.~\eqref{lnPDgamma} and \eqref{Ebar} the (multiplicative)
contribution of this term to $P(D|\gamma)$ is
\begin{equation}
\boxed{
P(D|\gamma) \propto \prod_{I=1}^M \left({\sigma_I \over \sigma^{(P)}_I}
\right)\, .}
\label{PDgamma_sig_PrPo}
\end{equation}

As a reminder, the factor of $\sigma^{(P)}_I \ (= \gamma^{-1/2})$ in
Eq.~\eqref{sig_PrPo} or \eqref{PDgamma_sig_PrPo} comes from the
normalization of the prior distribution in Eq.~\eqref{PFGauss},
while the factor of
$\sigma_I$, the width of the distribution of posterior distribution in fit
parameter $I$, comes from integrating over the fit parameters in Eq.~\eqref{PDmu}
(see also Eq.~\eqref{PDmu_a}). We see that the third term provides a penalty
given by Eq.~\eqref{sig_PrPo} for each parameter $I$ which is not much affected by
the regularization, i.e.\ $\gamma \ll \lambda_I^{(0)}$.
In this way, we prevent the minimum of the cost function
being at $\gamma = 0$, which would just give the maximum likelihood result.

\begin{center}
\begin{figure}
\includegraphics[width=10cm]{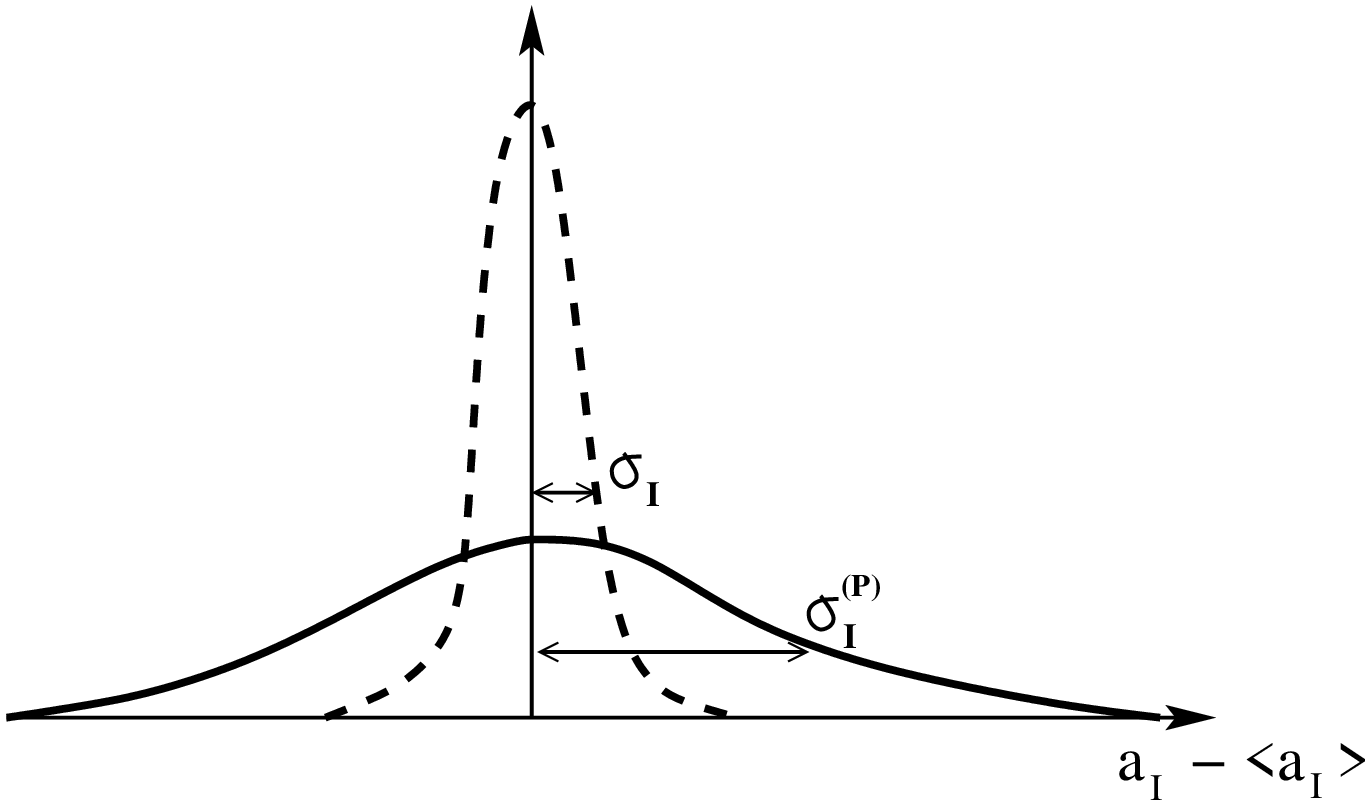}
\caption{
The posterior (dashed line) and prior (solid line)
distributions for a combination of fit parameters
corresponding to the $I$-th eigenvector, $a_I$, of the covariance matrix. The
width of the prior distribution, $\sigma_I^{(P)}$ is equal to $1/\gamma^{1/2}$
(for all $I$). The width of the posterior distribution, $\sigma_I$, is equal to
$1/(\gamma + \lambda_I^{(0)})^{1/2}$ where $\lambda_I^{(0)}$ is the $I$-th
eigenvalue of the matrix $U$ (the inverse of the covariance matrix) in the
absence of the hyperparameter $\gamma$. Parameters which are not much affected
by the regularization due to $\gamma$ have
$\lambda_I^{(0)} \gg \gamma$, so
$\sigma_I \ll \sigma_I^{(P)}$, while parameters which are
affected by the regularization have $\lambda_I^{(0)} \ll \gamma$, which gives
$\sigma_I \simeq \sigma_I^{(P)}$. The third term in Eq.~\eqref{Ebar} gives a
penalty of $2 \ln\left( \sigma_I^{(P)} / \sigma_I\right)$, see
Eq.~\eqref{sig_PrPo},
to the cost function $E$ for each
parameter in the fit which is not significantly altered by regularization.
}
\label{fig:post_prior}
\end{figure}
\end{center}

\item
The fourth and last term in Eq.~\eqref{Ebar} depends on the data but
is independent of the fit parameters or the
hyperparameter $\gamma$ and so will be omitted in our subsequent
discussion.
\end{enumerate}

We emphasize that the maximum posterior (MAP) values of the fitting parameters,
$\hat{a}_\alpha$, depend on the hyperparameter $\gamma$. However, when we
minimize the cost function $\overline{E}(\{\hat{a}_\alpha\},\gamma)$
in Eq.~\eqref{Ebar}
with respect to $\gamma$ we can neglect this dependence because the
$\hat{a}_\alpha$ are precisely those values where $\partial E_0 / \partial
a_\alpha$ equal zero (and the difference between $\overline{E}$ and $E_0$ in
Eq.~\eqref{E_E0} does not depend on the $\hat{a}_\alpha$).
Hence minimizing Eq.~\eqref{Ebar} with respect to $\gamma$
we find that the optimal choice for $\gamma$ is given by the self-consistent
solution of
\begin{equation}
\boxed{
\gamma \, \sum_{\alpha=1}^M \hat{a}_\alpha^2 = \sum_{I=1}^M 
{\lambda^{(0)}_I \over \gamma + \lambda^{(0)}_I} \, .
}
\label{mu}
\end{equation}
We remind the reader that the fit parameters $\hat{a}_\alpha$ depend on
$\gamma$, but the eigenvalues $\lambda^{(0)}_I$ are independent
of the $\hat{a}_\alpha$
and $\gamma$.
We denote by $\hat{\gamma}$
the value of $\gamma$ which satisfies Eq.~\eqref{mu}.

Note that we really want the probability for the model, and in
Eq.~\eqref{PMD'} this is not given exactly by $P(\hat{\gamma}|D)$ but has an extra
factor of $\hat{\gamma}^{-1}$. Furthermore, we will neglect the last term in
Eq.~\eqref{Ebar} since it is independent of fit parameters or $\gamma$. Hence,
for different values of $M$ we will compare the values of
\begin{equation}
P(M|D) \propto \exp \left(-{1\over 2}\,E(\{\hat{a}_\alpha\}, \hat{\gamma}) \right) 
\end{equation}
where
\begin{equation}
E(\{\hat{a}_\alpha\}, \hat{\gamma}) = 
\sum_{i=1}^N \left({y_i - \sum_{\alpha=1}^M \hat{a}_\alpha X_\alpha(x_i)
\over \sigma_i} \right)^2 + \hat{\gamma}\,  \sum_{\alpha=1}^M
\hat{a}_\alpha^2 +
\sum_{I=1}^M
\ln \left({\lambda_I^{(0)} + \hat{\gamma}  \over \hat{\gamma}}\right)
+ 2 \ln \hat{\gamma}\, .
\label{E}
\end{equation}


Our Bayesian procedure to avoid overfitting is therefore as follows:
\begin{enumerate}
\item
\label{one}
Choose a value for $M$ and initial value for $\gamma$.
\item
Determine the optimal fitting parameters $\hat{a}_\alpha$
from Eq.~\eqref{ahat}, and hence also the eigenvalues $\lambda^{(0)}_I$.
\label{two}
\item
Substitute these values into the RHS of 
Eq.~\eqref{mu} to get a new value for $\gamma$. Go
to \ref{two} and iterate to convergence to determine the optimal value
$\hat{\gamma}$.
\item
Repeat for different values of $M$ and
compare the values of the cost function $E(\{\hat{a}_\alpha\}, \hat{\gamma})$
in Eq.~\eqref{E}.
The optimal choice for $M$ is at the minimum of $E$.
\end{enumerate}

The Bayesian procedure we have described has some undesirable features.
In the maximum likelihood approach, the
value of $\chi^2$ correctly remains the same if a translation or
scaling of either the $x$ or $y$ axes takes place, and the fit itself is the
same apart from the trivial change of variables. However, the cost function in
the Bayesian analysis does not have these invariances. For example, if we
just add a
constant to all the data the fit should remain the same except that the constant is
added to fit parameter $a_0$. However, the coupling of the regularization
parameter $\gamma$ to $a_0$ in the second
term in Eq.~\eqref{Ebar} means that the fit is changed in a non-trivial way in
the Bayesian analysis.
Because of this lack of invariance, it may be useful to
transform the data so that the $x$ and $y$ variables lie between $-1$ and $1$,
say,
before doing the Bayesian analysis, and, indeed, we shall find it necessary to
do this. 

Even if all the axes are transformed in this way, we still find some strange
features in the self-constency condition, Eq.~\eqref{mu}, used to determine
$\hat{\gamma}$.  The $\lambda_I^{(0)}$ are the eigenvalues of the matrix $U$ in
Eq.~\eqref{Uab_general}.
The elements of $U$ get very large if the error bars on the data
are small. We find that in this case the optimal value of $\gamma$ does not
get correspondingly large, with the result that
$\lambda_I^{(0)} \gg \hat\gamma$ even for several parameters $I$ which are
not well determined by the data. This disagrees with the interpretation in
Ref.~\cite{bishop:06} that those fit parameters $I$ with $\lambda_I^{(0)} \gg
\hat\gamma$ \textit{are} those which \textit{are} well determined by the data.

\begin{center}
\begin{figure}
\includegraphics[width=8cm]{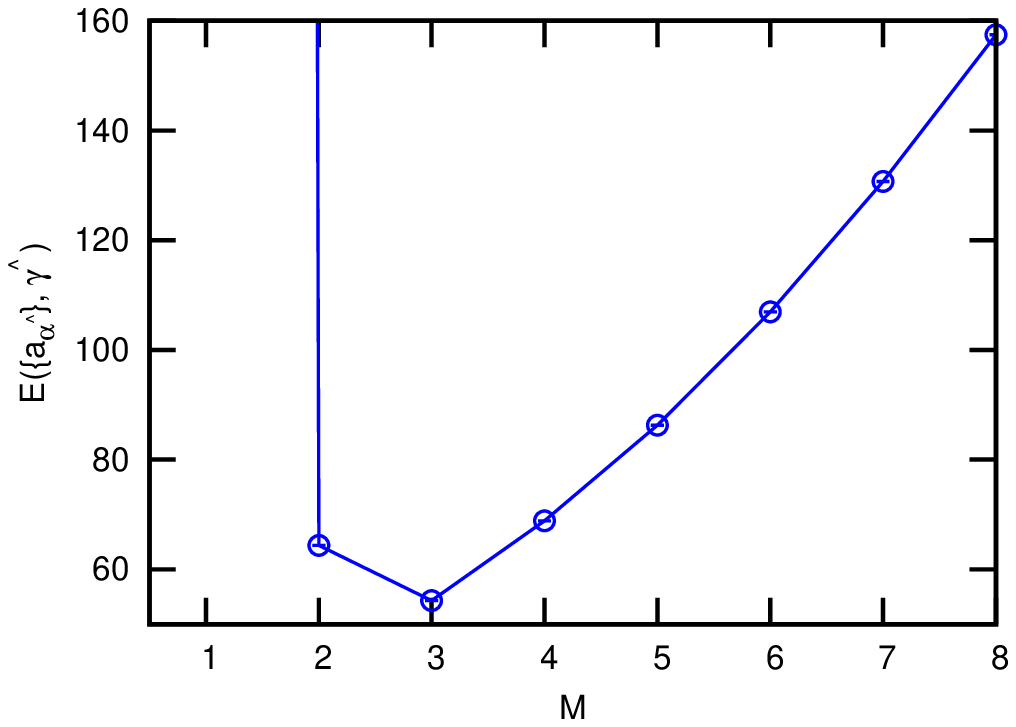}
\includegraphics[width=8cm]{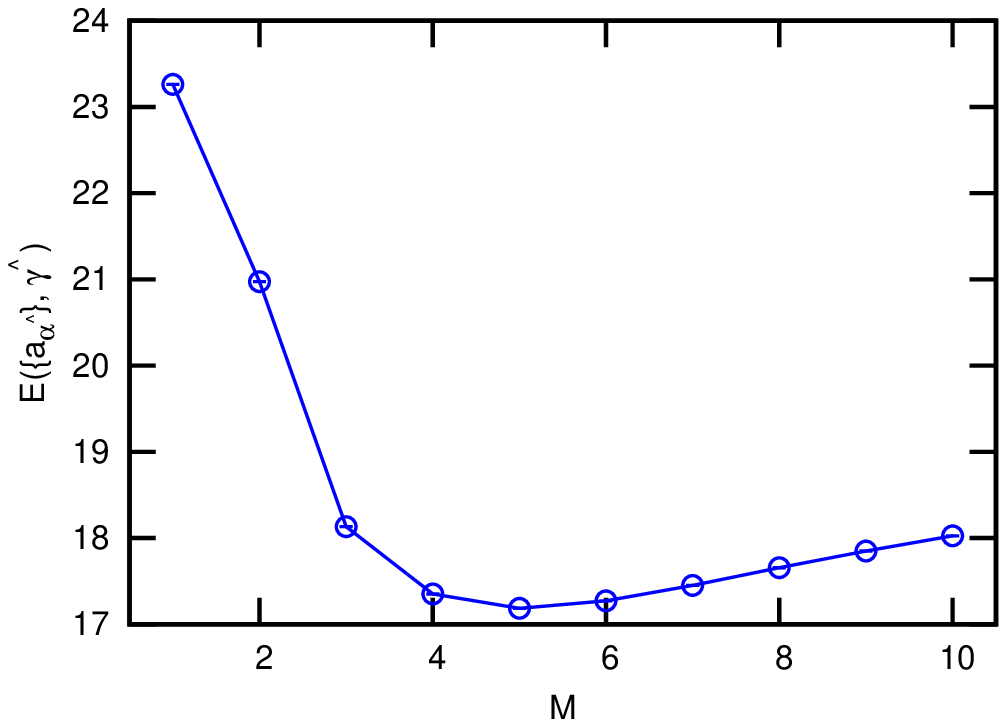}
\caption{
{\bf Left:}
The optimized cost function $E(\{\hat{a}_\alpha\}, \hat{\gamma})$
in Eq.~\eqref{E} 
plotted against $M$ 
for the data in Fig.~\ref{fig:lin_or_quad}.
\noindent {\bf Right:} The same quantity but for the data
in Fig.~\ref{fig:N10}. For both data sets the optimal choice for $M$ (where
$E$ is a minimum) should equal three. Hence this method works for the data in
Fig.~\ref{fig:lin_or_quad} but not for that in Fig.~\ref{fig:N10}.
}
\label{fig:cost_lin_or_quad}
\end{figure}
\end{center}

In Fig.~\ref{fig:cost_lin_or_quad} we plot the optimized
cost function 
$E(\{\hat{a}_\alpha\}, \hat{\gamma})$ given by Eq.~\eqref{E},
for the two different data sets, as a
function of $M$. Recall that both sets were determined from parabolas plus
noise, and so the optimal choice of $M$ should be $3$. This 
is correctly reproduced for the
data in Fig.~\ref{fig:lin_or_quad} but not for the data in Fig~\ref{fig:N10}.

\begin{center}
\begin{figure}
\includegraphics[width=8cm]{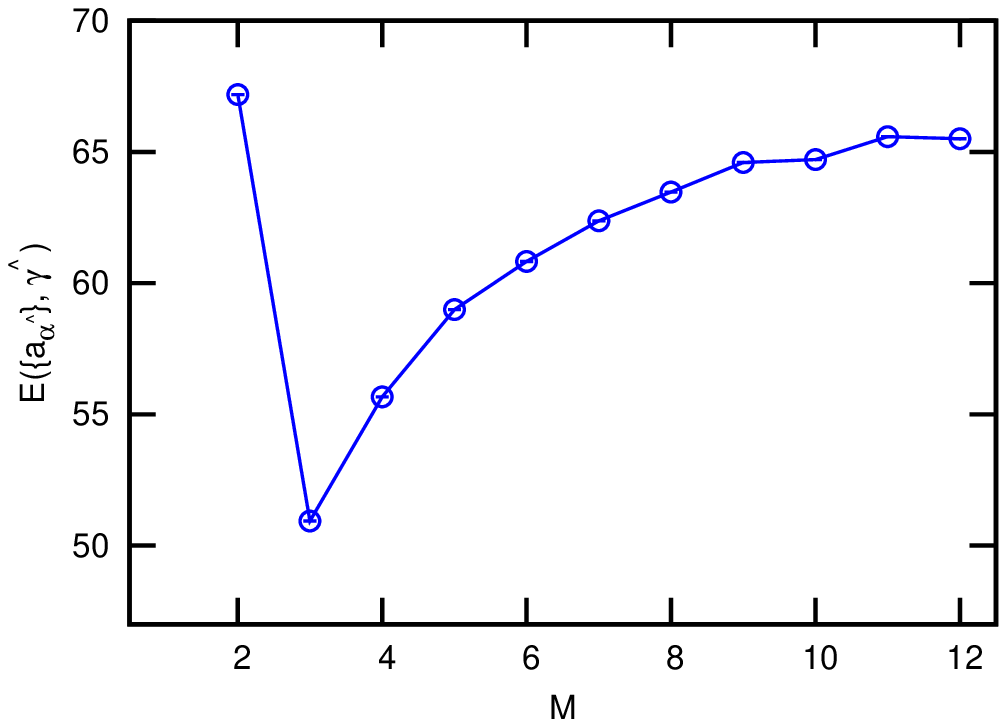}
\includegraphics[width=8cm]{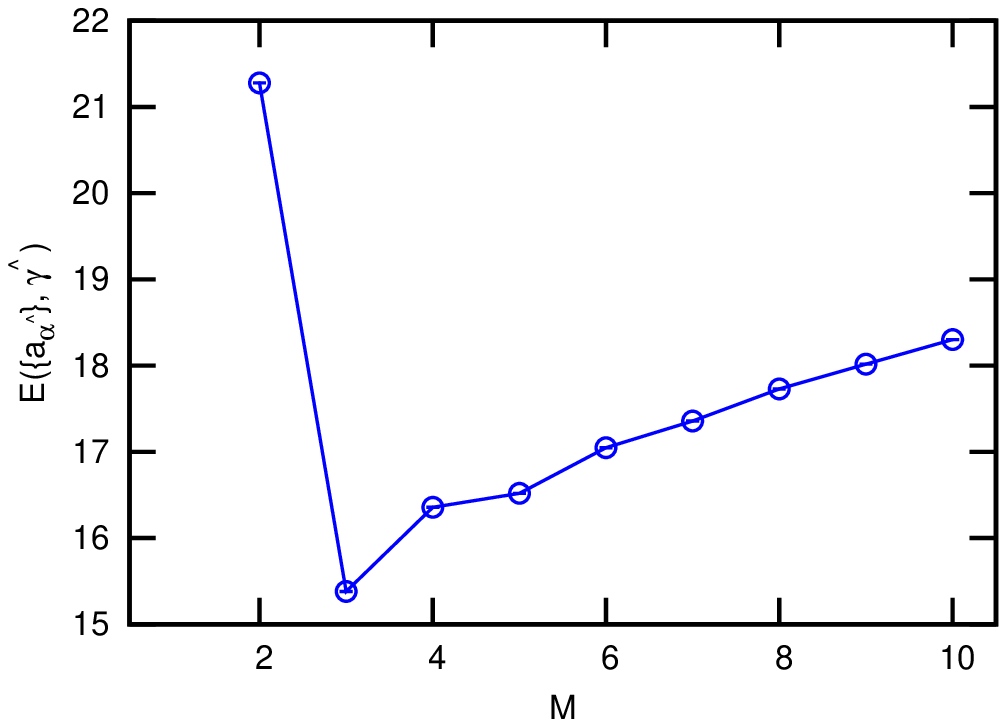}
\caption{
{\bf Left:}
The optimized cost function $E(\{\hat{a}_\alpha\}, \hat{\gamma})$
in Eq.~\eqref{E} 
is plotted versus $M$ 
when the data in Fig.~\ref{fig:lin_or_quad} is scaled so that the $x$ and $y$
variables lie between $-1$ and $1$.
\noindent {\bf Right:} The same quantity but for the data
in Fig.~\ref{fig:N10}. Unlike the unscaled analysis in
Fig.~\ref{fig:cost_lin_or_quad}, this correctly gives a minimum at $M=3$ for
both data sets. 
}
\label{fig:cost_lin_or_quad_shift}
\end{figure}
\end{center}

From the above discussion we suspect that the problem with the data in
Fig.~\ref{fig:N10} is due to lack of invariance of
the method to shift and scaling of the axes.  We have therefore performed the
same analysis but with the $x$ and $y$ variables scaled 
to lie between $-1$ and $1$ and
show the results in Fig.~\ref{fig:cost_lin_or_quad_shift}. Now, the minimum is
correctly found at $M= 3$ for both data sets.

\subsubsection{Conclusions for Model Selection}
In the maximum likelihood
approach one computes the goodness of fit factor $Q$ and looks for a peak as a
number of fit parameters, $M$, see the right hand panel in
Figs.~\ref{fig:chi2vM} and \ref{fig:N10b}.
Although $\chi^2$ monotonically
deceases with increasing $M$, $Q$ depends on $\chi^2$ per degree of freedom
and the latter decreases with increasing $M$. Hence there is a penalty on
increasing $M$ which can only be compensated for if there is a
\textit{substantial} decrease in $\chi^2$. If $M$ is too large, which is
the
over-fitting regime, $\chi^2$ only decreases slightly with $M$ and so $Q$
decreases. (Eventually $Q$ may increase again since the function fits the data
perfectly for $M = N$.)

In the Bayesian approach, one applies a regularization parameter $\gamma$ and
determines its optimal value. From this a cost function function is found
whose minimum value is the optimal choice for $M$. This worked ``off the
shelf'' for one set of data studied, see Fig.~\ref{fig:cost_lin_or_quad},
but to have it work for both sets of data it was necessary to scale the $x$
and $y$ coordinates to the range $-1$ to $1$, see
Fig.~\ref{fig:cost_lin_or_quad_shift}. The Bayesian approach is more
complicated than maximum likelihood method and evidently
has to be applied with care. It
can be generalized, for example, by allowing different hyperparameters for
each fit variable, but at a substantial \textit{additional} complexity.


\newpage
\appendix

\section{Central Limit Theorem}
\label{sec:clt}

In this appendix we give a proof of the central limit theorem.

We assume a distribution that falls off sufficiently fast at
$\pm \infty$ that the mean and variance are finite.
This \textit{excludes}, for example,
the Lorentzian distribution:
\begin{equation}
P_{\rm Lor} = {1 \over \pi}
{1 \over 1+x^2} \, .
\end{equation}
A common distribution which \textit{does} have a finite mean and variance
is the Gaussian distribution,
\begin{equation}
P_{\rm Gauss} = {1 \over \sqrt{2 \pi}\,  \sigma} \exp\left[-{(x-\mu)^2 \over 2
\sigma^2}\right] \, .
\label{Gauss}
\end{equation}
Using standard results for Gaussian integrals you should be able to show
that the distribution is normalized and that the mean and standard deviation
are $\mu$ and $\sigma$ respectively. We note that the probability that that a
Gaussian random variable is more than $c\, \sigma$, where $c$
is a constant, away from the
mean is given by
\begin{align}
P(|x - \mu| > c \, \sigma) =& {2 \over \sqrt{2 \pi}\, \sigma} \int_{\mu+c \sigma}^\infty 
\exp\left[-(x - \mu)^2 / (2 \sigma^2)\right] , \nonumber \\
&= {2 \over \sqrt{\pi}} \int_{c / \sqrt{2}}^\infty e^{-t^2} \, dt \nonumber \\
&= \text{erfc}(c / \sqrt{2}) \, ,
\label{erfc}
\end{align}
where $\text{erfc}$ is the complementary error function~\cite{press:92}.

Consider a distribution, \textit{not necessarily Gaussian}, with a
finite mean and distribution.
We pick $N$ independent and identically distributed random numbers $x_i$
from such a distribution and form the sum
$$
X = \sum_{i=1}^N x_i.
$$
distribution.

The determination of the distribution of $X$, which we call
$P_N(X)$,
uses the Fourier transform of $P(x)$,
called the ``characteristic function'' in the context of probability theory.
This is defined by
$$
Q(k) = \int_{-\infty}^\infty P(x) e^{i k x} \, d x \, . 
$$
Expanding out the exponential we can write $Q(k)$ in terms of the moments of
$P(x)$
$$
Q(k) = 1 + i k\langle x \rangle + {(i k)^2 \over 2!} \langle x^2 \rangle +
{(i k)^3 \over 3!} \langle x^3 \rangle + \cdots \, .
$$
It will be convenient in what follows to write $Q(k)$ as an exponential, i.e.
\begin{eqnarray}
Q(k) & = & \exp \left[ \ln \left(1 +
i k\langle x \rangle + {(i k)^2 \over 2!} \langle x^2 \rangle +
{(i k)^3 \over 3!} \langle x^3 \rangle + \cdots \right) \right] \nonumber \\
& = & \boxed{ \exp\left[ i k \mu  + {(i k)^2 \sigma^2 \over 2!} +
{c_3(i k)^3 \over 3!} +
{c_4 (i k)^4 \over 4!} + \cdots  \right] \, ,}
\label{cumulant}
\end{eqnarray}
where $c_3$ involves third and lower moments, $c_4$ involves fourth and
lower moments, and so on. The $c_n$ are called \textit{cumulant}
averages. 

For the important case of a Gaussian, 
the Fourier transform is obtained by ``completing the square''.
The result is that the Fourier transform of a
Gaussian is also a Gaussian, namely,
\begin{equation}
\boxed{
Q_{\rm Gauss}(k) = \exp\left[ i k \mu -{k^2 \sigma^2 \over 2} \right] \,  ,}
\label{Qgauss}
\end{equation}
showing that the higher order cumulants, $c_3, c_4$, etc. in
Eq.~(\ref{cumulant}) \textit{all vanish} for a Gaussian.

The distribution $P_N(x)$ can be expressed as
\begin{equation}
P_N(x) = \int_{-\infty}^\infty P(x_1) d x_1 \, \int_{-\infty}^\infty P(x_2) d x_2 \, 
\cdots
\int_{-\infty}^\infty P(x_N) d x_N 
\, \delta (X - \sum_{i=1}^N x_i)   \, .
\end{equation}
We evaluate this by using the
integral representation of the delta function
\begin{equation}
\delta(x) = {1 \over 2 \pi} \int_{-\infty}^\infty e^{i k x} \, d k \, ,
\end{equation}
so
\begin{align}
P_N(X)
&= \int_{-\infty}^\infty {d k \over 2 \pi}
\int_{-\infty}^\infty P(x_1) d x_1 \, \int_{-\infty}^\infty P(x_2) d x_2 \, 
\cdots
\int_{-\infty}^\infty P(x_N) d x_N \, 
\exp[i k (x_1 + x_2 + \cdots x_N - X)]  \\
&= \int_{-\infty}^\infty {d k \over 2 \pi} Q(k)^N e^{-i k X} \, ,
\label{inv_FT}
\end{align}
showing that the Fourier transform of $P_N(x)$, which we call $Q_N(k)$, is
given by
\begin{equation}
\boxed{Q_N(k) = Q(k)^N \, . }
\label{fourier_N}
\end{equation}
Consequently
\begin{equation}
Q_N(k) = 
\exp\left[ i k N\mu -{N k^2 \sigma^2 \over 2} + {N c_3(i k)^3 \over 4!} +
{N c_4 (i k)^4 \over 4!} + \cdots  \right]  \, .
\label{cumulant_N}
\end{equation}

Comparing with Eq.~(\ref{cumulant}) we see that 
\begin{quotation}
\noindent the mean of the
distribution of the sum of $N$ independent and identically distributed random
variables (the coefficient of $-i k$ in the exponential)
is $N$ times the mean of the distribution
of one variable, and the variance of the distribution of the sum 
(the coefficient of $-k^2/2!$) is $N$ times the variance of the
distribution of one variable.
\end{quotation}
These are general statements applicable for
\textit{any} $N$ and have already been derived in Sec.~\ref{sec:basic}.

However, if $N$ is \textit{large} we can now go further.
The distribution $P_N(X)$ is the inverse transform of $Q_N(k)$, see
Eq.~\eqref{inv_FT}, so
\begin{equation}
P_N(X)  = 
{1 \over 2\pi} \int_{-\infty}^\infty
\exp\left[ -i k X'-{N k^2 \sigma^2 \over 2!} + N{c_3(i k)^3 \over 3!} +
{N c_4 (i k)^4 \over 4!} + \cdots  \right]  \, d k \, ,
\label{invtrans}
\end{equation}
where
\begin{equation}
X' = X - N \mu \, .
\label{x'}
\end{equation}
Looking at the $-N k^2 / 2$ term
in the exponential in
Eq.~(\ref{invtrans}), we see that the integrand is significant for 
$ k < k^\star$, where $N \sigma^2 (k^\star)^2 = 1$, and negligibly small for
$k \gg k^\star$.
However, for $0 < k < k^\star$ the
higher order
terms in Eq.~(\ref{invtrans}), (i.e.\ those of order $k^3, k^4$ etc.) are very
small since $N (k^\star)^3 \sim N^{-1/2}, N (k^\star)^4 \sim N^{-1}$ and so on.
Hence the terms of higher order than $k^2$ in Eq.~(\ref{invtrans}), do not
contribute for large $N$ and so
\begin{equation}
\lim_{N \to \infty} P_N(X) 
= {1 \over 2\pi} \int_{-\infty}^\infty
\exp\left[ -i k X'-{N k^2 \sigma^2 \over 2} \right] \, d k \,  .
\label{invtransG}
\end{equation}
In other words, for large $N$ the distribution is the Fourier transform of a
Gaussian, which, as we know, is also a Gaussian. Completing the square in
Eq.~(\ref{invtransG}) gives
\begin{eqnarray}
\lim_{N \to \infty} P_N(X)  & = &
{1 \over 2\pi} \int_{-\infty}^\infty
\exp\left[-{N \sigma^2 \over 2 } \left(k - {i X' \over N \sigma^2}\right)^2
\right] \, d k \  \exp\left[ -{(X')^2  \over 2  N \sigma^2} \right] \nonumber \\
& = &
\boxed{
{1 \over \sqrt{2 \pi N} \, \sigma} \exp\left[ -{(X-N\mu)^2  \over 2 N \sigma^2}
\right] \, ,}
\label{clt}
\end{eqnarray}
where, in the last line, we used Eq.~(\ref{x'}).
This is a Gaussian with mean $N \mu$ and variance $N \sigma^2$. 
Equation (\ref{clt}) is the
\fbox{central limit theorem} in statistics. It tells us that,
\begin{quotation}
\noindent for $N
\to\infty$, the distribution of the sum of $N$ independent and identically
distributed random
variables is a
\textit{Gaussian} whose mean is
$N$ times the mean, $\mu$, of the
distribution of one variable, and whose variance is $N$ times the
variance of the distribution of one variable, $\sigma^2$, 
\textit{independent of the
form of the distribution of one variable}, $P(x)$, provided only that $\mu$
and $\sigma$ are finite.
\end{quotation}

The central limit theorem is of such generality that it is extremely important. 
It is the reason why the Gaussian distribution has such a preeminent
place in the theory of statistics. 

Note that if the distribution of the individual $x_i$ is Gaussian, then the
distribution of the sum of $N$ variables is \textit{always} Gaussian, even for
$N$ small. This follows from Eq.~\eqref{fourier_N} and the fact that
the Fourier transform of a Gaussian is a Gaussian.

In practice, distributions that we meet in nature, have a much broader tail
than that of the Gaussian distribution, which falls off very fast at large
$|x-\mu|/\sigma$. As a result, even if the distribution of the sum
approximates well a Gaussian in 
the central region for only
modest values of $N$, it might take a much larger value of $N$ to beat down the
weight in the tail to the value of the Gaussian. Hence, even for
moderate values of $N$, the probability of a deviation greater than $\sigma$
can be significantly larger than that of the Gaussian distribution which is
32\%. This caution will be important in Sec.~\ref{sec:fit} when we discuss the
quality of fits.

\begin{center}
\begin{figure}
\includegraphics[width=11cm]{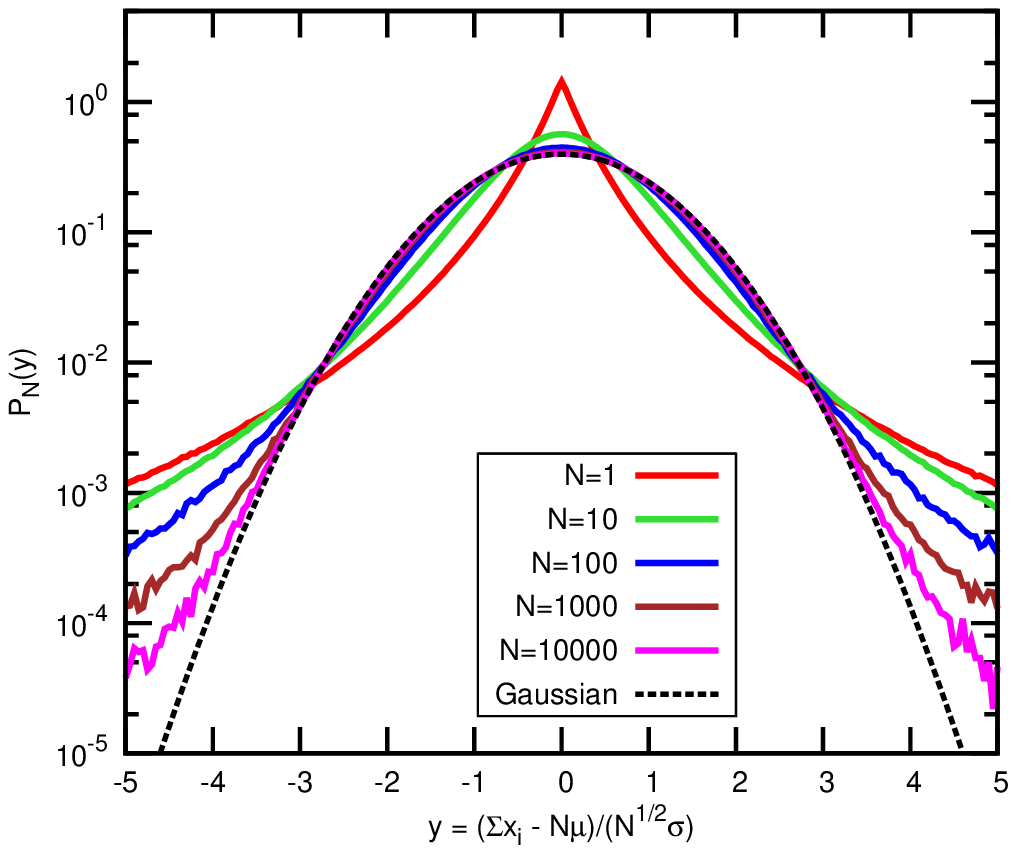}
\caption{
Figure showing the approach to the central limit theorem for the distribution
in Eq.~\eqref{dist_long}, which has mean, $\mu$, equal to 0, and standard
deviation, $\sigma$, equal to 1. The horizontal axis is the sum of $N$ random
variables divided by $\sqrt{N}$ which, for all $N$, has zero mean and 
standard deviation unity. For large $N$ the distribution approaches a
Gaussian. However, convergence is non-uniform, and is extremely slow in the
tails. Note the log vertical scale which is necessary to
display the weight in the tails.
}
\label{Fig:converge_to_clt}
\end{figure}
\end{center}
We will illustrate the slow convergence of the distribution of the sum to a
Gaussian in Fig.~\eqref{Fig:converge_to_clt}, in which the distribution of
the individual variables $x_i$ is
\begin{equation}
P(x) = {3 \over 2}\, {1 \over (1 + |x|)^4} \, .
\label{dist_long}
\end{equation}
This has mean 0 and standard deviation 1, but moments higher than the second
do not exist because the integrals diverge. For large $N$ the distribution
approaches a Gaussian, as expected, but convergence is very slow in the tails.

\section{The number of degrees of freedom}
\label{sec:NDF}

We assume Gaussian noise on the data and consider first
a straight line fit, so we have to determine
the values of $a_0$ and $a_1$ which minimize Eq.~\eqref{chisq_sline}. The $N$ terms
in Eq.~\eqref{chisq_sline} are
not statistically independent at the minimum because the values of $a_0$ and
$a_1$, given by Eq.~\eqref{sline}, depend on the data points $(x_i,
y_i, \sigma_i)$.

Consider the ``residuals'' defined by
\begin{equation}
\epsilon_i = {y_i - a_0 - a_1 x_i \over \sigma_i} \, .
\end{equation}
If the model were exact and we use the exact values of the parameters $a_0$
and $a_1$ the $\epsilon_i$ would be independent and each have a Gaussian
distribution with zero mean and standard deviation unity.
However, choosing the
\textit{best-fit} values of $a_0$ and $a_1$ \textit{from the data} according to
Eq.~\eqref{sline} implies that
\begin{subequations}
\begin{align}
\sum_{i=1}^N {1\over \sigma_i}\, \epsilon_i &= 0\, ,\\
\sum_{i=1}^N {x_i \over \sigma_i}\, \epsilon_i &= 0\, ,
\end{align}
\end{subequations}
which are are two \textit{linear constraints} on the $\epsilon_i$. This means that we only
need to specify $N-2$ of them to know them all. In the $N$ dimensional
space of the $\epsilon_i$ we have eliminated two directions, so there can be no Gaussian
fluctuations along them. However the other $N-2$
dimensions are unchanged, and will have the same Gaussian fluctuations as
before.  Thus $\chi^2$ has the distribution of a sum of squares of $N-2$
Gaussian random variables. We can intuitively understand why there are
$N-2$ degrees of freedom 
rather than $N$ by considering the case of
$N=2$. The fit goes perfectly through the two
points so one has $\chi^2=0$ exactly. This implies that there
are zero degrees of freedom since, on average,
each degree of freedom adds 1 to $\chi^2$.

Clearly this argument can be generalized to any fitting function which depends
\textit{linearly} on $M$ fitting parameters, assuming Gaussian noise on the data. 
The result is that $\chi^2$ has 
the distribution of a sum of squares of $\NDOF = N-M$ Gaussian random
variables, in which the quantity $\NDOF$ is called the ``number of
degrees of freedom''. 

Even if the fitting function depends non-linearly on the parameters, this last
result is often taken as a reasonable approximation.

\section{The chi-squared distribution and the goodness of fit parameter $\textbf{Q}$}
\label{sec:Q}

The $\chi^2$ distribution for $m$ degrees of freedom is the distribution of
the sum of $m$ independent random variables with a Gaussian distribution with
zero mean and standard deviation unity.  To determine this we write the
distribution of the $m$ variables $x_i$ as
\begin{equation}
P(x_1, x_2, \cdots, x_{m})\, dx_1 dx_2 \cdots dx_{m}
=
{1 \over (2 \pi)^{m/2}} \, e^{-x_1^2/2} \, e^{-x_2^2/2} \cdots e^{-x_{m}^2/2} \,
dx_1 dx_2 \cdots dx_{m} \, .
\end{equation}
Converting to polar coordinates, and integrating
over directions, we find the
distribution of the radial variable to be
\begin{equation}
\widetilde{P}(r) \, dr = 
{S_{m} \over (2 \pi)^{m/2}} \, r^{m-1}\, e^{-r^2/2} \, dr \, ,
\label{Pr}
\end{equation}
where $S_{m}$ is the surface area of a unit $m$-dimensional sphere. To determine
$S_{m}$ we integrate Eq.~\eqref{Pr} over $r$, noting that $\widetilde{P}(r) $ is
normalized, which gives
\begin{equation}
S_{m} = {2 \pi^{m/2} \over \Gamma(m/2)} \, ,
\label{Sm}
\end{equation}
where $\Gamma(x)$ is the Euler gamma function
defined by
\begin{equation}
\Gamma(x) = \int_0^\infty t^{x-1} \, e^{-t}\, dt \, . 
\label{gamma}
\end{equation}
From Eqs.~\eqref{Pr} and \eqref{Sm} we have
\begin{equation}
\widetilde{P}(r) = {1 \over 2^{m/2-1} \Gamma(m/2)} \, r^{m-1} e^{-r^2/2} \, .
\end{equation}
This is the distribution of $r$ but we want the distribution of $\chi^2 \equiv
\sum_i x_i^2 = r^2$. To avoid confusion of notation we write $X$ for $\chi^2$,
and define the $\chi^2$ distribution for $m$ variables as $P^{(m)}(X)$. We
have
$P^{(m)}(X) \, dX = \widetilde{P}(r) \, dr$ so the $\chi^2$ distribution for
$m$ degrees of freedom is
\begin{align}
P^{(m)}(X) &= {\widetilde{P}(r) \over dX / dr}  \nonumber \\
&\boxed{ = {1 \over 2^{m/2} \Gamma(m/2)} \, X^{(m/2)-1}\,
e^{-X/2} \qquad (X > 0) \, .}
\label{chisq-dist}
\end{align}
The $\chi^2$ distribution is zero for $X
< 0$. Using Eq.~\eqref{gamma} and the property of the
gamma function that $\Gamma(n+1) = n \Gamma(n)$ one can show that
\begin{subequations}
\begin{align}
\int_0^\infty P^{(m)}(X)\, d X &= 1 \, , \\
\langle X \rangle \equiv \int_0^\infty X\, P^{(m)}(X)\, d X &= m \, , \label{mean} \\
\langle X^2 \rangle \equiv \int_0^\infty X^2\, P^{(m)}(X)\, d X &= m^2 +
2m \, ,
\quad \mbox{so }\\
\langle X^2 \rangle - \langle X \rangle^2 &= 2 m  \, \label{var} .
\end{align}
\end{subequations}
Furthermore, the peak of the distribution is at $X = m - 2$ (for $m > 2$).

From Eqs.~\eqref{mean} and \eqref{var} we see that typically $\chi^2$ lies in
the range $m - \sqrt{2 m}$ to $m + \sqrt{2
m}$.
For large $m$ the distribution
approaches a Gaussian according to the central limit theory discussed in
Appendix~\ref{sec:clt}. Typically one focuses on the
value of $\chi^2$ per degree freedom
since this should be around unity for all $m$.

\begin{figure}
\begin{center}
\includegraphics[width=12cm]{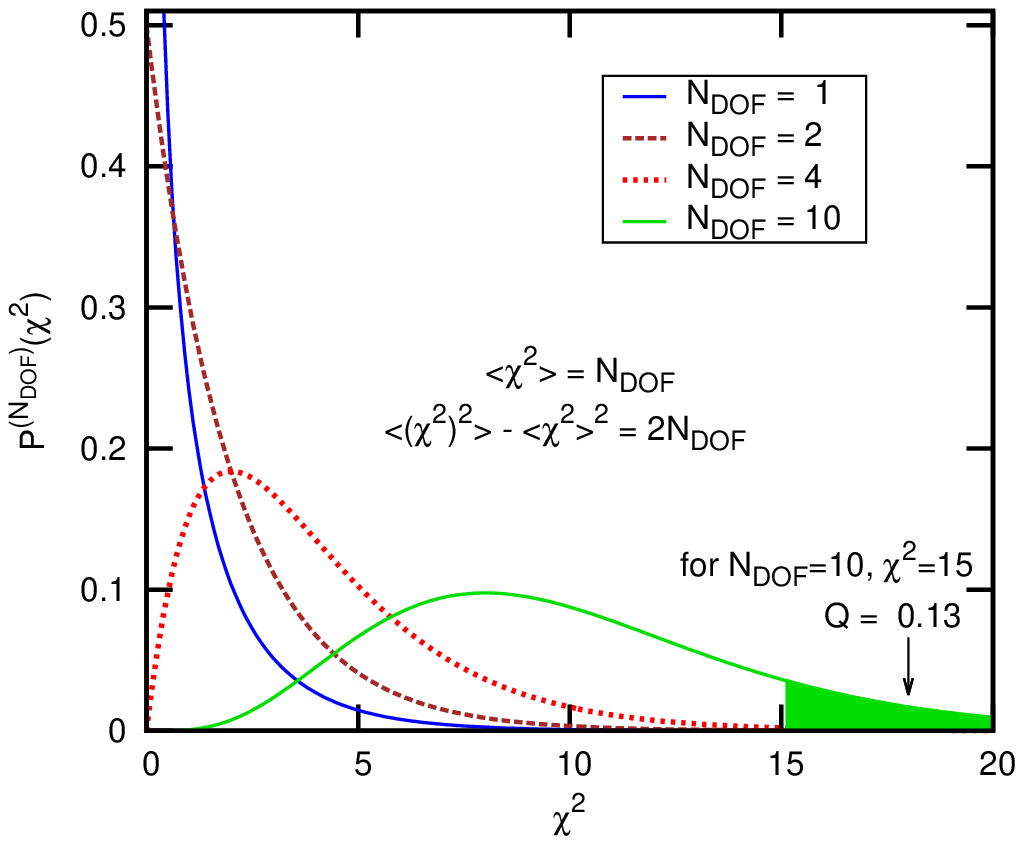}
\caption{The $\chi^2$ distribution for several values of $N_{\rm DOF}$ the number
of degrees of freedom. The mean and standard deviation depend on $N_{\rm DOF}$ in
the way specified. The goodness of fit parameter $Q$, defined in
Eq.~\eqref{Q_expression2}, depends on the values of
$N_{\rm DOF}$ and $\chi^2$, and is the probability that $\chi^2$ could 
have the specified value or larger by random chance. 
The area of the shaded region in the figure is the value of $Q$ for
$N_{\rm DOF}=10, \chi^2 = 15$. Note that the \textit{total}
area under each of the curves
is unity because they represent probability distributions.  
\label{fig:chisq}
}
\end{center}
\end{figure}

The goodness of fit parameter is the probability that the specified value of
$\chi^2$, or greater, could occur by random chance.  From Eq.~\eqref{chisq-dist}
it is given by
\begin{align}
Q &=  {1 \over 2^{m/2} \Gamma(m/2)} \, \int_{\chi^2}^\infty\, X^{(m/2)-1}\,
e^{-X/2} \, d X\, , \\ 
& \boxed{ =  {1 \over \Gamma(m/2)} \, \int_{\chi^2/2}^\infty\, y^{(m/2)-1}\,  e^{-y} \, dy\, , }
\label{Q_expression}
\end{align}
which is known as an incomplete gamma function. Code to generate the
incomplete gamma function is 
given in Numerical Recipes~\cite{press:92}. There is also
a built-in function to generate the
goodness of fit parameter in the \texttt{scipy}
package of \texttt{python} and in the graphics program \texttt{gnuplot}, see
the scripts in Appendix \ref{sec:scripts}.

The $\chi^2$ distribution for several value of $m \equiv N_{\rm DOF}$ is plotted in
Fig.~\eqref{fig:chisq}. 
The mean and variance are given by
Eqs.~\eqref{mean} and \eqref{var}.
For large $m$, according to the central limit theorem, the $\chi^2$
distribution becomes a Gaussian.
\vspace{-0.2cm}

Note that $Q=1$ for $\chi^2 = 0$ and
$Q\to 0$ for $\chi^2 \to\infty$.
Remember that $m$ is the number
of degrees of freedom, called $\NDOF$ elsewhere in these notes, and that
$\NDOF = N - M$, where $N$ is the number of
data points and $M$ is the number of fit parameters.

\section{Asymptotic standard error and how to get correct error bars from
gnuplot}
\label{sec:ase}
Sometimes one does not have error bars on the data. Nonetheless, one can still
use $\chi^2$ fitting to get an \textit{estimate} of those errors (assuming
that they are all equal) and thereby also get an error bar on the fit
parameters. The latter is called the ``asymptotic standard error''.  Assuming
the same error bar $\sigma_\text{ass}$ for all points, we determine
$\sigma_\text{ass}$ from the requirement that $\chi^2$ per degree of freedom
is precisely one, i.e.\ its mean value according to Eq.~\eqref{mean}.  This
gives
\begin{equation}
1 = {\chi^2 \over \NDOF} = {1 \over \NDOF} \, \sum_{i=1}^N 
\left( \, {y_i - f(x_i) \over \sigma_\text{ass} } \, \right)^2 \, ,
\end{equation}
or, equivalently,
\begin{equation}
\boxed{
\sigma_\text{ass}^2 = {1 \over \NDOF} \, \sum_{i=1}^N 
\left(y_i - f(x_i)\right)^2 \, .}
\label{sigma_ass}
\end{equation}
The error bars on the fit parameters are then obtained from
Eq.~\eqref{error_params}, with
the elements of $U$ given by Eq.~\eqref{Uab} in which $\sigma_i$ is replaced
by $\sigma_\text{ass}$. Equivalently, one can
set the $\sigma_i$ to unity in determining $U$ from
Eq.~\eqref{Uab}, and 
estimate the error on the fit parameters from 
\begin{equation}
\qquad\qquad\qquad\boxed{
\sigma^2_\alpha = \left(U\right)^{-1}_{\alpha\alpha}
\, \sigma^2_\text{ass}\, ,} \quad\text{(asymptotic standard error)}.
\label{assterr}
\end{equation}

A simple example of the use of the asymptotic standard error in a situation
where we don't know the error on the data points, is fitting to a
constant, i.e.\ \textit{determining the average of a set of data}, which
we already discussed in detail in Sec.~\ref{sec:averages}. In this case we have
\begin{equation}
U_{00} = N, \qquad v_0 = \sum_{i=1}^N y_i ,
\end{equation}
so the only fit parameter is
\begin{equation}
a_0 = {v_0 \over U_{00}} = {1\over N} \, \sum_{i=1}^N y_i = \overline{y} ,
\end{equation}
which gives, naturally enough, the average of the data points,
$\overline{y}$. The number of degrees of freedom is $N-1$, since there is one 
fit parameter, so
\begin{equation}
\sigma_\text{ass}^2 = {1 \over N - 1} \, \sum_{i=1}^N 
\left(y_i - \overline{y}\right)^2 \, ,
\end{equation}
and hence the square of the error on $a_0$ is given, from Eq.~\eqref{assterr}, by
\begin{equation}
\sigma^2_0 = {1 \over U_{00}}\, \sigma^2_\text{ass} =
{1 \over N ( N - 1)} \, \sum_{i=1}^N \left(y_i - \overline{y}\right)^2 \, ,
\end{equation}
which is precisely the expression for the error in the mean of a 
set of data given in Eq.~\eqref{finalans2}.

I now mention that a popular plotting program, \texttt{gnuplot},
which also does fits but unfortunately presents error bars on the fit parameters incorrectly
if there are error bars on the data.
Whether or not there are error bars on the points, \texttt{gnuplot} gives the
``asymptotic standard error'' on the fit parameters. \texttt{Gnuplot}
calculates the elements of $U$ correctly from Eq.~\eqref{Uab} including the
error bars, but then apparently also determines an ``assumed error'' from an
expression like Eq.~\eqref{sigma_ass} but including the error bars, i.e.

\begin{equation}
\sigma_\text{ass}^2 = {1 \over \NDOF} \, \sum_{i=1}^N 
\left(\ {y_i - f(x_i) \over \sigma_i}\ \right)^2 \  = \ {\chi^2 \over
\NDOF}, \qquad \text{(\texttt{gnuplot})}\, .
\end{equation}
Hence \texttt{gnuplot}'s $\sigma^2_\text{ass}$ is just the chi-squared per degree of
freedom. The error bar (squared) quoted by \texttt{gnuplot} is $
\left(U\right)^{-1}_{\alpha\alpha} \, \sigma^2_\text{ass}$, as in
Eq.~\eqref{assterr}. However, this is wrong since the error bars on the data 
points have \textit{already}
been included in calculating the elements of $U$, so the error on the fit
parameter $\alpha$ should really be
$\left(U\right)^{-1}_{\alpha\alpha}$. Hence,
\begin{quotation}
\noindent to get correct error
bars on fit parameters from \texttt{gnuplot} when there
are error bars on the points, you have to divide
\texttt{gnuplot}'s 
asymptotic standard errors by
the square root of the
chi-squared per degree of freedom (which gnuplot calls \texttt{FIT\_STDFIT}
and, fortunately, computes correctly). 
\end{quotation}
I have checked this statement by comparing with results from Numerical Recipes
routines, and also, for straight-line fits, by my own implementation of the
formulae. It is curious that I found no hits on this topic when 
Googling the internet. Can no one else have come across this
problem? Correction of \texttt{gnuplot} error bars is implemented in the
\texttt{gnuplot} scripts in Appendix \ref{sec:scripts}

The need to correct \texttt{gnuplot}'s error bars applies to linear as well
as non-linear models.

I recently learned that error bars on fit parameters given
by the routine \texttt{curve\_fit} of \texttt{python} also have to be
corrected in the same way. This is shown in two of the python scripts in
appendix~\ref{sec:scripts}. Curiously, a different python fitting
routine, \texttt{leastsq}, gives the error bars correctly.

\section{The distribution of fitted parameters determined from simulated datasets}
\label{sec:proof}

In this section we derive the equation for the distribution of fitted
parameters determined from simulated datasets, Eq.~\eqref{theorem}, assuming
an arbitrary linear model, see Eq.~\eqref{general_lin}. Projecting on to
a single fitting parameter, as above, this corresponds to
the lower figure in Fig.~\ref{Fig:distofa1}.

We have \textit{one} set of $y$-values, $y_i^{(0)}$, for which the fit
parameters are
$\vec{a}^{(0)}$. We then
generate an \textit{ensemble} of
simulated data sets, $y_i^S$, assuming the data has Gaussian noise with standard deviation
$\sigma_i$ centered on the actual data values $y_i^{(0)}$.
We ask for the probability that the fit to one of
the simulated data sets has parameters $\vec{a}^S$. 

This probability distribution is given by
\begin{equation}
P(\vec{a}^S) = \prod_{i=1}^N \left\{ {1 \over \sqrt{2 \pi} \sigma_i} \,
\int_{-\infty}^\infty \, d y_i^S\,
\exp\left[-{\left(y_i^S - y_i^{(0)}\right)^2 \over
2 \sigma_i^2}\right]\, \right\} \,
\prod_{\alpha=1}^M \delta \left( \sum_\beta U_{\alpha\beta} a_\beta^S -
v_\alpha^S\right) \, \det U \, ,
\label{dist_of_a}
\end{equation}
where the factor in curly brackets is (an integral over) the probability
distribution of the data points $y_i^S$, and the delta functions project out
those sets of data points which have a particular set of fitted parameters,
see
Eq.~\eqref{lin_eq}.
The factor of
$\det U$ is a Jacobian to normalize the distribution.
Using the integral representation of the delta function, and writing explicitly the
expression for $v_\alpha$ from Eq.~\eqref{v_general}, one has
\begin{align}
P(\vec{a}^S) =& \prod_{i=1}^N \left\{ {1 \over \sqrt{2 \pi} \sigma_i} \,
\int_{-\infty}^\infty \, d y_i^S\, \exp\left[-{\left(y_i^S - y_i^{(0)}
\right)^2 \over 2 \sigma_i^2}\right]\, \right\}
\times \qquad\qquad\qquad \\
& \ \prod_{\alpha=1}^M \left( {1 \over 2 \pi} \, \int_{-\infty}^\infty d k_\alpha
\exp\left[i k_\alpha\left( \sum_\beta U_{\alpha\beta} a_\beta^S -
\sum_{i=1}^N {y_i^S\, X_\alpha(x_i) \over \sigma_i^2} \right)\right] \right) \,
\det U \, .
\end{align}
We carry out the $y$ integrals by ``completing the square'',
\begin{align}
P(\vec{a}^S) = \prod_{\alpha=1}^M \left( {1 \over 2 \pi} \,
\int_{-\infty}^\infty d k_\alpha \right) \,
\prod_{i=1}^N \left\{ {1 \over \sqrt{2 \pi} \sigma_i} \,
\int_{-\infty}^\infty \, d y_i^S\, \exp\left[-{\left(y_i^S - y_i^{(0)}
 + i \vec{k}\cdot \vec{X}(x_i) \right)^2
\over 2 \sigma_i^2}\right] \right\} \times \\
\exp\left[ -{1 \over 2 \sigma_i^2} \, \left(\,
\left(\vec{k}\cdot\vec{X}(i)\right)^2 
+2 i \left(\vec{k} \cdot \vec{X}(x_i)\right) \, y_i^{(0)} 
 \right) \right] \times
\exp\left[i \sum_{\alpha,\beta} k_\alpha\, U_{\alpha\beta}\, a_\beta^S\right] \,
\det U \, .
\end{align}
Doing the $y^S$-integrals, the factors in curly brackets are equal to unity.
Using Eqs.~\eqref{Uab_general} and \eqref{v_general} and the fact that the
$U_{\alpha\beta}$ are independent of the $y_i^S$,
we then get
\begin{equation}
P(\vec{a}^S) = \prod_{\alpha=1}^M \left( {1 \over 2 \pi} \,
\int_{-\infty}^\infty d k_\alpha \right) \,
\exp\left[ -{1 \over 2} \sum_{\alpha,\beta} k_\alpha\, U_{\alpha\beta}\, k_\beta
+ i \sum_{\alpha,\beta} k_\alpha\,
\delta v_\alpha^S \right]
\, \det U \, ,
\end{equation}
where
\begin{equation}
\delta v_\beta^S  \equiv v_\beta^S - v^{(0)}_\beta \, ,
\end{equation}
with
\begin{equation}
v_\alpha^{(0)} = 
\sum_{i=1}^N {y_i^{(0)} \, X_\alpha(x_i) \over \sigma_i^2} \, .
\end{equation}
We do
the $k$-integrals by working in the basis in which $U$ is diagonal.
The result is
\begin{equation}
P(\vec{a}^S) = {\left( \det U \right)^{1/2} \over (2\pi)^{M/2}} \, 
\exp\left[-{1 \over 2}\, \sum_{\alpha,\beta} \delta v_\alpha^S
\left(U^{-1}\right)_{\alpha\beta}
\delta v_\beta^S \right]
\, .
\end{equation}
Using Eq.~\eqref{lin_eq} and the fact that $U$ is symmetric we get our 
final result
\begin{equation}
\boxed{
P(\vec{a}^S) = {\left( \det U \right)^{1/2} \over (2\pi)^{M/2}} \, 
\exp\left[-{1 \over 2}\, \sum_{\alpha,\beta} \delta a_\alpha^S\,
U_{\alpha\beta}\, \delta a_\beta^S \right]}
\, ,
\label{P_of_a}
\end{equation}
which is Eq.~\eqref{theorem}, including the normalization constant in front of
the exponential.

\section{The distribution of fitted parameters from repeated sets of measurements}
\label{sec:proof2}

In this section we derive the equation for the distribution of fitted
parameters determined in the hypothetical situation that
one has many actual data sets.  Projecting on to a single fitted parameter,
this corresponds to the upper panel in Fig.~\ref{Fig:distofa1}.

The exact value of the data is $y_i^\text{true} = 
\vec{a}^\text{true} \cdot \vec{X}(x_i)$, see Eq.~\eqref{general_lin},
and the distribution of the $y_i$
in an actual data set, which differs from $y_i^\text{true}$
because of noise, has a distribution, assumed Gaussian here, centered on
$y_i^\text{true}$ with standard deviation $\sigma_i$. Fitting each of these
real data sets, the probability distribution for the fitted parameters is given
by
%
\begin{equation}
P(\vec{a}) = \prod_{i=1}^N \left\{ {1 \over \sqrt{2 \pi} \sigma_i} \,
\int_{-\infty}^\infty \, d y_i\,
\exp\left[-{\left(y_i - \vec{a}^\text{true} \cdot \vec{X}(x_i)\right)^2 \over
2 \sigma_i^2}\right]\, \right\} \,
\prod_{\alpha=1}^M \delta \left( \sum_\beta U_{\alpha\beta} a_\beta -
v_\alpha\right) \, \det U \, ,
\label{dist_of_a2}
\end{equation}
see Eq.~\eqref{dist_of_a} for an explanation of the various factors.
Proceeding as in Appendix~\ref{sec:proof} we have
\begin{align}
P(\vec{a}) = \prod_{i=1}^N \left\{ {1 \over \sqrt{2 \pi} \sigma_i} \,
\int_{-\infty}^\infty \, d y_i\, \exp\left[-{\left(y_i - \vec{a}^\text{true}
\cdot \vec{X}(x_i)\right)^2 \over 2 \sigma_i^2}\right]\, \right\}
\times \qquad\qquad\qquad \\
\prod_{\alpha=1}^M \left( {1 \over 2 \pi} \, \int_{-\infty}^\infty d k_\alpha
\exp\left[i k_\alpha\left( \sum_\beta U_{\alpha\beta} a_\beta -
\sum_{i=1}^N {y_i\, X_\alpha(x_i) \over \sigma_i^2} \right)\right] \right) \,
\det U \, ,
\end{align}
and doing the $y$- integrals by completing the square gives
\begin{align}
P(\vec{a})&= \prod_{\alpha=1}^M \left( {1 \over 2 \pi} \,
\int_{-\infty}^\infty d k_\alpha \right) \times
\\
&\exp\left[ -{1 \over 2 \sigma_i^2} \, \left(\,
\left(\vec{k}\cdot\vec{X}(i)\right)^2 
+2 i \left(\vec{k} \cdot \vec{X}(x_i)\right) \, \left(\vec{a}^\text{true} \cdot
\vec{X}(x_i)\right) \right) \right] \times
\exp\left[i \sum_{\alpha,\beta} k_\alpha\, U_{\alpha\beta}\, a_\beta\right] \,
\det U \, .
\end{align}
Using Eq.~\eqref{Uab_general} we then get
\begin{equation}
P(\vec{a}) = \prod_{\alpha=1}^M \left( {1 \over 2 \pi} \,
\int_{-\infty}^\infty d k_\alpha \right) \,
\exp\left[ -{1 \over 2} \sum_{\alpha,\beta} k_\alpha\, U_{\alpha\beta}\, k_\beta
+ i \sum_{\alpha,\beta} k_\alpha\, U_{\alpha\beta}\, \delta a_\beta\right]
\, \det U \, ,
\end{equation}
where
\begin{equation}
\delta a_\beta  \equiv a_\beta - a^\text{true}_\beta \, ,
\end{equation}
and we used Eq.~\eqref{Uab_general}.
The $k$-integrals are done by working in the basis in which $U$ is diagonal.
The result is
\begin{equation}
\boxed{
P(\vec{a}) = {\left( \det U \right)^{1/2} \over (2\pi)^{M/2}} \, 
\exp\left[-{1 \over 2}\, \sum_{\alpha,\beta} \delta a_\alpha\,
U_{\alpha\beta}\, \delta a_\beta \right]}
\, .
\label{P_of_a2}
\end{equation}
In other words, the distribution of the fitted parameters obtained
from many sets of actual data, about the
\textit{true} value $\vec{a}^\text{true}$ is a Gaussian. Since we are
assuming a linear model, the matrix of coefficients $U_{\alpha\beta}$ is a
constant, and so the distribution in Eq.~\eqref{P_of_a2}
is the \textit{same} as in Eq.~\eqref{P_of_a}. Hence
\begin{quotation}
\noindent For a linear model with Gaussian noise,
the distribution of fitted parameters, obtained
from simulated data sets, relative to \textit{value from the one actual data set}, 
is the same as the distribution of parameters
from many actual data sets relative to \textit{the true value}, see
Fig.~\ref{Fig:distofa1}.
\end{quotation}
This result is also valid for a non-linear model
if the range of parameter values needed is sufficiently small 
that the model can be represented by an effective one. It is usually assumed to
be a reasonable approximation even if this condition is not fulfilled.


\section{Fitting correlated data}
\label{corr_data}

Consider $N$ data points $(x_i, y_i, \sigma_i), i = 1, 2, \cdots N$. 
Correlations among the $y$-values are described by a
matrix $C$ where
\begin{equation}
C_{ij} \equiv \langle \delta y_i  \delta y_j \rangle = 
\langle y_i y_j \rangle  - \langle y_i \rangle \, \langle y_j \rangle \, .
\end{equation}
In this section we assume a linear model for ease of notation,
but the generalization to a
non-linear model is straightforward.
Assuming Gaussian noise,
the probability distribution for
the data which gives these correlations is
\begin{equation}
P(\{y\}) = {1 \over (2 \pi)^{N/2} ( \det C)^{1/2}} \, \exp\left[-{1 \over 2}\sum_{i, j}
\left(y_i - \sum_\alpha a_\alpha X_\alpha(x_i)\right) \,
\left( C^{-1}\right)_{ij} \, 
\left(y_j - \sum_\beta a_\beta X_\beta(x_j)\right) \right] \, ,
\label{Py}
\end{equation}
where we have used the following properties of Gaussian integrals:
\begin{align}
\label{eq1}
\int_{-\infty}^\infty dy_1 \int_{-\infty}^\infty dy_2 \cdots \int_{-\infty}^\infty dy_N 
\exp\left[-{1\over 2} \sum_{i, j} y_i A_{ij} y_j \right] &=
{ \left(2 \pi\right)^{N/2} \over (\det A)^{1/2}} \, , \\
{\displaystyle  \int_{-\infty}^\infty dy_1 \int_{-\infty}^\infty dy_2 \cdots \int_{-\infty}^\infty dy_N 
\, y_k y_l \, \exp\left[-{1\over 2} \sum_{i, j} y_i A_{ij} y_j \right]  \over
\displaystyle \int_{-\infty}^\infty dy_1 \int_{-\infty}^\infty dy_2 \cdots \int_{-\infty}^\infty dy_N 
\exp\left[-{1\over 2} \sum_{i, j} y_i A_{ij} y_j \right]} &=
\left(A^{-1}\right)_{kl} \, ,
\label{eq2}
\end{align}
in which we assumed that the matrix $A$ is positive definite, i.e.~all its
eigenvectors are positive. Equations \eqref{eq1} and \eqref{eq2} are obtained by
doing a change of variables in the $N$-dimensional space of the
$y$-s to new variables which are in the direction of the
eigenvectors of $A$. 
The Jacobian of the transformation is unity
because it is the determinant of the (orthogonal) matrix, $U$, which the
diagonalizes the real, symmetric matrix $A$.
The integrals are now independent and can be easily performed. In
Eq.~\eqref{eq1} the product of the eigenvalues has been written as the
determinant. In Eq.~\eqref{eq2} we have noted that if $A = U D U^{-1}$, where
$D$ is the diagonal matrix with eigenvalues on the diagonal, then $A^{-1} = U
D^{-1} U^{-1}$ (and, since $U$ is orthogonal, $U^{-1} = U^T$, the transpose).

Since least-squares is equivalent to maximum likelihood we have to maximize
the probability in Eq.~\eqref{Py}. This is equivalent to minimizing a
``\textit{cost function}'' which is minus (two times) the exponent in
Eq.~\eqref{Py},
i.e.
\begin{equation}
\boxed{
\sum_{i, j}
\left(y_i - \sum_\alpha a_\alpha X_\alpha(x_i)\right) \,
\left( C^{-1}\right)_{ij} \, 
\left(y_j - \sum_\beta a_\beta X_\beta(x_j)\right) \, .}
\label{min_gen}
\end{equation}
In the absence of correlations $C_{ij} = \sigma_i^2\, \delta_{ij}, 
\left(C^{-1}\right)_{ij} = \delta_{ij} / \sigma_i^2$ and we recover the earlier
expression for $\chi^2$ in Eq.~\eqref{chisq_gen}. Minimizing
Eq.~\eqref{min_gen}
with respect to
the $a_\alpha$ we get equations of the same form as before, namely
\begin{equation}
\sum_{\beta=1}^M U_{\alpha\beta} \, a_\alpha = v_\beta\, ,
\end{equation}
but with different expressions for $U$ and $v$, namely
\begin{subequations}
\begin{align}
&\boxed{
U_{\alpha\beta} = \sum_{i, j} X_\alpha(x_i) \left(C^{-1}\right)_{ij}
X_\beta(x_j),} \\
&\boxed{
v_{\alpha}  
= \sum_{i, j} X_\alpha(x_i) \left(C^{-1}\right)_{ij} y_j \, .}
\end{align}
\label{Uv_corr}
\end{subequations}
rather
than Eqs.~\eqref{Uab_general} and \eqref{v_general}. The covariance
matrix of the parameters is still given by Eq.~\eqref{Covab}.

In practice, though, we rarely have enough information on the correlations
between data points for Eqs.~\eqref{min_gen}--\eqref{Uv_corr} to be useful. 

\renewcommand\baselinestretch{0.97}
\section{Scripts for some data analysis and fitting tasks}
\label{sec:scripts}

In this appendix I give sample scripts using perl, python and gnuplot for
some basic data analysis and fitting tasks. I include output from the scripts
when acting on certain datasets which are available on the web.

Note ``\texttt{this\_file\_name}'' refers to the name of the script being
displayed (whatever you choose to call it.)

\subsection{Scripts for a jackknife analysis}

The script reads in values of $x$ on successive lines of the input file and
computes $\langle x^4\rangle / \langle x^2\rangle^2$, including an error bar
computed using the jackknife method.
\subsubsection{Perl}
\begin{verbatim}
#!/usr/bin/perl
#
#  Usage: "this_file_name data_file"
#  (make the script executable; otherwise you have to preface the command with "perl")
#
$n = 0;
$x2_tot = 0; $x4_tot = 0;
#
# read in the data
#
while(<>)   # Note this very convenient perl command which reads each line of
            # of each input file in the command line
{
    @line = split;
    $x2[$n] = $line[0]**2;
    $x4[$n] = $x2[$n]**2;
    $x2_tot += $x2[$n];
    $x4_tot+= $x4[$n];
    $n++;
}
#
# Do the jackknife estimates
#
for ($i = 0; $i < $n; $i++)
{
    $x2_jack[$i] = ($x2_tot - $x2[$i]) / ($n - 1);
    $x4_jack[$i] = ($x4_tot - $x4[$i]) / ($n - 1);
}
$x2_av = $x2_tot / $n;    # Do the overall averages
$x4_av = $x4_tot / $n;
$g_av = $x4_av / $x2_av**2;

$g_jack_av = 0; $g_jack_err = 0; # Do the final jackknife estimate
for ($i = 0; $i < $n; $i++)
{
    $dg = $x4_jack[$i] / $x2_jack[$i]**2;
    $g_jack_av += $dg;
    $g_jack_err += $dg**2;
}
$g_jack_av /= $n;
$g_jack_err /= $n;
$g_jack_err = sqrt(($n - 1) * abs($g_jack_err - $g_jack_av**2));

printf " Overall average is   %8.4f\n", $g_av;
printf " Jackknife average is %8.4f +/- %6.4f \n", $g_jack_av, $g_jack_err;
\end{verbatim}
Executing this file on the data in 
\href{http://young.physics.ucsc.edu/bad-honnef/data.HW2}
{\texttt{\underline{http://young.physics.ucsc.edu/bad-honnef/data.HW2}}}
gives
\begin{verbatim}
 Overall average is     1.8215
 Jackknife average is   1.8215 +/- 0.0368 
\end{verbatim}

\subsubsection{Python}
\begin{verbatim}
#
# Program written by Matt Wittmann
#
# Usage: "python this_file_name data_file"
#
import fileinput
from math import *

x2 = []; x2_tot = 0.
x4 = []; x4_tot = 0.
for line in fileinput.input():  # read in each line in each input file.
                                # similar to perl's while(<>)
    line = line.split()
    x2_i = float(line[0])**2
    x4_i = x2_i**2
    x2.append(x2_i)             # put x2_i as the i-th element in an array x2
    x4.append(x4_i)
    x2_tot += x2_i
    x4_tot += x4_i
n = len(x2)                     # the number of lines read in
#
# Do the jackknife estimates
#
x2_jack = []
x4_jack = []
for i in xrange(n):
    x2_jack.append((x2_tot - x2[i]) / (n - 1))
    x4_jack.append((x4_tot - x4[i]) / (n - 1))

x2_av = x2_tot / n              # do the overall averages
x4_av = x4_tot / n
g_av = x4_av / x2_av**2
g_jack_av = 0.; g_jack_err = 0.

for i in xrange(n):             # do the final jackknife averages
    dg = x4_jack[i] / x2_jack[i]**2
    g_jack_av += dg
    g_jack_err += dg**2

g_jack_av /= n
g_jack_err /= n
g_jack_err = sqrt((n - 1) * abs(g_jack_err - g_jack_av**2))

print " Overall average is   %8.4f" % g_av
print " Jackknife average is %8.4f +/- %6.4f" % (g_jack_av, g_jack_err)
\end{verbatim}
The output is the same as for the perl script.

\subsection{Scripts for a straight-line fit}
\label{sec:lin}
\subsubsection{Perl, writing out the formulae by hand}
\begin{verbatim}
#!/usr/bin/perl
#
#  Usage: "this_file_name data_file"
#  (make the script executable; otherwise preface the command with "perl")
#
#  Does a straight line fit to data in "data_file" each line of which contains
#  data for one point, x_i, y_i, sigma_i
#
$n = 0;
while(<>)    # read in the lines of data
{
    @line = split;  # split the line to get x_i, y_i, sigma_i
    $x[$n] = $line[0]; 
    $y[$n] = $line[1]; 
    $err[$n] = $line[2]; 
    $err2 = $err[$n]**2;    # compute the necessary sums over the data
    $s += 1 / $err2;
    $sumx  += $x[$n] / $err2 ;
    $sumy  += $y[$n] / $err2 ;
    $sumxx += $x[$n]*$x[$n] / $err2 ;
    $sumxy += $x[$n]*$y[$n] / $err2 ;
    $n++;
}

$delta = $s * $sumxx - $sumx * $sumx ;  # compute the slope and intercept
$c = ($sumy * $sumxx - $sumx * $sumxy) / $delta ;
$m = ($s  *  $sumxy - $sumx * $sumy) / $delta ;
$errm = sqrt($s / $delta) ;
$errc = sqrt($sumxx / $delta) ;

printf ("slope     = %10.4f +/- %7.4f \n", $m, $errm); # print the results
printf ("intercept = %10.4f +/- %7.4f \n\n", $c, $errc);

$NDF = $n - 2; # the no. of degrees of freedom is n - no. of fit params
$chisq = 0;    # compute the chi-squared
for ($i = 0; $i < $n; $i++)
{
    $chisq += (($y[$i] - $m*$x[$i] - $c)/$err[$i])**2;
}
$chisq /= $NDF; 
printf ("chi squared / NDF = %7.4lf \n", $chisq);
\end{verbatim}
Acting with this script on the data in
\href{http://young.physics.ucsc.edu/bad-honnef/data.HW3}
{\texttt{\underline{http://young.physics.ucsc.edu/bad-honnef/data.HW3}}}
gives
\begin{verbatim}
slope     =     5.0022 +/-  0.0024 
intercept =     0.9046 +/-  0.2839 

chi squared / NDF =  1.0400
\end{verbatim}

\subsubsection{Python, writing out the formulae by hand}
\begin{verbatim}
#
# Program written by Matt Wittmann
#
# Usage: "python this_file_name data_file"
#
# Does a straight-line fit to data in "data_file", each line of which contains
# the data for one point, x_i, y_i, sigma_i
#
import fileinput
from math import *

x = []
y = []
err = []
s = sumx = sumy = sumxx = sumxy = 0.

for line in fileinput.input():   # read in the data, one line at a time
    line = line.split()          # split the line
    x_i = float(line[0]);   x.append(x_i)
    y_i = float(line[1]);   y.append(y_i)
    err_i = float(line[2]); err.append(err_i)
    err2 = err_i**2
    s += 1 / err2               # do the necessary sums over data points
    sumx    += x_i / err2
    sumy    += y_i / err2
    sumxx   += x_i*x_i / err2
    sumxy   += x_i*y_i / err2

n = len(x)                      # n is the number of data points
delta = s * sumxx - sumx * sumx # compute the slope and intercept
c = (sumy * sumxx - sumx * sumxy) / delta
m = (s * sumxy - sumx * sumy) / delta
errm = sqrt(s / delta)
errc = sqrt(sumxx / delta)

print "slope     = %10.4f +/- %7.4f " % (m, errm)
print "intercept = %10.4f +/- %7.4f \n" % (c, errc)

NDF = n - 2                    # the number of degrees of freedom is n - 2
chisq = 0.

for i in xrange(n):            # compute chi-squared
    chisq += ((y[i] - m*x[i] - c)/err[i])**2;

chisq /= NDF
print "chi squared / NDF = %7.4lf " % chisq
\end{verbatim}
The results are identical to those from the perl script.
\subsubsection{Python, using a built-in routine from scipy}
\begin{verbatim}
#
# Python program written by Matt Wittmann
#
# Usage: "python this_file_name data_file"
#
# Does a straight-line fit to data in "data_file", each line of which contains
# the data for one point, x_i, y_i, sigma_i. 
#
# Uses the built-in routine "curve_fit" in the scipy package. Note that this
# requires the error bars to be corrected, as with gnuplot
#
from pylab import *
from scipy.optimize import curve_fit

fname = sys.argv[1] if len(sys.argv) > 1 else 'data.txt'
x, y, err = np.loadtxt(fname, unpack=True)  # read in the data
n = len(x)

p0 = [5., 0.1]  # initial values of parameters
f = lambda x, c, m: c + m*x                 # define the function to be fitted
                                            # note python's lambda notation
p, covm = curve_fit(f, x, y, p0, err)       # do the fit
c, m = p
chisq = sum(((f(x, c, m) - y)/err)**2)      # compute the chi-squared
chisq /= n - 2                              # divide by no.of DOF
errc, errm = sqrt(diag(covm)/chisq)         # correct the error bars

print "slope     = %10.4f +/- %7.4f " % (m, errm)
print "intercept = %10.4f +/- %7.4f \n" % (c, errc)
print "chi squared / NDF = %7.4lf " % chisq
\end{verbatim}
The results are identical to those from the above scripts.
\subsubsection{Gnuplot}
\begin{verbatim}
#
# Gnuplot script to plot points, do a straight-line fit, and display the
# points, fit, fit parameters, error bars, chi-squared per degree of freedom,
# and goodness of fit parameter on the plot.
#
# Usage: "gnuplot this_file_name"
#
# The data is assumed to be a file "data.HW3", each line containing
# information for one point (x_i, y_i, sigma_i). The script produces a
# postscript file, called here "HW3b.eps".
#
set size 1.0, 0.6
set terminal postscript portrait enhanced font 'Helvetica,16'
set output "HW3b.eps"
set fit errorvariables  # needed to be able to print error bars
f(x) = a + b * x        # the fitting function
fit f(x) "data.HW3" using 1:2:3 via a, b # do the fit
set xlabel "x"
set ylabel "y"
ndf = FIT_NDF               # Number of degrees of freedom
chisq = FIT_STDFIT**2 * ndf # chi-squared
Q = 1 - igamma(0.5 * ndf, 0.5 * chisq) # the quality of fit parameter Q
#
#  Below note how the error bars are (a) corrected by dividing by
#  FIT_STDFIT, and (b) are displayed on the plot, in addition to the fit
#  parameters, neatly formatted using sprintf.
#
set label sprintf("a = %7.4f +/- %7.4f", a, a_err/FIT_STDFIT) at 100, 400
set label sprintf("b = %7.4f +/- %7.4f", b, b_err/FIT_STDFIT) at 100, 330
set label sprintf("{/Symbol c}^2 = %6.2f", chisq) at 100, 270
set label sprintf("{/Symbol c}^2/NDF = %6.4f", FIT_STDFIT**2) at 100, 200
set label sprintf("Q = %9.2e", Q) at 100, 130
plot \                    # Plot the data and fit
"data.HW3" using 1:2:3 every 5 with errorbars notitle pt 6 lc rgb "red" lw 2, \
f(x) notitle lc rgb "blue" lw 4 lt 1
\end{verbatim}
The plot shows the result of acting with this gnuplot script on
the data in
\href{http://young.physics.ucsc.edu/bad-honnef/data.HW3}
{\texttt{\underline{http://young.physics.ucsc.edu/bad-honnef/data.HW3}}}.
The results agree with those of the other scripts.
\begin{center}
\begin{figure}
\includegraphics[width=11cm]{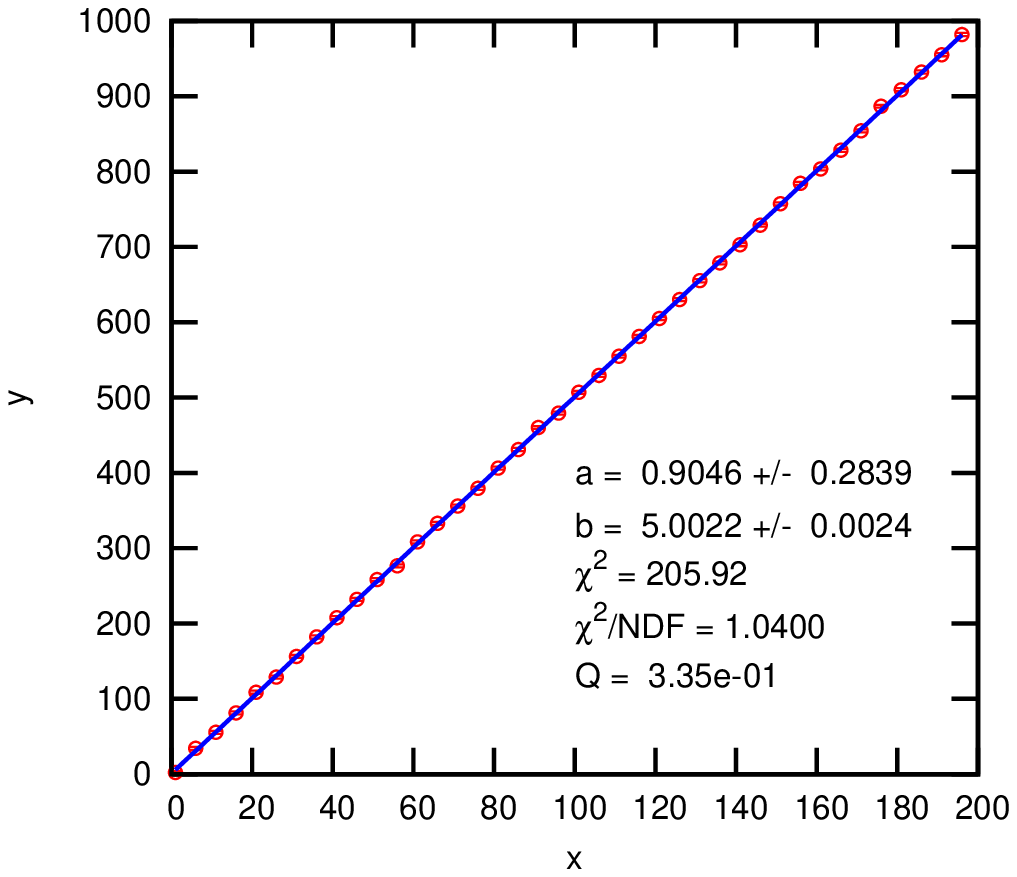}
\caption{Plot showing the data used and the resulting fit to a linear model
discussed in Sec.~\ref{sec:lin}.}
\end{figure}
\end{center}

\subsection{Scripts for a fit to a non-linear model}
\label{sec:nonlin}
We read in lines of data each of which contains three entries $x_i, y_i$ and
$\sigma_i$.  These are fitted to the form
\begin{equation}
y = T_c + A / x^\omega \, ,
\end{equation}
to determine the best values of $T_c, A$ and $\omega$.

\subsubsection{Python}
\begin{verbatim}
#
# Python program written by Matt Wittmann
#
# Usage: "python this_file_name data_file"
#
# Does a fit to the non-linear model
#
# y = Tc + A / x**w
#
# to the data in "data_file", each line of which contains the data for one point,
# x_i, y_i, sigma_i. 
#
# Uses the built-in routine "curve_fit" in the scipy package. Note that this
# requires the error bars to be corrected, as with gnuplot
#
from pylab import *
from scipy.optimize import curve_fit
from scipy.stats import chi2

fname = sys.argv[1] if len(sys.argv) > 1 else 'data.txt'
x, y, err = np.loadtxt(fname, unpack=True)  # read in the data
n = len(x)                                  # the number of data points

p0 = [-0.25, 0.2, 2.8]                      # initial values of parameters
f = lambda x, Tc, w, A: Tc + A/x**w         # define the function to be fitted
                                            # note python's lambda notation
p, covm = curve_fit(f, x, y, p0, err)       # do the fit
Tc, w, A = p
chisq = sum(((f(x, Tc, w, A) - y)/err)**2)  # compute the chi-squared
ndf = n -len(p)                             # no. of degrees of freedom
Q = 1. - chi2.cdf(chisq, ndf)               # compute the quality of fit parameter Q
chisq = chisq / ndf                         # compute chi-squared per DOF
Tcerr, werr, Aerr = sqrt(diag(covm)/chisq)  # correct the error bars

print 'Tc = %10.4f +/- %7.4f' % (Tc, Tcerr)
print 'A  = %10.4f +/- %7.4f' % (A, Aerr)
print 'w  = %10.4f +/- %7.4f' % (w, werr)
print 'chi squared / NDF = %7.4lf' % chisq
print 'Q  = %10.4f' % Q
\end{verbatim}
When applied to the data in
\href{http://young.physics.ucsc.edu/bad-honnef/data.HW4}
{\texttt{\underline{http://young.physics.ucsc.edu/bad-honnef/data.HW4}}}
the output is
\begin{verbatim}
Tc =    -0.2570 +/-  1.4775
A  =     2.7878 +/-  0.8250
w  =     0.2060 +/-  0.3508
chi squared / NDF =  0.2541
Q  =     0.9073
\end{verbatim}

\subsubsection{Gnuplot}
\begin{verbatim}
#
# Gnuplot script to plot points, do a fit to a non-linear model
# 
# y = Tc + A / x**w
#
# with respect to Tc, A and w, and display the points, fit, fit parameters,
# error bars, chi-squared per degree of freedom, and goodness of fit parameter
# on the plot.
#
# Here the data is assumed to be a file "data.HW4", each line containing
# information for one point (x_i, y_i, sigma_i). The script produces a
# postscript file, called here "HW4a.eps".
#
set size 1.0, 0.6
set terminal postscript portrait enhanced
set output "HW4a.eps"
set fit errorvariables       # needed to be able to print error bars
f(x) = Tc + A / x**w         # the fitting function
set xlabel "1/x^{/Symbol w}"
set ylabel "y"
set label "y = T_c + A / x^{/Symbol w}" at 0.1, 0.7
Tc = 0.3                      # need to specify initial values
A = 1
w = 0.2
fit f(x) "data.HW4" using 1:2:3 via Tc, A, w # do the fit
set xrange [0.07:0.38]
g(x) = Tc + A * x
h(x) = 0 + 0 * x
ndf = FIT_NDF                # Number of degrees of freedom
chisq = FIT_STDFIT**2 * ndf  # chi-squared
Q = 1 - igamma(0.5 * ndf, 0.5 * chisq) # the quality of fit parameter Q
#
#  Below note how the error bars are (a) corrected by dividing by
#  FIT_STDFIT, and (b) are displayed on the plot, in addition to the fit
#  parameters, neatly formatted using sprintf.
#
set label sprintf("T_c = %5.3f +/- %5.3f",Tc, Tc_err/FIT_STDFIT) at 0.25, 0.33
set label sprintf("{/Symbol w} = %5.3f +/- %5.3f",w, w_err/FIT_STDFIT) at 0.25, 0.27
set label sprintf("A = %5.2f +/- %5.2f",A, A_err/FIT_STDFIT) at 0.25, 0.21
set label sprintf("{/Symbol c}^2 = %5.2f", chisq) at 0.25, 0.15
set label sprintf("{/Symbol c}^2/NDF = %5.2f", FIT_STDFIT**2) at 0.25, 0.09
set label sprintf("Q = %5.2f", Q) at 0.25, 0.03
#
# Plot the data and the fit
#
plot "data.HW4" using (1/$1**w):2:3 with errorbars notitle lc rgb "red" lw 3 pt 8 ps 1.5, \
g(x) notitle lc rgb "blue" lw 3 lt 2 , \
h(x) notitle lt 3 lw 4
\end{verbatim}
The plot shows the result of acting with this gnuplot script on 
the data at 
\href{http://young.physics.ucsc.edu/bad-honnef/data.HW4}
{\texttt{\underline{http://young.physics.ucsc.edu/bad-honnef/data.HW4}}}
The results agree with those of the python script above.
\begin{center}
\begin{figure}
\includegraphics[width=11cm]{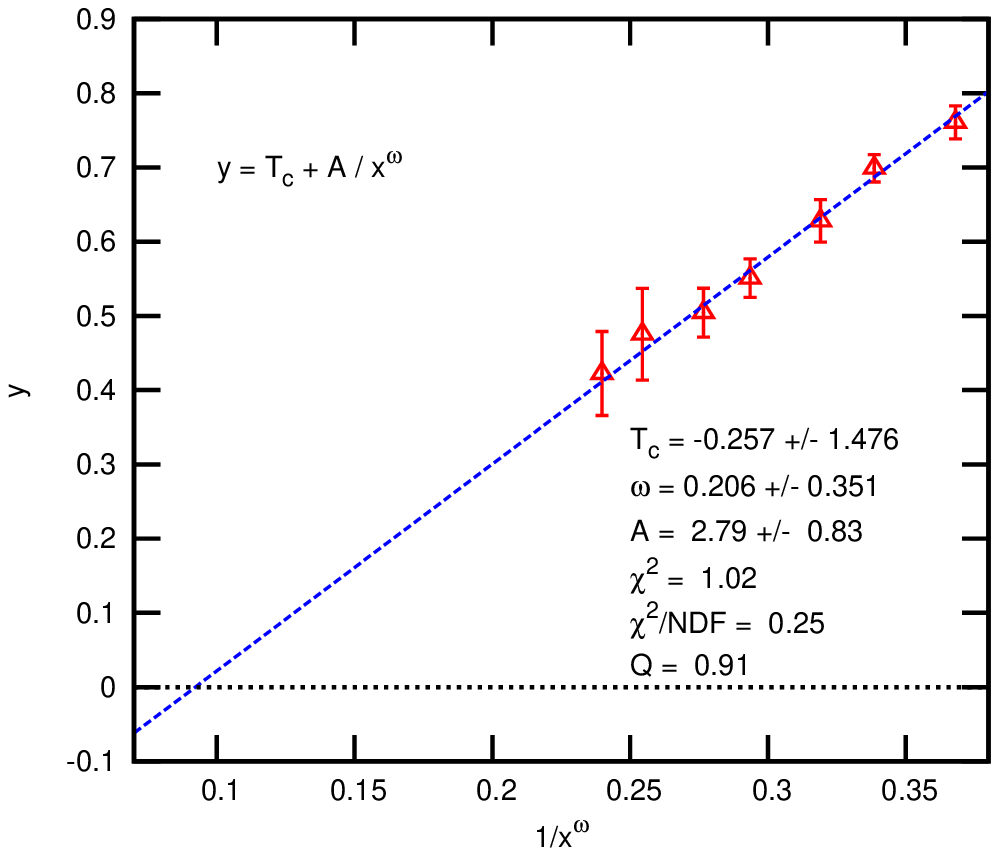}
\caption{Plot showing the data used and the resulting fit to a non-linear model
discussed in Sec.~\ref{sec:nonlin}.}
\end{figure}
\end{center}
The quoted error bars in $T_c$ are clearly ridiculous and arise because the
code gives symmetric error bars whereas the variation of $\chi^2$ about the
minimum is very asymmetric, as sketched in the right panel of
Fig.~\ref{fig:chi2}. It would be better to get asymmetric error bars for $T_c$
by determining $\chi^2$ as a function of $T_c$, while optimizing with
respect to the other
parameters, and then estimating the
values of $T_c$ where $\Delta
\chi^2 = 1$, see Fig.~\ref{fig:chi2} and the discussion in
Sec.~\ref{sec:conf_limits}.
The interested student is invited to do this. Even better would
be to do the bootstrap analysis discussed in Sec.~\ref{sec:resample}
but this requires the raw data, that is to say the $N_i$
$y$-values for each data point $i$ which, when averaged, give
the results for $y_i$ and $\sigma_i$ used in the fit. Unfortunately the
raw data is not available in this case.

\acknowledgments
I'm grateful to Alexander Hartmann for inviting me to give a lecture at the
Bad Honnef School on ``Efficient Algorithms in Computational Physics'' in
September 2012, which provided the motivation to write up a first version of
these notes, and also for a several very helpful discussions.

In addition I would like to thank Matt Wittmann for helpful discussions about fitting
and data analysis using \texttt{python} and for permission to include his
python codes. I am also grateful to Wittmann and Christoph Norrenbrock for
helpful comments on an earlier version of the manuscript. This work is
supported by the Humboldt Foundation, and by the NSF through grant DMR-1207036.

\newpage
\bibliography{refs}

\end{document}